\documentclass[a4paper,11pt]{article}
\pdfoutput=1
\usepackage{jcappub}
\usepackage[T1]{fontenc}

\newcommand{\be}{\begin{equation}}
\newcommand{\ee}{\end{equation}}
\newcommand{\bea}{\begin{eqnarray}}
\newcommand{\eea}{\end{eqnarray}}
\newcommand{\bk}{{\mathbf k}}
\newcommand{\bp}{{\mathbf p}}
\newcommand{\GW}{{_{\rm GW}}}

\title{\boldmath Gravitational wave production from preheating: parameter dependence}

\author[a]{Daniel G. Figueroa\,}
\author[b]{, Francisco Torrent\'i\,}

\affiliation[a]{\,CERN, Theory Division, 1211 Geneva, Switzerland}
\affiliation[b]{\,Instituto de F\'isica Te\'orica IFT-UAM/CSIC, Universidad Aut\'onoma de Madrid, Cantoblanco 28049 Madrid, Spain.}

\emailAdd{daniel.figueroa@cern.ch}
\emailAdd{f.torrenti@csic.es}

\abstract{Parametric resonance is among the most efficient phenomena generating gravitational waves (GWs) in the early Universe. The dynamics of parametric resonance, and hence of the GWs, depend exclusively on the resonance parameter $q$. The latter is determined by the properties of each scenario: the initial amplitude and potential curvature of the oscillating field, and its coupling to other species. Previous works have only studied the GW production for fixed value(s) of $q$. We present an analytical derivation of the GW amplitude dependence on $q$, valid for any scenario, which we confront against numerical results. By running lattice simulations in an expanding grid, we study for a wide range of $q$ values, the production of GWs in post-inflationary preheating scenarios driven by parametric resonance. We present simple fits for the final amplitude and position of the local maxima in the GW spectrum. Our parametrization allows to predict the location and amplitude of the GW background today, for an arbitrary $q$. The GW signal can be rather large, as $h^2\Omega_{\rm GW}(f_p) \lesssim 10^{-11}$, but it is always peaked at high frequencies $f_p \gtrsim 10^{7}$ Hz. We also discuss the case of spectator-field scenarios, where the oscillatory field can be e.g.~a curvaton, or the Standard Model Higgs.}

\begin{document}
\maketitle
\flushbottom

\section{Introduction}

A phase of quasi-exponential expansion, inflation, represents the leading framework to explain the initial conditions of the Universe~\cite{Ade:2015lrj}. The inflationary period is typically parametrized in terms of a scalar field, the {\it inflaton}. In the simplest approach, the inflaton sustains a slow-roll regime, in which its potential energy dominates over its kinetic energy. As a consequence of this, the Universe inflates. The inflaton slowly accelerates as it rolls down its potential, and when its kinetic energy eventually becomes significant, the slow-roll regime terminates. Following inflation, the {\it reheating} stage converts the energy available into different particle species; an event that represents the origin of (most of) the matter in the Universe. When the created particles eventually 'thermalize' and dominate the total energy budget, the Universe enters into the standard radiation era. The reheating process represents, in a way, the 'Bang' of the {\it hot Big Bang} paradigm.

If the inflaton potential exhibits a monomial shape at the stages following immediately after inflation, the inflaton starts oscillating around the minimum of its potential, typically with a large amplitude. These oscillations can give rise to parametric resonance~\cite{Traschen:1990sw,Kofman:1994rk}, a phenomenon by which particle species coupled to the inflaton, are created in energetic bursts. In the case of bosonic species, the production of particles is resonant, and the energy transferred grows exponentially within few inflaton oscillations~\cite{Traschen:1990sw,Kofman:1994rk,Shtanov:1994ce,Kaiser:1995fb,Kofman:1997yn,Greene:1997fu,Kaiser:1997mp,Kaiser:1997hg}. In the case of fermions, Pauli blocking prevents resonance from developing, though there is also a significant transfer of energy~\cite{Greene:1998nh,Greene:2000ew,Peloso:2000hy,Berges:2010zv}. The production of particle species in this way after inflation -- either of fermions or bosons --, represent what is meant by a 'preheating' stage. For recent reviews on parametric resonance and preheating mechanisms in general, see~\cite{Allahverdi:2010xz,Amin:2014eta}.

Inflationary preheating is not the only context where parametric resonance is developed in the early Universe. If there is a light spectator field present during inflation, it naturally develops a large amplitude during inflation thanks to its quantum fluctuations. Thus, some time after inflation, the spectator field eventually starts oscillating around the minimum of its potential, potentially developing parametric resonance of other field species coupled to it. This is the case e.g.~of the curvaton~\cite{Enqvist:2001zp,Lyth:2001nq,Moroi:2001ct,Mazumdar:2010sa,Enqvist:2008be, Enqvist:2012tc, Enqvist:2013qba, Enqvist:2013gwf}, or of the Standard Model Higgs\footnote{Note that this differs from the case of Higgs-Inflation~\cite{Bezrukov:2007ep,Bezrukov:2010jz}, where the post-inflationary decay of the SM Higgs via parametric resonance~\cite{Bezrukov:2008ut,GarciaBellido:2008ab,Figueroa:2009jw,Figueroa:2014aya}, belongs to the context of preheating, as the Higgs rather plays there the role of the inflaton, instead of a spectator field.}~\cite{DeSimone:2012qr,Enqvist:2013kaa,Enqvist:2014tta,Herranen:2015ima,Figueroa:2014aya,Figueroa:2015rqa,Enqvist:2015sua,Figueroa:2016dsc}. 

Independently of the context, we will often refer to the oscillatory field as the 'mother' field, and to the created species as the 'daughter' fields. Particle production of daughter fields via parametric resonance, corresponds in fact, to a non-perturbative, non-linear, and out-of-equilibrium phenomenon. Due to this, the violent excitation of field species via parametric resonance is expected to produce large scalar metric perturbations~\cite{Bassett:1998wg,Bassett:1999mt,Bassett:1999ta,Finelli:2000ya,Chambers:2007se,Bond:2009xx}, and a significant amount of gravitational waves~\cite{Khlebnikov:1997di,Easther:2006gt,Easther:2006vd,GarciaBellido:2007dg,GarciaBellido:2007af,Dufaux:2007pt,Dufaux:2008dn,Figueroa:2011ye,Bethke:2013aba,Bethke:2013vca,Figueroa:2016ojl}. Our aim in this paper is precisely to parametrize the production of gravitational waves (GWs) from parametric resonance in the early Universe\footnote{Note that we do not consider the case of 'oscillons', which correspond to stable field configurations formed whenever a field oscillates around the minimum of its potential, as long as the potential shape meets certain circumstances, see e.g.~\cite{Copeland:1995fq,Amin:2011hj}. For the GW production from oscillons see~\cite{Zhou:2013tsa,Antusch:2016con}.}. Our present work is actually a natural continuation of our previous work~\cite{Figueroa:2016wxr}, which we will refer to as {\it Paper I}. There we presented a study of the parameter-dependence of the mother and daughter fields' dynamics, in different preheating and spectator-field scenarios where parametric resonance is naturally expected to occur. In this paper we present an analysis of the parameter dependence of the GW production from parametric resonance. We focus on the paradigmatic cases of preheating after chaotic inflation models, though our results can be extended to other cases, as long as they exhibit a potential with a monomial shape during the stages following inflation. We also discuss, to a lesser extent, the case of parametric resonance from inflationary spectator fields. In particular we explain their inability to produce a large amount of GWs.  

We have characterized the GW production from parametric resonance during all its relevant stages, from the initial linear growth of the daughter field fluctuations, through the intermediate non-linear stage, till the relaxation towards a stationary distribution. We have parametrized the GW spectra by surveying the relevant circumstances and parameters in each case. Let us note that, even though GW production during inflationary preheating has been widely considered in the literature, there still lacks a systematic parametrization of the GW backgrounds produced as a function of the different parameters\footnote{There exists nonetheless a parameter-fit analysis of the GW production in Hybrid preheating~\cite{Dufaux:2008dn}, but this corresponds to a spinodal instability of the daughter field modes, not to parametric resonance.}. In this work we fill in this gap. We have used massively parallelized lattice simulations, obtaining simple fits to the most significant quantities, like the characteristic peak scales and associated amplitudes of the matter and GW spectra.

The paper is divided as follows. In Section~\ref{sec:Analytics} we first present an estimation of the amplitude of the GW background produced during parametric resonance, based on the analytic understanding of the linear stage of the daughter field(s) excitation. In Section~\ref{sec:lattice} we present our results from numerical lattice simulations of preheating, for a quartic inflaton potential in Section~\ref{sec:lphi4-results}, and for a quadratic inflaton potential in Section~\ref{sec:m2phi2-results}. In Section \ref{sec:SubFields} we present a discussion about models where the mother field represents only a sub-dominant energy component of the Universe, like in spectator field scenarios. In Section \ref{sec:Summary} we list together the most relevant fitted formulas obtained from our numerical analysis. In Section~\ref{sec:discussion} we summarize and discuss some implications of our results. In the appendix~\ref{app:Lattice-formulation} we present details of the lattice formulation we used. 

All through the paper we consider $\hbar = c = 1$ units, and represent the reduced Planck mass by {\small$m_p^2 = {1/{8\pi G}} \simeq 2.44\cdot10^{18}$ GeV}. We take a flat background with Friedmann-Lemaitre-Robertson-Walker (FLRW) metric {\small$ds^2 =  dt^2 - a^2 (t) dx^i dx^i$}, where {\small$a(t)$} is the scale factor, and {\small$t$} the cosmic time. 

\section{Gravitational waves from parametric resonance. Analytic estimation.}
\label{sec:Analytics}

We want to study the GWs created by the excitation of some field(s) undergoing parametric resonance. Specifically, we consider an (initially) homogeneous field $\phi$ oscillating around the minimum of its potential $V(\phi)$, coupled to another scalar field $X$ with coupling $g^2\phi^2 X^2$. The classical equations of motion (EOM) describing the dynamics of this system are
\be \ddot \phi - \frac{1}{a^2} \nabla^2\phi + 3 H \dot \phi + g^2  X^2 \phi  + \frac{\partial V(\phi)}{\partial \phi} = 0 \ , \hspace{0.5cm} \ddot X -  \frac{1}{a^2} \nabla^2 X + 3 H \dot X + g^2 \phi^2 X = 0  \label{eq:generic-eom} \ ,\ee
where $H \equiv {\dot{a}/a}$ is the Hubble rate. As $\phi$ is considered to be initially homogeneous, we can initially forget about the gradient term $a^{-2} \nabla^2\phi$ in the EOM. We can also forget initially about the backreaction of the field $X$ over the oscillations of $\phi$, since the excitation of $X$ (and hence the growth of its amplitude) builds up in time only after a certain number of $\phi$ oscillations. Hence, we ignore also the term $g^2  X^2 \phi$. For a monomial potential $V(\phi) \propto \phi^n$, the solution for $\phi$ under the previous circumstances, admits an oscillatory solution as~\cite{Turner:1983he} 
\begin{equation}\label{eq:ApproxVarPhi}
\phi(t) \approx \Phi(t)F(t)\,,
\end{equation}
with $\Phi(t) \equiv \Phi_{\rm i}(t/t_{\rm i})^{-2/n}$ a decreasing amplitude from some initial time $t_{\rm i}$, and $F(t)$ an oscillatory function. The details of $\Phi(t)$ and $F(t)$ depend of course on the specific choice of potential. For power-law potentials 
\be V(\phi) = {1\over n}\lambda M^{4-n}\phi^n \ , \label{eq:powerlaw-pot}\ee 
with $\lambda$ a dimensionless coefficient and $M$ some mass scale, $F(t)$ is in fact not periodic (except for $n = 2$). Yet the frequency of oscillations changes only relatively slowly in time as $\Omega_{\rm osc} \equiv \sqrt{d^2V/d\phi^2} = \sqrt{\lambda} M^{2-n/2}\Phi^{(n/2-1)} \equiv \omega_{*}(t/t_{\rm i})^{1-2/n}$, with $\omega_{*} \equiv \sqrt{\lambda} M^{2-n/2}\Phi_{\rm i}^{(n/2-1)}$. We can therefore use the initial field amplitude $\Phi_{\rm i}$ and angular frequency $\omega_{*}$, to define natural field and space-time variables as 
\begin{eqnarray}\label{eq:naturalVariables}
\vec{x} \rightarrow \vec{y} \equiv \omega_{*}\vec{x}\,,~~~~~~ t \rightarrow z \equiv \omega_{*}\tau\,,~~~~\tau \equiv \int {dt\over a(t)}\,,\\
\phi \rightarrow \varphi \equiv a(t){\phi\over\Phi_{\rm i}}\,,~~~~~~ X \rightarrow X_c \equiv a(t)X \,,\nonumber \hspace*{1.5cm}
\end{eqnarray}
which, with the exception of $X_c$, are all dimensionless. The EOM of the $X_c$ field reads
\be
{d^2 X_c \over dz^2} + \left(q \varphi^2 - \nabla_{\hspace*{-1mm}y}^2\right) X_c = {1\over a}{d^2a\over dz^2}X_c \label{eq:conformal-eom} \ ,\ee
where $q$ is the so called {\it resonance parameter}, defined like
\be\label{eq:ResParam}
q \equiv {g^2\Phi_{\rm i}^2\over \omega_{*}^2}\,.
\ee
Note that for certain potentials, the definition for the resonance parameter conventionally includes a numerical factor of $\sim \mathcal{O}(1)$ multiplying the dimensionless ratio in Eq.~(\ref{eq:ResParam}). This is the case, for example, of the quadratic potential $V(\phi) = m^2\phi^2/2$, with $m$ some mass scale. In this case, the resonance parameter is usually defined as $q \equiv {g^2\Phi_{\rm i}^2/4m^2}$, introducing the extra factor ${1/4}$ to match the corresponding definition in the {\it Mathieu} equation, see~\cite{Kofman:1997yn}. On the contrary, for a quartic potential $V(\phi) = \lambda\phi^4/4$, Eq.~(\ref{eq:ResParam}) gives $q = {g^2/\lambda}$, matching exactly the resonance parameter definition in the {\it Lam\'e} equation~\cite{Greene:1997fu}. Of course, this is purely conventional, and what really matters is just the dimensionless ratio $\propto {g^2(\Phi_{\rm i}/\omega_{*})^2}$ captured in Eq.~(\ref{eq:ResParam}).

In most of the relevant situations in the early Universe where parametric resonance takes place, the oscillatory field $\phi$ is considered to be initially a homogeneous classical field configuration, whereas the field $X_c$ is considered to be a quantum field, initially in vacuum. The scalar field $X_c$ can be promoted into a quantum operator by means of the standard quantization procedure
\begin{equation}
\label{eq:ChiQuant}
X_c (\mathbf{x},t) \equiv a(t)X(\mathbf{x},t) = \int\frac{d  \mathbf{k}}{\left(2\pi\right)^{3}}e^{-i\mathbf{k\cdot x}}\left[\hat{a}_{\mathbf{k}}X^{(c)}_{\mathbf{k}}(t)+\hat{a}_{-\mathbf{k}}^{\dagger}{X^{(c)}_{\mathbf{k}}}^{*}(t)\right],
\end{equation}
where the creation/annihilation operator satisfies the canonical commutation relations 
\begin{equation}\label{eq:commutation}
[\hat{a}_{\mathbf{k}},\hat{a}_{\mathbf{k'}}^{\dagger}]=\left(2\pi\right)^{3}\delta^{(3)}(\mathbf{k-k'}),
\end{equation}
with other commutators vanishing. The (initial) vacuum state is defined as usual as $\hat{a}_{\mathbf{k}}|0\rangle = 0$. Given our choice of Eqs.~(\ref{eq:ChiQuant}), (\ref{eq:commutation}), we note that the mode functions $X^{(c)}_{\mathbf{k}}$ have dimensions of (Energy)$^{-1/2}$. From Eq.~(\ref{eq:conformal-eom}) we obtain the EOM for the latter as
\begin{equation}\label{eq:ParamResEqFourier}
{d^2\over dz^2}X^{(c)}_{\mathbf{k}} + \left(\kappa^{2} + q \varphi^2\right)X^{(c)}_{\mathbf{k}} \simeq 0\,,~~~~~ \kappa \equiv {k\over \omega_{*}}\,,
\end{equation}
where we have discarded a term $\propto {1\over a}{d^2a\over dz^2}$ in this equation, as it is negligible at sub-horizon scales $\kappa^2 \gg {1\over a}({da\over dz})^2 \sim {1\over a}{d^2a\over dz^2}$. Given the oscillatory nature of $\varphi$, Eq.~(\ref{eq:ParamResEqFourier}) can exhibit unstable solutions of the type $X^{(c)}_{\mathbf{\kappa}} \sim e^{\mu_q({\kappa}) z}$, with $\mu_q({\kappa})$ some complex exponent. For certain values of $\lbrace q, \kappa\rbrace$, $\mathfrak{Re} [\mu_{\kappa}] > 0$, causing an exponential growth of the given field mode amplitude. It is precisely this unstable behavior, occurring only within finite-momenta 'resonance bands' with $\mathfrak{Re} [\mu_{\kappa}] > 0$, that we call parametric resonance. For a detailed description of the phenomena of parametric resonance in the early Universe, see~\cite{Amin:2014eta}; for a parameter-fit analysis based on numerical simulations, see {\it Paper I}~\cite{Figueroa:2016wxr}.

\subsection{Spectrum of gravitational waves}
\label{subsec:GWspectrum}

The exponential growth of the $X^{(c)}_{\mathbf{k}}$ modes experiencing parametric resonance, generates a significant anisotropic stress $\Pi_{ij} \sim \partial_i X_c \partial_j X_c$, which in turn creates GWs, as we will see next.  Gravitational waves correspond to the transverse and traceless (TT) degrees of freedom of metric perturbations,
\begin{equation}
ds^{2}= a^{2}(\tau)\left(-d\tau^{2} + \delta_{ij}+h_{ij}\right)dx^{i}dx^{j}\ ,
\end{equation} 
where $d\tau = dt/a(t)$ is the conformal time, and $h_{ij}$ verify the conditions $\partial_{i}h_{ij} = 0$ (transversality) and $h_{i}^{i} = 0$ (tracelessness). Linearizing the Einstein equations lays down the EOM for the generation and propagation of GWs in a FLRW background, 
\begin{eqnarray}
{{h}}_{ij}'' + 2\mathcal{H} {{h}}_{ij}' - \nabla^2 h_{ij} = {2\over m_p^2}\,\Pi_{ij}^{\rm TT} \ , \label{eq:GWeom}
\end{eqnarray}
where $' \equiv {d\over d\tau}$ represents derivatives with respect to conformal time, and we have defined $\mathcal{H} \equiv a'/a$ as the comoving Hubble rate. The source of GWs, $\Pi_{ij}^{\rm TT}$, is the TT-part of the anisotropic stress of the system, defined as
\begin{equation}
\label{StressTensorChi}
\Pi_{ij}^{{\rm TT}} \equiv \left\lbrace \partial_{i}X\,\partial_{j}X \right\rbrace^{{\rm TT}} = {1\over a^2}\left\lbrace \partial_{i}X_c \,\partial_{j}X_c \right\rbrace^{{\rm TT}}\ .
\end{equation}
The anisotropic stress should really be sourced by the gradients of all excited fields, including the mother field term $\partial_i \phi \partial_j \phi$. However, as mentioned above, we take the mother field as approximately homogeneous at initial times, so we ignore such term for the moment. The contribution of the mother field will be however included -- automatically -- in the lattice simulations that we will present in Section~\ref{sec:lattice}.

Obtaining the TT-part of a tensor in configuration space amounts to a non-local operation, so it is more convenient to work in Fourier space, where a geometrical TT-projection can be easily built. The EOM of the GWs in Fourier space reads
\begin{eqnarray}\label{eq:GWeq}
{h}_{ij}''(\bk,\tau)+2\mathcal{H}{h}'_{ij}(\mathbf{k},\tau) + k^2 h_{ij}(\mathbf{k},\tau) = {2\over m_p^2} \Pi_{lm}^{\rm TT } (\mathbf{k},\tau)\ .
\end{eqnarray}
In momentum space, the ${\rm TT}$ operation is defined as $\Pi_{lm}^{\rm TT}(\mathbf{k},\tau) \equiv \Lambda_{ij,lm}(\hat\bk)\Pi_{lm}(\mathbf{k},\tau)$, with $\Pi_{ij}(\mathbf{k},\tau)$ the Fourier transform of $\Pi_{ij}(\mathbf{x},\tau)$, and $\Lambda_{ij,lm}(\hat\bk)$ a projector defined as\footnote{Note that when we define the analogous TT-projector on a lattice grid for numerical simulations, this requires a different definition than that in Eq.~(\ref{eq:projector}) for the continuum, see Appendix~\ref{app:Lattice-formulation}, or Ref.~\cite{Figueroa:2011ye} for more details.}
\begin{eqnarray}\label{eq:projector}
\Lambda_{ij,lm}(\mathbf{\hat k}) \equiv P_{il}(\hat\bk)
P_{jm}(\hat\bk) - {1\over2} P_{ij}(\hat\bk)
P_{lm}(\hat\bk),\,~~~P_{ij} = \delta_{ij} - \hat k_i \hat k_j\,,~~~ \hat k_i = k_i/k \ .
\end{eqnarray}

The energy density spectrum of a stochastic (isotropic) background of GW (at subhorizon scales) takes the form
\be \frac{d \rho_{_{\rm GW}}}{d \log k } = \frac{k^3 m_p^2}{8 \pi^2 a^2} \mathcal{P}_{h'}(k,\tau) \label{eq:ThetaGW} \ , \ee
where $\langle h'(\mathbf{k},\tau) h^{*'}(\mathbf{k'},\tau)\rangle = (2 \pi)^3 \mathcal{P}_{h'} (\kappa,\tau) \delta^{(3)} (\mathbf{k} - \mathbf{k'})$. This can be written as an explicit function of the source matter fields as (see $e.g.$~\cite{Figueroa:2012kw}) 
\begin{equation}\label{eq:GW_spectra(Pi)}
\frac{d\rho_{\mathrm{\GW}}}{d\log k}\left(k,\tau\right) =  \frac{1}{4\pi^2a^4(\tau)}\,{k^3\over m_p^2}\int_{\tau_{i}}^{\tau}d\tau'\int_{\tau_{i}}^{\tau}d\tau''\,a(\tau')\,a(\tau'')\cos[k(\tau'-\tau'')]\,\Pi^2(k,\tau',\tau''),
\end{equation}
where $\Pi^2$ is the Unequal-Time-Correlator (UTC) of the source of $\Pi_{ij}^{\rm TT}$, defined as 
\begin{eqnarray}\label{eq:UTC}
\langle \text{0}| \Pi_{ij}^{{\rm TT}}(\mathbf{k},\tau)\Pi_{ij}^{{\rm TT}^{\hspace*{0.2mm}*}}\hspace*{-0.7mm}(\mathbf{k'},\tau') |0\rangle \equiv (2\pi)^3 \Pi^2(k,\tau,\tau') \delta^{(3)}(\bk-\bk)\,.
\end{eqnarray}
Substituting the quantized field Eq.~(\ref{eq:ChiQuant}) into Eq.~(\ref{StressTensorChi}), leads to the expression in Fourier space,
\begin{eqnarray}
&& \Pi_{ij}^{{\rm TT}}(\mathbf{k},\tau) = \\ 
&& {\Lambda_{ij,lm}(\hat{k})\over (2\pi)^{3}a^2(\tau)}\int\hspace*{-0.1cm}d\mathbf{p}~p_{l}p_{m}\left(\hat{a}_{\mathbf{p}}X^{(c)}_{\mathbf{p}}(\tau)+\hat{a}_{-\mathbf{p}}^{\dagger}{X^{(c)^{*}}_{\mathbf{p}}}(\tau)\right)\left(\hat{a}_{\mathbf{k-p}}X^{(c)}_{\mathbf{k-p}}(\tau)+\hat{a}_{-\mathbf{\left(k-p\right)}}^{\dagger}{X^{(c)^{*}}_{\mathbf{k-p}}}(\tau)\right) \ . \nonumber
\end{eqnarray}
The only combinations of creation/annihilation operators which contribute to the expectation value in Eq.~(\ref{eq:UTC}), turn out to be
\begin{equation}\label{eq:survivingCombinationBosons}
	\langle \text{0} |\hat{a}_{\mathbf{p}}\hat{a}_{\mathbf{k-p}}\hat{a}_{\mathbf{q}}^{\dagger}\hat{a}_{\mathbf{k'-q}}^{\dagger} |0\rangle =\left(2\pi\right)^{6}\big[\delta^{(3)}(\mathbf{k-p-q}) +\delta^{(3)}(\mathbf{p-q})\big]\delta^{(3)}(\mathbf{k-k'}),
\end{equation}
\begin{equation}\label{eq:survivingCombinationBosonsII}
\langle \text{0}|\hat{a}_{\mathbf{p}}\hat{a}_{\mathbf{-(k-p)}}^{\dagger}\hat{a}_{\mathbf{q}}\hat{a}_{-(\mathbf{k'-q})}^{\dagger}|0\rangle = (2\pi)^{6}\delta^{(3)}(\mathbf{k})\delta^{(3)}(\mathbf{k'}-\bk),
\end{equation}
where we have used the commutation rule Eq.~(\ref{eq:commutation}). Since the second term Eq.~(\ref{eq:survivingCombinationBosonsII}) can be re-written as proportional to $\delta^{(3)}(\mathbf{k})\delta^{(3)}(\mathbf{k'})$, it does not contribute to $\Pi^{2}(k,t,t')$ at finite momenta $k = k' \neq {0}$. Thus, only the term Eq.~(\ref{eq:survivingCombinationBosons}) contributes to the final expression of the UTC, which reads
\begin{equation}
\Pi^{2}(k,\tau,\tau') = \frac{1}{4\pi^{2}a^{2}(\tau)a^{2}(\tau')}\int dp\, d\theta\, p^{6}\sin^{5}\theta\,X^{(c)}_{\mathbf{p}}(\tau)X^{(c)}_{\mathbf{k-p}}(\tau){X^{(c)^{*}}_{\mathbf{k-p}}}(\tau'){X^{(c)^{*}}_{\mathbf{p}}}(\tau'),
\end{equation}
where we have used the result $\Lambda_{ij,lm}(\hat{k})\big(p_{i}(k-p)_{j}(k-p)_{l}p_{m}+p_{i}(k-p)_{j}p_{l}(k-p)_{m}\big)$ $= p^{4}\sin^{4}\theta$, with $\theta$ the angle between $\bp$ and $\bk$. The spectrum of GWs Eq.~(\ref{eq:GW_spectra(Pi)}) is finally given by
\begin{eqnarray}
\frac{d\rho_{_{\rm GW}}}{d\log k}\left(k,t\right) 
= \frac{Gk^{3}}{2\pi^{3}}\int dp\, d\theta\, p^{6}\sin^{5}\theta\,\left(\left|I_{(c)}(k,p,\theta,\tau)\right|^{2}+\,\left|I_{(s)}(k,p,\theta,\tau)\right|^{2}\right)  \ , \label{eq:specGW}
\end{eqnarray}
with
\begin{equation}\label{eq:IsIc}
I_{(c)} \equiv \int_{\tau_{i}}^{\tau}\frac{d\tau'}{a(\tau')}\cos(k\tau')X^{(c)}_{\mathbf{\mathbf{k-p}}}(\tau')X^{(c)}_{\mathbf{\mathbf{p}}}(\tau')\,,~~I_{(s)} \equiv \int_{\tau_{i}}^{\tau}\frac{d\tau'}{a(\tau')}\sin(k\tau')X^{(c)}_{\mathbf{\mathbf{k-p}}}(\tau')X^{(c)}_{\mathbf{\mathbf{p}}}(\tau') \ .
\end{equation}

It is perhaps worth making a small digression at this point, as vacuum expectation values like the UTC in Eq.~(\ref{eq:UTC}), require regularization of ultraviolet divergences. This has not been much of an issue addressed in the literature, because in lattice simulations of bosonic fields, the large-momentum modes causing the divergence are not captured in the simulations. However, if we were to include all the infinite tower of bosonic modes, all the way up to $|{\bf k}| \rightarrow \infty$, we would encounter an UV divergence. As the lattice spacing is finite, so it is the highest momentum captured in a lattice simulation. The excited modes in a bosonic sector develop typically a huge hierarchy of amplitudes between infrared (IR) and ultraviolet (UV) modes, with the UV amplitudes very suppressed compared to the IR. It is therefore customary to simply measure the slope simulated in the captured UV range, and then extrapolate such slope to the modes ranging from the highest mode captured in the simulation up to $|{\bf k}| \rightarrow \infty$. The would-be UV divergences are then automatically removed. Hence, despite the fact that one formally requires a regularization procedure to obtain a finite UTC, in practice there is no real need. This is certainly the case for the preheating scenarios that we study in this paper, where the UV 'tails' of the field distributions are exponentially suppressed compared to the dominant mode amplitudes\footnote{In the case of GW production from fermionic parametric excitation~\cite{Enqvist:2012im,Figueroa:2013vif,Figueroa:2014aya}, due to Pauli blocking there is no such a hierarchy of amplitudes between the IR and the UV modes. Therefore, in the case of fermions, one must necessarily deal with regularization, as indeed shown in detail in~\cite{Figueroa:2013vif}.}. 

Coming back to GW production, let us derive how a GW background generated in the early universe redshifts till today. Since GWs decouple immediately after production, we simply need to redshift appropriately the frequency and amplitude of the spectrum computed at the end of the GW generation. Let us denote by $t_{\rm i}$ the initial time at the onset of GW production, $t_{\rm f}$ as the end of GW production, $t_{_{\rm RD}}$ as the first moment when the Universe becomes radiation dominated (RD), and finally $t_o$ as the present time. The ratio between the scale factors at the end of inflaton and today can be written as
\be {a_i\over a_o} = {a_i \over a_{_{\rm RD}} } {a_{_{\rm RD}} \over a_o} =  \epsilon_{\rm i}^{1/4}\left(g_{s,o}\over g_{s,_{\rm RD}}\right)^{1\over3}\left(g_{o}\over g_{_{\rm RD}}\right)^{-{1\over4}}\left({\rho_o\over\rho_{\rm i}}\right)^{1\over4}\ . \ee
In the second equality, we have used that during the thermal phase of the Universe, $\rho \propto g_t T^4$ and $a T \propto g_{s,t}^{-1/3}$, with $g_{s,t}$ and $g_t$  the entropic and relativistic degrees of freedom respectively at a time $t$, and $T$ the temperature. We have also used that between $t_{\rm i}$ and $t_{_{\rm RD}}$, the energy density scales as $\rho \propto a^{-3 (1 + \omega )} $ with $w = p/\rho$ the effective equation of state (pressure-to-density ratio) of the Universe. We have introduced the factor
\be\label{eq:epsilonParameter}
\epsilon_{\rm i} \equiv \left({a_{\rm i}\over a_{_{\rm RD}}}\right)^{(1-3w)}\,,
\ee
which quantifies the (averaged) expansion rate of the Universe between $t_{\rm i}$ and $t_{_{\rm RD}}$. Taking into account that $g_{s,t} \sim g_{t}$, we 
see that $\left(g_{s,o}/g_{s,_{\rm RD}}\right)^{1/3}\left(g_{o}/g_{_{\rm RD}}\right)^{-{1/4}} \sim \left(g_{o}/g_{_{\rm RD}}\right)^{1/12} \sim \mathcal{O}(1)$ [$\approx 1.77$ if $g_{o}/g_{_{\rm RD}} = 10^3$, $\approx 1.47$ if $g_{o}/g_{_{\rm RD}} = 10^2$]. Putting all together, and using the energy density of relativistic species today $\rho_o \approx 2\cdot 10^{-15} eV^4$, the frequency today associated to a co-moving mode $k$ of a GW background created in the early universe between $t_{\rm i}$ and $t_{\rm f}$, reads
\begin{eqnarray} f \equiv \left(a_{\rm i}\over a_o\right){k\over2\pi} &=& \epsilon_{_{\rm i}}^{1/4}\left(\frac{g_{s,\rm{o}}}{g_{s,\rm{RD}}}\right)^{1\over 3}\left(\frac{g_{\rm{o}}}{g_{\rm{RD}}}\right)^{-{1\over 4}}\left(\frac{\rho_o}{\rho_{\rm i}}\right)^{1\over 4}\frac{k}{2\pi} \nonumber\\
&\simeq& \epsilon_{\rm i}^{1/4}\left(\frac{k}{\rho_{\rm i}^{1/4}}\right) \times 8\cdot10^{9}~\mathrm{Hz} \ .\label{eq:ftoday}
\end{eqnarray}

On the other hand, the spectral amplitude of the GW background today, normalized to the actual critical energy density $\rho_c$, can be obtained as
\begin{eqnarray}\label{eq:GWtoday}
h^{2}\Omega_{\GW} \equiv \frac{h^2}{\rho_c} \frac{d \rho_{_{\rm GW}}}{d \log k} &=& h^2\Omega_{\rm rad}\left(a_{{\rm f}}\over a_{_{\rm RD}}\right)^{1-3w}\left(g_{s,o}\over g_{s,_{\rm RD}}\right)^{4\over3}\left(g_{_{\rm RD}}\over g_{o}\right)\Omega_{_{\rm GW}}^{({\rm f})} \nonumber \\ 
&\simeq& h^{2}\Omega_{\mathrm{rad}}\left(\frac{g_{o}}{g_{_{\rm RD}}}\right)^{1/3}\times \epsilon_{\rm i}\left(a_{{\rm f}}\over a_{_{\rm i}}\right)^{1-3w}\Omega_{_{\rm GW}}^{({\rm f})} \nonumber \\
&\approx&  \mathcal{O}(10^{-6}) \times 4\epsilon_{\rm i}\left(a_{{\rm f}}\over a_{_{\rm i}}\right)^{1-3w}\Omega_{_{\rm GW}}^{({\rm f})} \ ,
\end{eqnarray}
where $\Omega_{_{\rm GW}}^{\rm (f)} \equiv {1\over\rho_{\rm f}}\left(d\rho_{\GW}\over d\log k\right)_{\rm f}$. For this derivation, in the second line we have used that $g_{s,t} \sim g_t$, and in the third line that  $h^{2}\Omega_{\mathrm{rad}} \simeq 4\cdot 10^{-5}$ and $\left(g_{o}/g_{\rm RD}\right)^{1/3}\sim\mathcal{O}(0.1)$.

If the Universe is in a RD phase already at the onset of GW production at $t_{\rm i}$ (i.e.~$t_{\rm RD} \leq t_{\rm i}$), then $w = {1\over3}$, and the expansion history factors in Eqs.~(\ref{eq:ftoday}) and (\ref{eq:GWtoday}) are simply $\epsilon_{\rm i} = 1$ and $ (a_{\rm f} / a_{\rm i} )^{1 - 3 w}= 1$.  This is the case, e.g.~for preheating from an inflaton with potential $V(\phi)\propto\phi^{4}$, as its an-harmonic oscillations produce effectively $w \simeq 1/3$~\cite{Turner:1983he}. However, if the Universe is in an expanding phase with $w < {1\over3}$ between $t_{\rm i}$ and $t_{_{\rm RD}}$, then there is always a frequency shift to the IR by a factor $\epsilon_{\rm i}^{1/4} < 1$, as well as an amplitude suppression by a factor $\epsilon_i(a_{{\rm f}}/a_{{\rm i}})^{(1 - 3 w)} = (a_{{\rm f }}/{a_{{\rm RD}}})^{(1 - 3 w)} \equiv \epsilon_{\rm f}  < 1$.  This is the case for example if the GW production takes place in a MD background ($w = 0$), like when the oscillations of an inflaton with potential $V(\phi)\propto\phi^{2}$ dictate the expansion of the universe~\cite{Turner:1983he}. If on the contrary, the equation of state is \textit{stiff} with $w > {1/3}$, then the frequency shifts to the UV, while the amplitude of the GW background is enhanced by a factor $\epsilon_i(a_{{\rm f}}/a_{{\rm i}})^{(1 - 3 w)} = (a_{{\rm f }}/{a_{{\rm RD}}})^{(1 - 3 w)} \equiv \epsilon_{\rm f} > 1$.

\subsection{Estimation of the GW production from parametric resonance}

Let us first of all express the GW spectrum Eq.~(\ref{eq:specGW}) in the natural variables defined in Eq.~(\ref{eq:naturalVariables}). The total fraction of energy density in GWs at the end of production $t = t_{\rm f}$ reads
\begin{eqnarray}\label{eq:AmpProd}
\Omega_{_{\rm GW}}^{\rm (f)} \equiv {1\over \rho_{\rm f} }\left({d\rho_{_{\rm GW}}\over d\log k}\right)_{\rm f} = \left(\omega_{*}\over m_p\right)^2\left(a_{\rm i}\over a_{\rm f}\right)^{1-3w}{\kappa^3 \mathcal{F}_{\rm f}(\kappa)\over 8\pi^4 \tilde \rho_{\rm i}} \ ,
~~~~~~~~\\
\label{eq:AmpProd2}
\mathcal{F}_{\rm f}(\kappa) \equiv \omega_*^2\int d\tilde p\, d\theta\, {\tilde p}^{6}\sin^{5}\theta \,\left(\left|\tilde I_{(c)}(\kappa,\tilde p,\theta,z_{\rm f})\right|^{2} + \,\left|\tilde I_{(s)}(\kappa,\tilde p,\theta,z_{\rm f})\right|^{2}\right)\,,
\end{eqnarray}
where $\kappa \equiv k/\omega_{*}$, $\tilde p \equiv p/\omega_{*}$, $z_{\rm f} \equiv \omega_{*} \tau_{\rm f}$, $\tilde \rho_{\rm i} \equiv \rho_{\rm i}/\omega_{*}^4$, and $\tilde I_{(x)}$ are the same functions as in Eq.~(\ref{eq:IsIc}), but written in terms of the dimensionless variables $z$, $\tilde p$, and $\kappa$. Note that we have also introduced the Planck mass $m_p \simeq 2.44\cdot10^{18}$ GeV, and used the fact that the energy density $\tilde\rho_{\rm f}$ at the end of GW production, can be expressed as a function of the initial energy density as $\tilde \rho_{\rm f} = \tilde \rho_{\rm i} (a_{\rm i}/a_{\rm f})^{3(1+w)}$, with $w$ the equation of state parameter between $z_{\rm i}$ and $z_{\rm f}$. 

A scalar field undergoing broad resonance with $q > 1$, experiences an excitation of the field modes up to a given (comoving) cut-off $\kappa \lesssim \kappa_*$~\cite{Kofman:1997yn,Greene:1997fu}
\begin{equation}
\label{eq:kF}
\begin{cases}
\kappa_{*}\sim q^{1/4} &, \,\, V(\phi)\propto\phi^{4}\vspace{0.3cm}\\
\kappa_{*}\sim(a/a_{\rm i})^{1/4}q^{1/4} &, \,\, V(\phi)\propto\phi^{2}
\end{cases}
\end{equation}
where $q$ is the resonance parameter in Eq.~(\ref{eq:ResParam}). In other words, the excitation of a field in broad resonance consists typically in the developing of large field amplitudes for modes with momentum inside a radius $\kappa \lesssim \kappa_{*}$. We will refer to this configuration as a 'Bose-sphere', outside which ($\kappa > \kappa_*$) the field occupation number vanishes, hence suppressing the GW production. Consequently, GWs will only be created inside the Bose-sphere $\kappa < \kappa_*$. Of course, after a number of oscillations of the initially homogeneous mother field, the excited daughter field backreacts into the former. This results in an excitation of finite modes of the mother field, breaking apart its homogeneous condition. From that moment onward, the two-field system becomes non-linear, and one expects higher modes $\kappa \gtrsim \kappa_*$ of the daughter field to be excited. This re-scattering effect broadens up the radius of the Bose-sphere as $\kappa_* \longrightarrow \alpha(q)\kappa_*$, with $\alpha(q)$ a function of $q$, typically of the order of $\mathcal{O}(1) \lesssim \alpha(q) \lesssim \mathcal{O}(10)$ (as measured in simulations). In our following estimation of the GW signal, we will ignore this late enhancement, and simply consider the GW production from the initially excited modes within the Bose-sphere. When confronting our analytic estimates against our numerical results, we will however adjust the parameters involved, taking into account the enhancement effect in the range of mode excitation of the daughter field (as this is done automatically by the lattice simulation).

If we look carefully at the integrand of $\mathcal{F}_{\rm f}$ in Eq.~(\ref{eq:AmpProd2}), we see an angular modulation $\sin^5\theta$, and more importantly, we observe that its amplitude grows with the internal momenta as $\propto \tilde p^6$. Given the structure of the $\tilde I_{(x)}$ functions, even though we can not really predict their dependence on $\tilde p$, we know that they should drop abruptly for $\tilde p > \kappa_*$. Thus, if we assume that $\tilde I_{(x)}$ changes only smoothly as a function of $\tilde p$ for $\tilde p < \kappa_*$, we then expect a growth of the integrand with $\tilde p$ until we hit the cut-off scale $\kappa_*$. Since the modes outside of the Bose-sphere $\tilde p > \kappa_*$ are not excited, the integrand amplitude will be suppressed (typically exponentially) for those momenta. We expect on general grounds, that there will be a peak in the GW spectrum at some scale $\kappa = \kappa_p$, located roughly around the maximally excited momentum of the scalar field spectrum, i.e. $\kappa_p \sim \kappa_*$. Then, using Eq.~(\ref{eq:kF}), the frequency today Eq.~(\ref{eq:ftoday}) of the GW peak, is expected to be 
\begin{eqnarray}
\label{eq:fpToday}
f_{p} &\simeq& 8\cdot10^{9}\epsilon_i^{1/4}\left(\frac{1}{\tilde{\rho}_{\rm i}}\right)^{\frac{1}{4}}\kappa_p~{\rm Hz}\nonumber\\
&\sim& 8\cdot10^{9}\left(\frac{\omega_*}{{\rho}_{\rm i}^{1/4}}\right)\,\epsilon_i^{\frac{1}{4}}\,q^{\frac{1}{4} + \eta}~{\rm Hz}
\times
\begin{cases}
1 &, ~V(\phi)\propto\phi^{4}\vspace{0.3cm}\\
\left(\frac{a_{\rm f}}{a_{\rm i}}\right)^{\frac{1}{4}} &, ~V(\phi)\propto\phi^{2} \ .
\end{cases}
\end{eqnarray}
In the second step, we have introduced a parameter $\eta$, quantifying the goodness of our analytical estimation, and in particular, of our assumption $\kappa_p \sim \kappa_* \propto q^{1/4}$. As our derivation ignores the enhancement effect $\kappa_* \to \alpha(q)\kappa_*$ mentioned before, it is likely that the scaling of $f_p$ as $\propto q^{1/4}$ does not hold, as $\alpha(q)$ is expected to be a function of $q$. Only when confronting Eq.~(\ref{eq:fpToday}) with our lattice simulations from Section~\ref{sec:lattice}, we will be able to quantify whether $\eta$ represents only a small correction to the $d\log f_p / d\log q \simeq 1/4$ behavior, i.e.~$|\eta| \ll 1/4$, or whether $d\log f_p / d\log q$ is completely different from $1/4$, and $|\eta| \gtrsim 1/4$.

Let us now work out an analytical estimate for the $q$-dependence of the GW background peak amplitude. We first need to figure out an analytic estimate for $\mathcal{F}_{\rm f}(\kappa_{p})$. Ignoring the angular dependence in the integrand of $\mathcal{F}_{\rm f}(\kappa_{p})$, which will contribute only to a $\mathcal{O}(1)$-modulation factor, we can actually write
\begin{equation}
\mathcal{F}_{\rm f}(k_{p}) \sim \omega_*^2\int d\tilde p\, {\tilde p}^{6} \,\left|\tilde I(\kappa,\tilde p,\tau_{\rm f})\right|^{2},
\end{equation}
where $\tilde{I}$ represents either of the $\tilde{I}_{(x)}$ functions, whose amplitude is expected to be of the same order, $|\tilde{I}_{(c)}| \sim |\tilde {I}_{(s)}|$. As only the modes within the Bose-sphere are excited, we can make the following $ansatz$
\begin{equation}\label{eq:Antsaz}
\omega_*\big|\tilde{I}\big| = C_q\,{\tilde p^{\,n}}\,\theta(\kappa_{p}-\tilde p),
\end{equation}
where $C_q$ is a dimensionless amplitude (possibly depending on $q$ as we will argue later), $n$ is an effective spectral index characterizing the dependence of $\tilde I$ with $\tilde p$, and $\theta(x)$ is the step function. We expect $\big|\tilde I\big|$ to decrease with growing $\tilde p$, so $n < 0$. This is because the larger the $|\tilde p|$, the faster the field mode functions oscillate in the integrand of $\big|\tilde I_{(x)}\big|$, and hence the smaller the amplitude of $\big|\tilde{I}_{(x)}\big|$ should be due to 'phase erasing' effects. Let us recall that the mode functions $X_k^{(c)}$ have dimensions of (energy)$^{-1/2}$, so the $\tilde I_{(x)}$ functions have dimensions of (energy)$^{-1}$, and hence the factor $\omega_*$ in the $lhs$ of Eq.~(\ref{eq:Antsaz}). The original $I_{(x)}$ functions~Eq.~(\ref{eq:IsIc}) have however dimensions of (energy)$^{-2}$, so the natural value for the spectral index, just based on dimensional considerations, is $n = -2$. Using Eq.~(\ref{eq:kF}), we expect then
\begin{equation}\label{eq:Antsaz_F*}
\mathcal{F}_{\rm f}(\kappa_{p}) \sim C_q^2\kappa_{*}^{3+4\delta} \,\simeq\, C_q^2q^{\frac{3}{4}+\delta}\times\begin{cases}
 ~~1 &, ~V(\phi)\propto\phi^{4}\vspace{0.3cm}\\
\left({a_{\rm f}\over a_{\rm i}}\right)^{\frac{3}{4}+\delta} &, ~V(\phi)\propto\phi^{2}\,,
\end{cases}
\end{equation}
where we have introduced the parameter $\delta$ through
\begin{equation}
n \equiv -2(1 - \delta)\,,
\end{equation}
to account for possible deviations with respect to our educated guess of $n = -2$. One can consider $\delta$ as a parameter characterizing the deviation of the truly high-momenta tail of $\tilde{I}$, with respect to the simple power-law characterization introduced in Eq.~(\ref{eq:Antsaz}), cut-off at $\kappa_p$ via the Heaviside function $\theta(\kappa_{p}-\tilde p)$.

In the case of $V(\phi) \propto \phi^2$ (or, for this matter, of any potential different than $V(\phi) \propto \phi^4$), we can reabsorb the scale factor dependence $({a_{\rm f}/a_{\rm i}})^{3/4+\delta}$ into the dimensionless amplitude $C_q$, i.e.~$C_q^2({a_{\rm f}/a_{\rm i}})^{3/4+\delta} \longrightarrow C_q^2$. We can then write $\mathcal{F}_{\rm f}(\kappa_{p}) \sim C_q^2q^{\frac{3}{4}+\delta}$ for any potential $V(\phi) \propto \phi^n$, understanding that $C_q$ characterizes not only the amplitude of the {\it ansatz} Eq.~(\ref{eq:Antsaz}), but also some (typically mild) dependence on the expansion history during GW production.

Using Eq.~(\ref{eq:Antsaz_F*}), we infer that the GW peak amplitude at $z= z_{\rm f}$ scales as
\begin{eqnarray}\label{eq:PeakAmplitude}
\Omega_{_{\rm GW}}^{({\rm f})}(\kappa_p) \equiv \left(a_{{\rm f}}\over a_{_{\rm i}}\right)^{3w-1}\left(\omega_{*}\over m_p\right)^2{\kappa_f^3 \mathcal{F}_{\rm f}(\kappa)\over 8\pi^4 \tilde \rho_{\rm i}} \sim {C_q^2\over 8\pi^4}\frac{\omega_*^{6}}{\rho_i m_{p}^{2}}\,q^{\frac{3}{2}+\delta}\,.
\end{eqnarray}

To obtain the final q-dependence of $\Omega_{\rm GW}^{\rm (f)}(\kappa_p)$, we still need to determine whether $C_q$ depends on $q$. For this, note that our ansatz in Eq.~(\ref{eq:Antsaz}) is equivalent to modelling the parametric daughter field excitation as a 'box'
\begin{equation}\label{eq:BoxModel}
|X^{(c)}_k|^2 = C_q\,k_*^{-1}\,\theta(k_*-k)\,,
\end{equation}
where the factor $1/k_*$ in the $rhs$ simply reflects the dimensions of $X^{(c)}_k$ as (energy)$^{-1/2}$. As discussed before, the theory of parametric resonance indicates that only the modes within a Bose-sphere $k \lesssim k_* \sim q^{1/4} \omega_*$ are excited, so the bigger the resonance parameter $q$, the larger the range of modes excited.  As one cannot transfer an arbitrary large amount of energy as $q$ increases, $C_q$ must be a decreasing function of $q$. 

To determine the $q$-dependence of $C_q$, we need however some extra information. In {\it Paper I} we defined the {\it decay} time as the moment when the energy transferred (from the mother field) into the daughter field has become sufficiently large, so that the amplitude and energy density of the mother field (that has previously decreased noticeably) does not evolve significantly anymore. See {\it Paper I} for more details about the definition and quantification of the different time scales involved in the problem. We expect therefore the fraction of energies of the daughter to the mother field, to be independent of $q$ at this time\footnote{The decay time scale depends however on $q$.}. As shown by detailed simulations in {\it Paper I}, this is precisely the case: the fraction of the daughter field energy components (say its kinetic and gradient energies) to the total energy of the system, seems to be roughly independent of $q$ (with some mild scatter) at the time of decay. 
 
Using this fact, we can characterize the daughter field spectrum using the box-modeling Eq.~(\ref{eq:BoxModel}), by considering for example its gradient energy as
\begin{eqnarray}
{1\over\omega_*^4}\langle (\vec\nabla X_{c})^2\rangle \sim \int {dk k^2\over\omega_*^4} |k X^{(c)}_k|^2 \sim C_q\int {dk k^4\over\omega_*^4} k_*^{-1}\theta(k_*-k) \sim C_q {k_*^4\over\omega_*^4} \propto const\,.
\end{eqnarray}
Using that $k_* \sim q^{1/4}\omega_*$, this leads to
\begin{eqnarray}\label{eq:Cq}
C_q \equiv C \,q^{-1}\,,
\end{eqnarray}
where $C$ is now a (dimensionless) constant independent of $q$. We could have used also the kinetic energy of the $X_c$ field, with identical conclusion. This means, for instance, that during the linear stage of parametric resonance, the peak of the daughter field spectrum $k^3|X^{(c)}_k|^2$ should scale as
\begin{eqnarray}\label{eq:PeakMatterField}
k_*^3|X^{(c)}_{k_*}|^2 \sim (q^{1/4}\omega_*)^3C_qk_*^{-1} \propto q^{-1/2}\,.
\end{eqnarray}
As we will show in Section \ref{sec:lattice}, our numerical simulations display precisely this behavior.

Finally, using Eq.~(\ref{eq:PeakAmplitude}) and Eq.~(\ref{eq:Cq}), we find that the GW peak amplitude at $z= z_{\rm f}$ is given by
\begin{eqnarray}\label{eq:PeakAmplitudeV2}
\Omega_{_{\rm GW}}^{({\rm f})}(\kappa_p) = {C^2\over 8\pi^4}\frac{\omega_*^{6}}{\rho_i m_{p}^{2}}\,q^{-\frac{1}{2}+\delta}\ .
\end{eqnarray}
Substituting this into Eq.~(\ref{eq:GWtoday}), the corresponding amplitude of the GW peak today is
\begin{eqnarray}\label{eq:PeakAmplitudeToday}
h^{2}\Omega_{_{\rm GW}} (f_p) &\sim& \mathcal{O} (10^{-9} ) \times \epsilon_{\rm i}\,C^{2}\frac{\omega_*^{6}}{\rho_i m_{p}^{2}}\,q^{-\frac{1}{2}+\delta}\ ,
\end{eqnarray}
where we have used ${1\over 8\pi^4} \sim 1.3\cdot 10^{-3}$, and absorbed the factor $1.3\cdot4\cdot(a_{\rm f}/a_{\rm i})^{1 -3 \omega}$ into the $C^2$ constant. 

The parameters $\eta$ and $\delta$ in Eqs.~(\ref{eq:fpToday})-(\ref{eq:PeakAmplitudeToday}) quantify the goodness of our peak parametrization as a function of $q$. In Section~\ref{sec:lattice} we will confront our formulae against the outcome of numerical simulations, sampling over a wide range of $q$ values within each scenario considered. 

Before we move into the results from lattice simulations, let us note one last thing: the analytic estimate derived so far is only valid for $q > 1$. If $q < 1$, the maximum momentum excited by the resonance does not longer grow as $k_* \propto q^{1/4}$, and the assumption that $\big|\tilde I\big|$ decreases smoothly as $\tilde p$ grows (for $\tilde p < \kappa_*$) is no longer valid. When $q < 1$, we are in a narrow resonance regime, and the structure of the excited momenta is much more complicated, as only specific set of modes are excited~\cite{Kofman:1997yn}. Therefore, we expect $\tilde I$ to behave differently with respect $\tilde p$. In particular, for $q < 1$, more and more excited modes enter within a sphere of increasingly larger radius $\tilde p$. Thus, we may expect $\big|\tilde I\big|$ to grows with $\tilde p$ or, equivalently, that $n > 0$ in Eq.~(\ref{eq:Antsaz}). In the case of $q < 1$, the $\delta$-parametrization of the deviations with respect $n = -2$, loses its original meaning, as $\delta$ is not expected anymore to be a small correction to the $\propto q^{-1/2}$ behavior. However, our simulations in Section \ref{sec:lattice} cannot capture the narrow resonance regime $q < 1$, so we cannot quantify whether Eq.~(\ref{eq:Antsaz}) is a good ansatz in this case. On the other hand, the most physically interesting cases in the early Universe are those where broad resonance $q > 1$ develops, so we stick to this circumstance from now on.

\section{Gravitational waves from parametric resonance. Numerical Results.} \label{sec:lattice}

In this section we perform lattice simulations of scenarios where the mother field dominates the energy budget of the Universe. We thus identify the mother field $\phi$ with the inflaton, so that the parametric resonance of the daughter field(s) $X$ coupled to $\phi$, represents the period of {\it Inflationary Preheating}. We consider single-field slow-roll scenarios where the inflaton has a monomial potential $V_{\rm inf} (\phi)$. Inflation ends when the slow-roll parameters become approximately of order unity, and the Hubble rate becomes smaller than the inflaton effective mass. The inflaton has in that moment a large vacuum expectation value, so its amplitude starts oscillating around the minimum of its potential. This induces a strong creation -- via parametric resonance -- of the particle species coupled to the inflaton, as long as their coupling strength is sufficiently big. As the inflaton and its decay products are the dominant energy components of the Universe, the time-evolution of the scale factor must be obtained by solving self-consistently the equation of motion of the fields, together with the Friedmann equations sourced by the fields' energy and pressure densities. We consider the two paradigmatic models of chaotic inflation, where the inflaton has either a quartic potential (Section \ref{sec:lphi4-results}) or a quadratic potential (Section \ref{sec:m2phi2-results}):
    \bea
     V_{\rm inf} (\phi) = \left\{ \begin{array}{ll}
        \frac{1}{4} \lambda \phi^4  , \hspace{1cm} & \lambda \approx 9 \times 10^{-14} , \vspace*{2mm}\\ 
         \frac{1}{2} m^2 \phi^2 , \hspace{1cm} & m \approx 6 \times 10^{-6} m_p \ .
        \end{array} \right. \label{eq:inflation-potentials}
    \eea
The strength of the parameters $\lambda$ and $m$ is fixed by the amplitude of the observed CMB anisotropies. Both scenarios of inflation are in fact challenged by recent CMB measurements~\cite{Ade:2015lrj}, the quartic case more severely. However, the simple addition of an non-minimal gravitational coupling between the inflaton $\phi$ and the Ricci curvature $R$, $\xi R\phi^2$ (with $\xi \sim -10^{-3}$), reconcile these scenarios with CMB observations~\cite{Tsujikawa:2013ila}. The addition of the non-minimal coupling changes, of course, the exact values of $m^2$ and $\lambda$, required to fit the scalar power spectrum amplitude at the CMB scales. This is however only an $\mathcal{O}(1)$ change, so we can safely maintain the values given in Eq.~(\ref{eq:inflation-potentials}), knowing that order-of-magnitude wise they are still valid. Besides, towards the end of inflation, the presence of such non-minimal coupling becomes negligible, and the inflaton dynamics can be described entirely by its monomial potential $V_{\rm inf}(\phi)$. In the quartic model, the (oscillation averaged) energy density of the inflaton scales as in a RD background, with $\rho_\phi \propto 1/a^4$. The scale factor evolves correspondingly as $a(t) \propto t^{1/2}$. In the quadratic model, the (oscillation averaged) energy density of the inflaton evolves as in a MD background, with $\rho_\phi \propto 1/a^3$. The scale factor evolves correspondingly as $a(t) \sim t^{2/3}$. 

In both scenarios we always consider a symmetric interaction $g^2\phi^2 X^2$ between the mother field $\phi$ and the daughter field $X$. This interaction is scale free, with $g^2$ a dimensionless coupling constant. This is particularly convenient from the point of view of the lattice, since any other form of interaction would require the introduction of a new mass scale. Besides, this interaction has been often assumed in the context of preheating, and it is the leading interaction term in the context of gauged daughter fields, as shown in~\cite{Figueroa:2015rqa,Lozanov:2016pac}. It is also interesting to note that this interaction does not lead to a tree level decay of the mother field into the daughter species, so all the transfer of energy from $\phi$ into $X$ will be due only to the non-perturbative effects characteristic of parametric resonance.

\subsection{Lattice simulations of preheating with quartic potential}
\label{sec:lphi4-results}

In this section we concentrate in the study of GW production during preheating when the inflaton has a quartic potential
\be V (\phi) = \frac{1}{4} \lambda \phi^4 \ .\ee
Following the discussion after Eq.~(\ref{eq:powerlaw-pot}), the initial frequency of oscillation is given by $\omega_* = \sqrt{\lambda} \phi_{\rm i}$, with $\phi_{\rm i}$ the initial value of the inflaton field (determined below). Hence, it is useful to define a new set of dimensionless field, spacetime, and momentum variables as
\be \varphi \equiv \frac{a}{\phi_{\rm i}} \phi \ , \hspace{0.3cm} \chi \equiv \frac{a}{\phi_{\rm i}} X \ , \hspace{0.3cm} z \equiv \sqrt{\lambda} \phi_{\rm i} \int_{t_i}^t \frac{dt'}{a(t')} \ , \hspace{0.3cm} \vec{y} \equiv \sqrt{\lambda} \phi_{\rm i} \vec{x} \ , \hspace{0.3cm}  \kappa \equiv \frac{k}{\sqrt{\lambda} \phi_{\rm i}}\ . \label{eq:lphi4-variables}\ee
In these units, the Eqs.~(\ref{eq:generic-eom}) for the inflaton and daughter field are
\bea \varphi'' -\frac{a''}{a} \varphi -  \nabla_y^2 \varphi + \left( \varphi^{2} + q \chi^2 \right) \varphi = 0  \ , \hspace{0.5cm} \chi'' -  \frac{a''}{a}\chi - \nabla_y^2 \chi + q \varphi^2 \chi = 0 \ , \label{eq:eom-lame}\eea
where $' \equiv d / dz$, $\nabla_y$ is the laplacian in terms of the natural spatial variables, and the resonance parameter $q$ is
\be q \equiv \frac{g^2}{\lambda} \label{eq:lphi4-qresdef} \ .\ee
When $\phi \gg m_p$, the inflaton slowly rolls down its potential, so that its potential energy drives the inflationary expansion. However, when $\phi \sim m_p$, this regime gets broken and preheating starts. The inflaton starts oscillating around the minimum of its potential with frequency $\sim \omega_*$, and each time the inflaton crosses zero, all particles coupled to it may be strongly created, depending on the strength of their coupling. We define the time $t_{\rm i}$ of the onset of the oscillatory regime when the inflaton effective mass becomes equal to the Hubble rate, i.e. when $H (t_{\rm i}) = \sqrt{\lambda} \phi_{\rm i}$.  Imposing the slow-roll condition to the field $\phi$ at very early times, and solving numerically the Friedmann and Klein-Gordon equations for the inflaton in a self-consistent way, we obtain \be\phi (t_{\rm i} ) \equiv \phi_{\rm i} \simeq 3.05 m_p \ , \hspace{0.5cm} \dot{\phi} (t_{\rm i} ) \simeq -3.54 \sqrt{\lambda} m_p^2 \ . \ee 
In our lattice code, we take $t_{\rm i}$ as the initial time of our simulations, and simulate the subsequent preheating stage and the associated production of GWs. 

Before we analyze the GW production, let us briefly review some known results about preheating in this model. As discussed in Section \ref{sec:Analytics}, at initial times the inflaton can be taken as a homogeneous oscillating field, and backreaction effects can be neglected. In this case, Eq.~(\ref{eq:ParamResEqFourier}) for the daughter field modes takes the form of the \textit{Lam\'e} equation, which admits solutions of the type $\chi_{\kappa} \propto e^{\mu_{\kappa} z}$, with $\mu_{\kappa}$ the so-called Floquet index \cite{Greene:1997fu}. For a specific $q$ in Eq.~(\ref{eq:eom-lame}), the Floquet index has a real positive solution $\mathfrak{Re} [\mu_{\kappa}] > 0$  for certain momenta $\kappa$ (with $\mathfrak{Re} [\mu_{\kappa}]\lesssim 0.23$), and hence for these modes the particle number $n_k \sim |\chi_{\kappa}|^2 \sim e^{2\mu_{\kappa} z}$ grows exponentially. This defines a series of resonance bands in the $(\kappa,q)$ plane, which are usually depicted as a stability/instability chart. There are two possible types of resonance structures, depending on the particular value that $q$ takes. On the one hand, for resonance parameters inside the intervals $q \in [1,3], [6,10], [15,21], \dots$, the dominant resonance band is of the type $0 < \kappa < \kappa_+$, i.e. modes are excited down to the minimum momentum $\kappa=0$. On the other hand, for $q \in (0,1),(3,6),(10,15) \dots$, the main resonance band is of the type $\kappa_- < \kappa < \kappa_+$, with $\kappa_- > 0$.

Of course, the description of the daughter field dynamics as dictated by the \textit{Lam\'e} equation breaks down eventually, when the backreaction effects from the daughter field onto the inflaton condensate become noticeably. To fully account for all the effects in the post-inflationary dynamics, we simulate the system with classical real-time lattice simulations. In {\it Paper I}, we did a systematical lattice study of this scenario, identifying two natural time scales, which depend on $q$. The first one is the {\it back-reaction} time $z_{\rm br}$, which signals the time at which the energy density of the daughter field has grown enough and starts affecting the oscillating mother field. This signals the onset of the decay of the inflaton amplitude and energy density. The second time scale is the decay time $z_{\rm dec}$, when the system has almost achieved an equipartition and stationary regime, and the inflaton energy no longer decreases. At $z_{\rm br}$ the energy is distributed rather 'undemocratically' between the different fields: the mother field still possesses more than $\sim 90 \%$ of the total energy, while the daughter field and the interaction energy account for the remaining $\sim 10 \%$. Those energy fractions are actually roughly independent of $q$. At $z_{\rm dec}$ the inflaton has already transferred a significant fraction of its energy, and it only represents approximately $\sim 40 \%$ of the total energy budget\footnote{So notice that the inflaton does never decay completely through parametric resonance, and in fact its energy still represents a sizeable part of the total budget.}. At this time, $\sim 55 \%$ of the energy is in the daughter field, while the remaining $\sim 5 \%$ is in form of interaction energy. Again, the energy fractions at $z_{\rm dec}$ are mostly independent of $q$.
\begin{figure}
      \begin{center} \includegraphics[width=14cm]{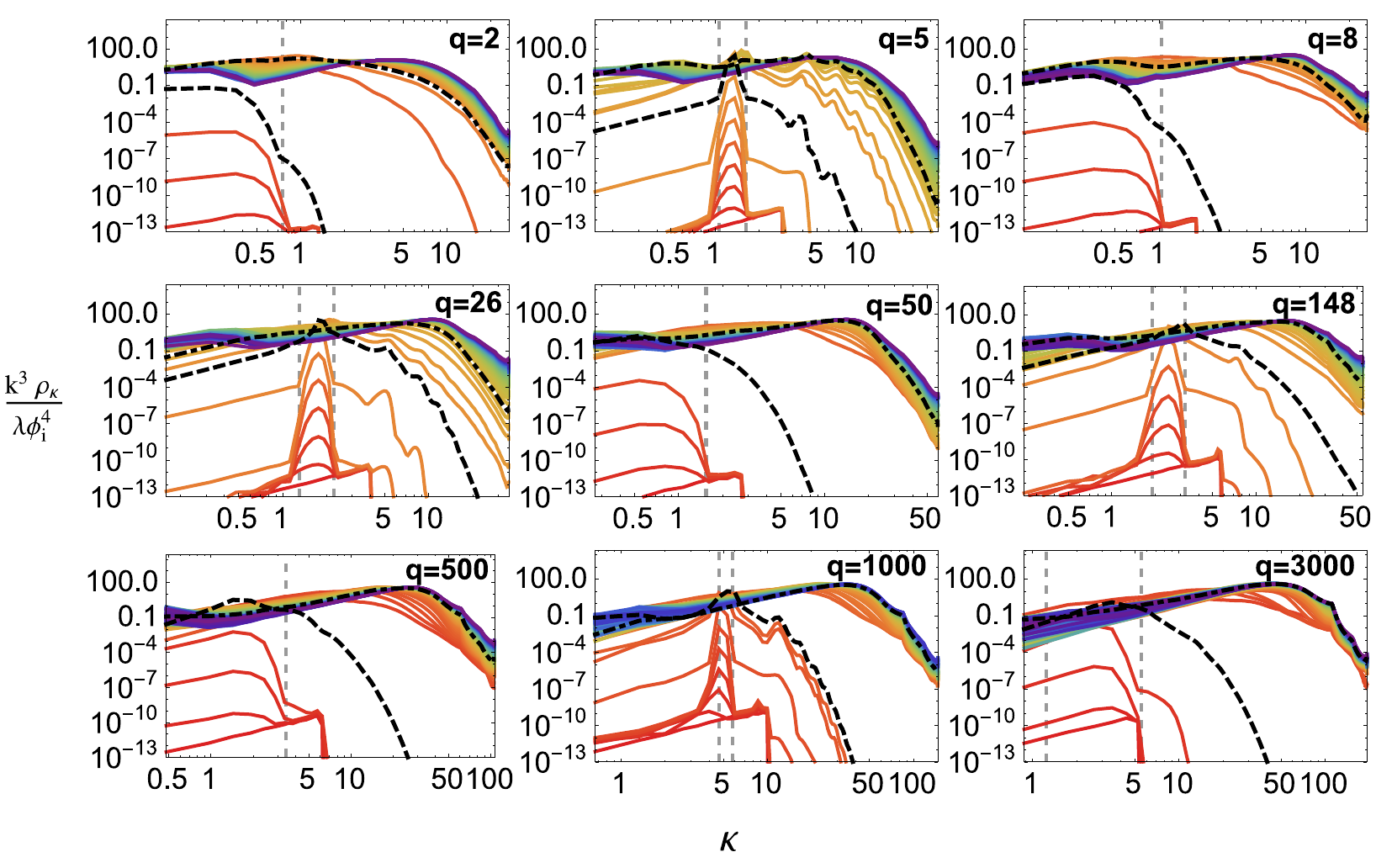} \end{center} 
      \caption{We show the time-evolution of the daughter-field energy density spectra $\kappa^3 \rho_{\kappa, \chi} / (\lambda \phi_{\rm i}^4)$ [Eq.~(\ref{eq:energy-density})] as a function of the momentum, for different values of the resonance parameter $q$. The spectra correspond to times $z= 0, 10, 20, \dots$, with red lines corresponding to early times, and purple lines to late times. In panels $q=5,26,148,1000,3000$, which correspond to cases with a main resonance band of the type $\kappa_- < \kappa < \kappa_+$, we indicate $\kappa_-$ and $\kappa_+$ with dashed, vertical lines. In the rest of panels, in which the main band has the form $0 < \kappa < \kappa_+$, we simply indicate the position of $\kappa_+$.  The values of $\kappa_{\pm}$ have been obtained from the numerical properties of the \textit{Lam\'e} equation. We also show with black dashed and black dot-dashed curves the spectra at times $z \approx z_{\rm br}$ and $z \approx z_{\rm dec}$ respectively. } 
            \label{fig:lphi4-spectra}
 \end{figure}
 
We have carried out several lattice simulations of the preheating process, including the tensor fields representing the GWs as extra dynamical fields. We have considered different values of the resonance parameter $q$ in the interval $0.4 < q < 5000$. Lower values cannot be simulated because the resonance bands are too narrow to be captured in the lattice, while any larger values cannot be considered due to a lack of a good UV coverage. This is explained in more detail in the Appendix B of {\it Paper I}. All the results presented in this section have been obtained from simulations with $N^3 = (256)^3$ points, and minimum momenta $\kappa_{\rm min} \sim \mathcal{O} (0.1)$, the specific number depending on the particular case. More details on this, as well as on the consistency of the results for different lattice parameters, can be found in Appendix \ref{app:Lattice-formulation}.

To see how the times $z_{\rm br}$ and $z_{\rm dec}$ are reflected in the spectra, see Fig.~\ref{fig:lphi4-spectra}. There we show the time-evolution of the energy density spectra of the daughter field, defined as 
\be k^3 \rho_{k, \chi} = \frac{\lambda \phi_{\rm i}^4}{2} \kappa^3 \left(  | \chi^{'}_{\kappa} |^2  + \omega_{\kappa,\chi}^{2} | \chi_{\kappa} |^2 \right) \ , \hspace{0.5cm} \omega_{\kappa,\chi} = \sqrt{\kappa^2 + q \varphi^2 - (a''/a)} \label{eq:energy-density} \ ,\ee
obtained from lattice simulations, and for different values of the resonance parameter $q$. For $q=5,26,148,1000,3000$, the main resonance band is of the form $\kappa_- < \kappa < \kappa_+$, while for $q=2,8,50,500$, the band is of the type $0 < \kappa < \kappa_+$. As expected, for initial times $z \lesssim z_{\rm br}$, the linear analysis is approximately valid, and the growth of the daughter field takes place mainly inside the resonance bands delimited with dashed, vertical lines. This generates a structure of peaks in the field spectra, due to the particular structure of resonance bands of the {\it Lam\'e} equation. However, for late times $z \gtrsim z_{\rm br}$, i.e. when the backreaction effects on the inflaton condensate are already significant, the spectra grow outside these bands, washing out the structure of peaks created during the initial stages. The daughter field populates modes of higher-momenta, due to the scattering among modes induced by the coupling between fields. Finally, when the stationary regime is achieved at times $z \gtrsim z_ {\rm dec}$, the spectra does not evolve appreciably anymore, and its amplitude reaches a final saturated value\footnote{In reality, the 'saturated' amplitudes will evolve smoothly at times $z \gg z_{\rm dec}$, as the field distributions adapt themselves on their way towards equilibrium~\cite{Figueroa:2016wxr}. However, during this regime no GWs are emitted, so we are not interested in this late stage.}.

 \begin{figure}
      \begin{center}
       \includegraphics[width=6.5cm]{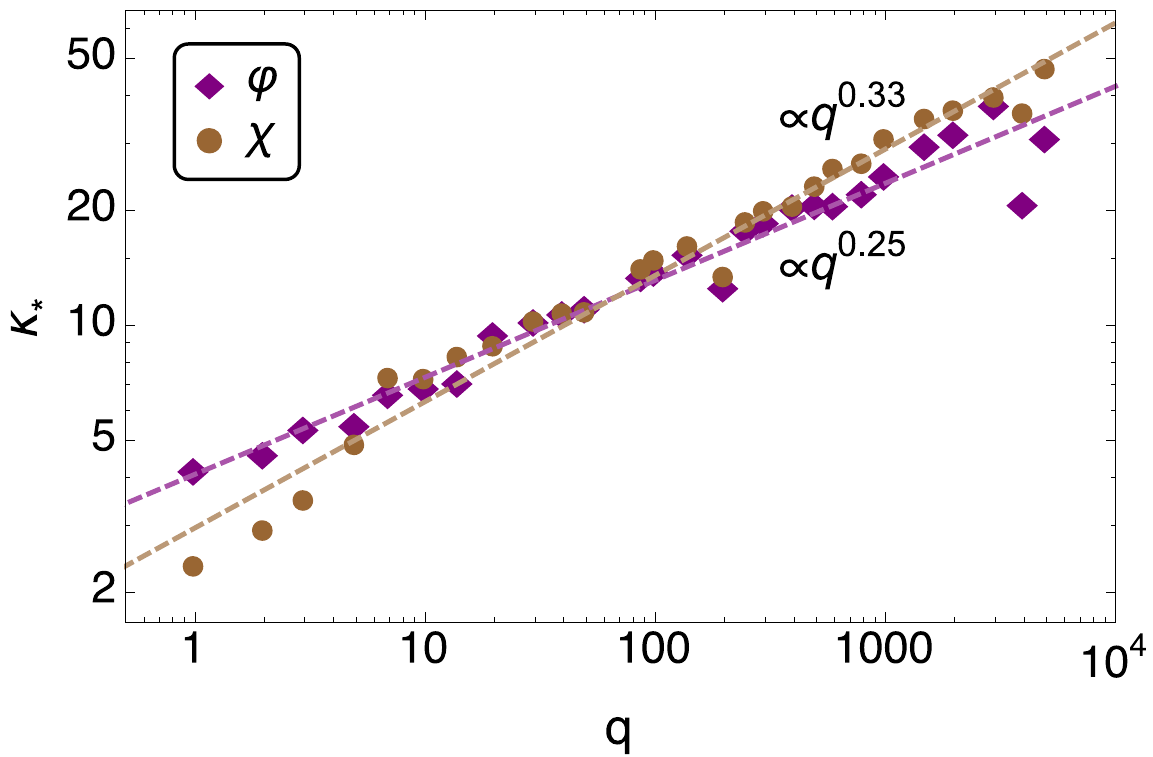} \,\,\,\,
        \includegraphics[width=7cm]{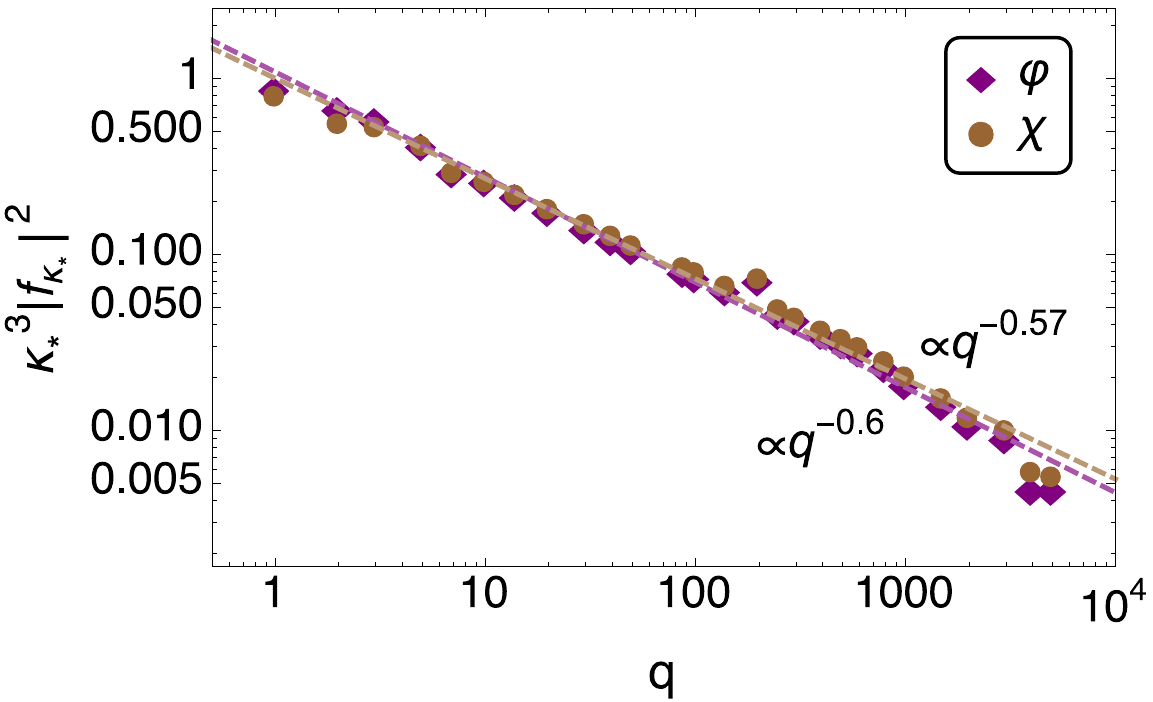} 
        \end{center}
      \caption{{\it Left:} Position of the peak $\kappa_*$ as a function of $q$, for the inflaton and daughter fields, when a final saturated amplitude has already been reached. {\rm Right:} Spectral amplitude $\kappa_*^3 |f_{\kappa_*}|^2$ at the peak position $\kappa_*$, for both the inflaton and the daughter fields ($f= \varphi, \chi$). In both panels, each point corresponds to a single lattice simulation, and the dashed straight lines correspond to the fits in Eq.~(\ref{eq:lphi4fits-MatterSpectra}).} 
            \label{fig:lphi4-fitspectra}
 \end{figure}
 
Finally, before we move into the analysis of the GW spectra, let us check whether the matter spectra obtained from lattice simulations obey the scaling with $q$ derived in Section~\ref{sec:Analytics}. Essentially, due to the structure of resonance bands, the position of the peak should scale as $\kappa_* \propto q^{1/4}$ [Eq.~(\ref{eq:kF})], while according to our calculations, the amplitude of such peak should scale as $\kappa_*^3 |\chi_{\kappa_*}|^2 \propto q^{-1/2}$ [Eq.~(\ref{eq:PeakMatterField})]. In Fig.~\ref{fig:lphi4-fitspectra} we plot both quantities as a function of $q$, extracted from our lattice simulations when the fields spectra have saturated. We obtain the following fits,
 \bea 
  {\rm Daughter~field}~\chi: \hspace{0.5cm} \kappa_* & \approx & 13 \left( \frac{q}{100} \right)^{0.33} \ , \hspace{0.5cm} \kappa_*^{3}{ |\chi_{\kappa_*} |^2}\approx  7 \cdot 10^{-2}   \left( \frac{q}{100} \right)^{-0.57} \ , \nonumber \\ 
 {\rm Mother~field}~\varphi: \hspace{0.5cm}\kappa_* &\approx & 13  \left( \frac{q}{100} \right)^{0.25} \ , \hspace{0.5cm} \kappa_*^{3} {| \varphi_{\kappa_*} |^2} \approx 7 \cdot 10^{-2}  \left( \frac{q}{100} \right)^{-0.60} \ .
 \label{eq:lphi4fits-MatterSpectra}
 \eea
The power-law scaling for the daughter field spectral peak, obtained from lattice simulations, coincides quite well with the theoretical prediction $\propto q^{-1/2}$ [Eq.~(\ref{eq:PeakMatterField})] , with a deviation of the mean exponent with respect the theoretical value of only $100\times{(0.57-0.5)\over0.5} \sim 14\%$. On the other hand, the theoretical location of the daughter field's peak at $\kappa_* \sim q^{1/4}$ [Eq.~(\ref{eq:kF})] is actually only realized with a correction (of the exponent) of $100\times{(0.33-0.25)\over0.25} \sim 30\%$. The fact that the daughter spectra deviate from the theoretical expectations should not be seen as surprising: strictly speaking, such predictions are only expected to be valid during the linear regime of the daughter fluctuations growth. The spectra fitted in Eq.~(\ref{eq:lphi4fits-MatterSpectra}), however, are measured at $z \gtrsim z_{\rm dec}$, so non-linear effects cannot be ignored. Therefore, it is rather natural to expect some deviation from the theoretically predicted scaling, and yet it is remarkable that the spectrum amplitudes scale approximately as predicted from the linear analysis.

It is also remarkable that the mother field $\varphi$ follows a similar scaling as the daughter field, even though we did not have a clear expectation in this respect. The (mean value of the) peak position of the mother field spectra is actually located exactly at $\sim q^{1/4}$, whereas the exponent of the spectral peak amplitude presents only a deviation of $100\times{(0.6-0.5)\over0.5} \sim 20\%$ with respect the theoretical predicted scaling. We think this is due to the coupling between the daughter and the mother fields: slightly before $z \simeq z_{\rm br}$, when the daughter field modes have already grown significantly (following the resonance pattern of the linear analysis), the same modes of the mother field are excited, thanks to the interaction term. This 'dragging' effect is clearly seen in the inflaton spectra, see e.g.~Fig 11 (top panels) in Appendix A of \textit{Paper I}.

 \subsubsection{Gravitational wave parametrization}\label{sec:lphi4-GWparam}
        
Let us discuss now the production of GWs during the preheating process just analyzed. Let us define first of all a rescaled tensor field as 
\begin{equation}
{\bar h}_{ij} \equiv a h_{ij}\,.
\end{equation}
Using this rescaling and those defined in Eq.~(\ref{eq:lphi4-variables}), the equation of motion of the GWs Eq.~(\ref{eq:GWeom}) takes the form
\be {\bar h}_{ij}^{''} - \nabla_y^2 {\bar h}_{ij} - \frac{a''}{a} {\bar h}_{ij} = \frac{2 \phi_{\rm i}^2}{a m_p^2} \left[ \partial_i \varphi \partial_j \varphi +  \partial_i \chi \partial_j \chi \right]^{\rm TT} \ . \label{eq:lphi4-GWeom} \ee
The total energy of the system $\rho_t$ (contributed by the matter fields, as the GWs are energetically very sub-dominant) can be written as
\be \rho_t (z)  = \frac{\lambda \phi_{\rm i}^4}{a^4} \times \left[  \frac{1}{2} \sum_{f=\chi,\varphi}  \left(f' - f \frac{a'}{a} \right) + \frac{1}{2} \sum_{f=\chi,\varphi} |\nabla_y f |^2 +  \frac{1}{2}  q \varphi^2 \chi^2 + \frac{1}{4} \varphi^4 \right] \equiv \frac{\lambda \phi_{\rm i}^4}{a^4} E_t \ .\ee
The spectrum of GWs in the continuum Eq.~(\ref{eq:ThetaGW}), normalized over the total energy density of the system, can be written as
\bea \Omega_{\rm GW} (\kappa, z) \equiv \frac{1}{\rho_t(z)} \frac{d \rho_{{\rm GW}}}{d \log k}(\kappa, z) = {m_p^2\over \phi_i^2}\frac{(\sqrt{\lambda}\phi_i\kappa)^3}{8 \pi^2 V E_t(z)} \int {d\Omega_k\over 4\pi} \left|({\bar h}'_{ij} - \mathcal{H} {\bar h}_{ij})(\kappa,z)\right|^2 \ . \label{eq:lphi4-GWspectra}
\eea
For the discrete version of the GW spectrum, see Appendix~\ref{app:Lattice-formulation}.

 \begin{figure}
      \begin{center}
       \includegraphics[width=7.5cm]{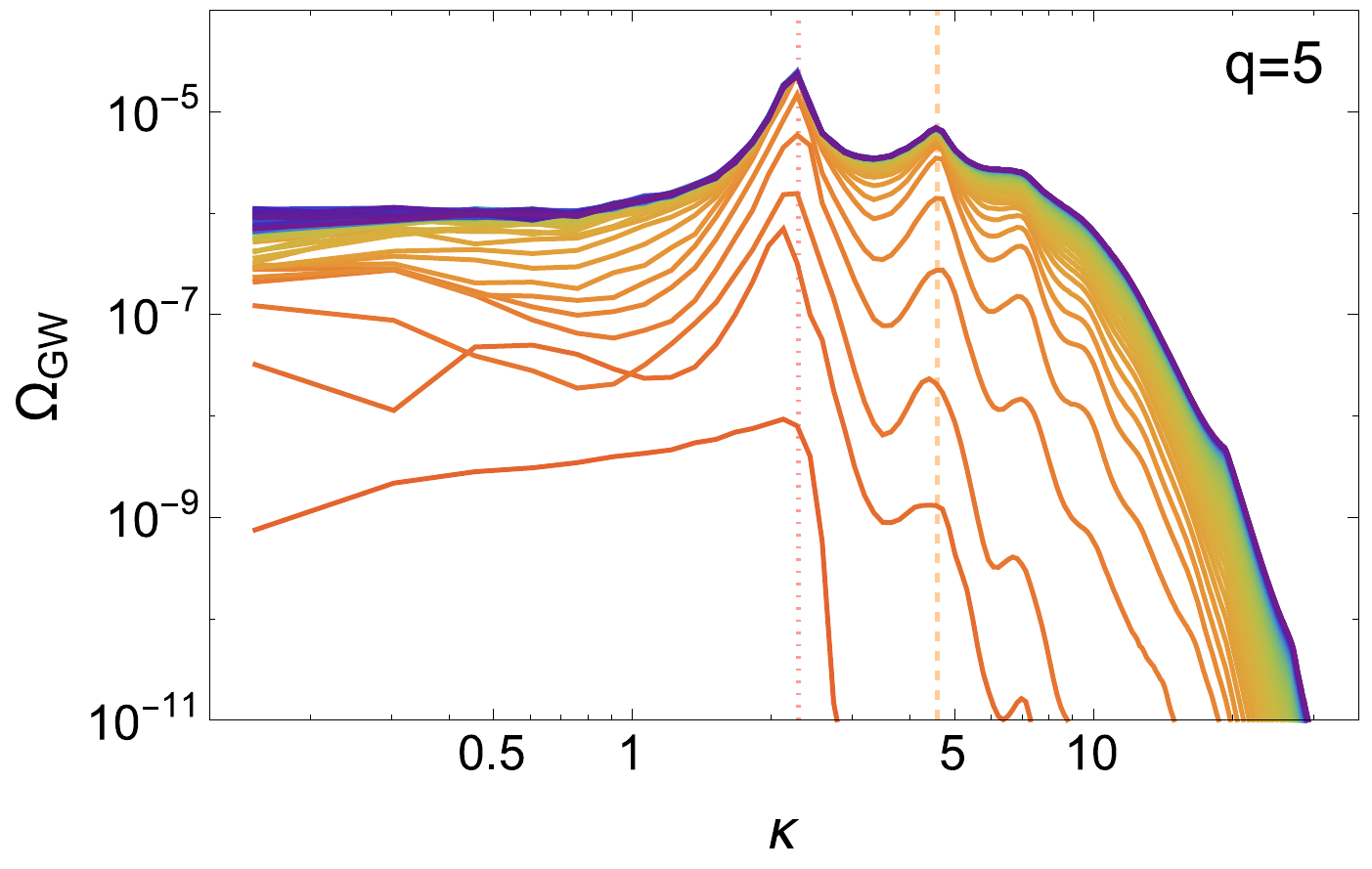} 
        \includegraphics[width=7.5cm]{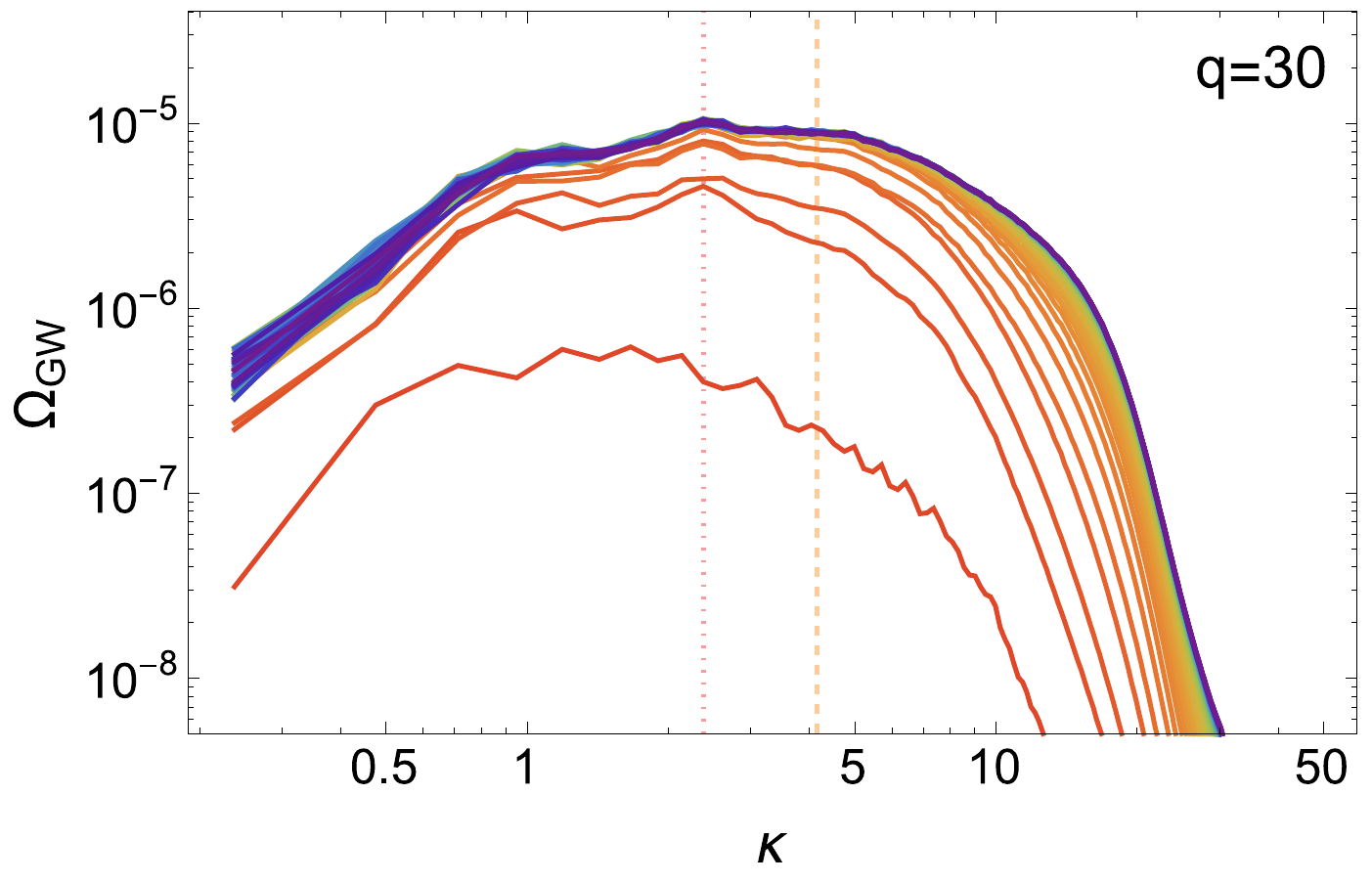} 
        \includegraphics[width=7.5cm]{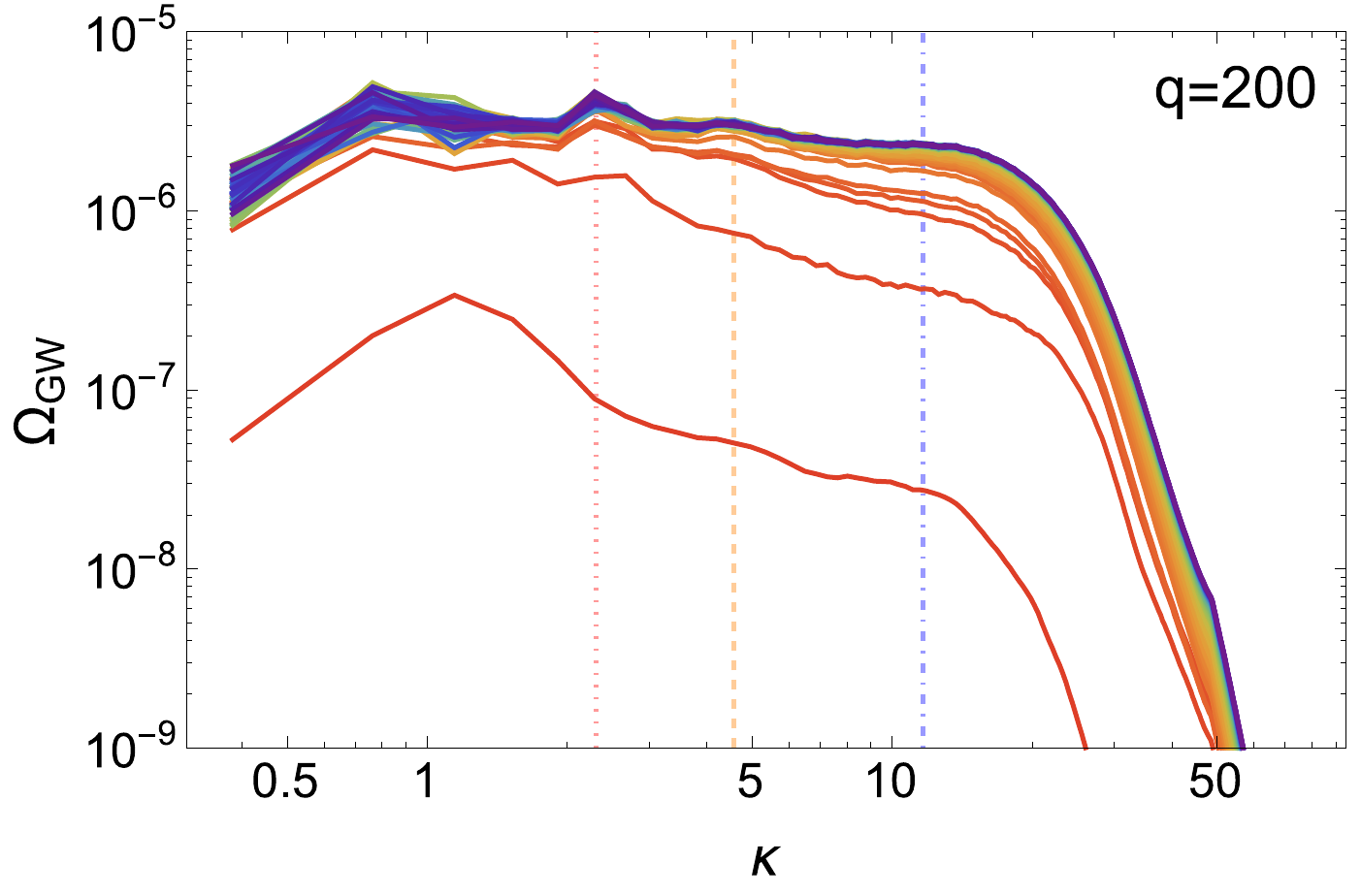} 
        \includegraphics[width=7.5cm]{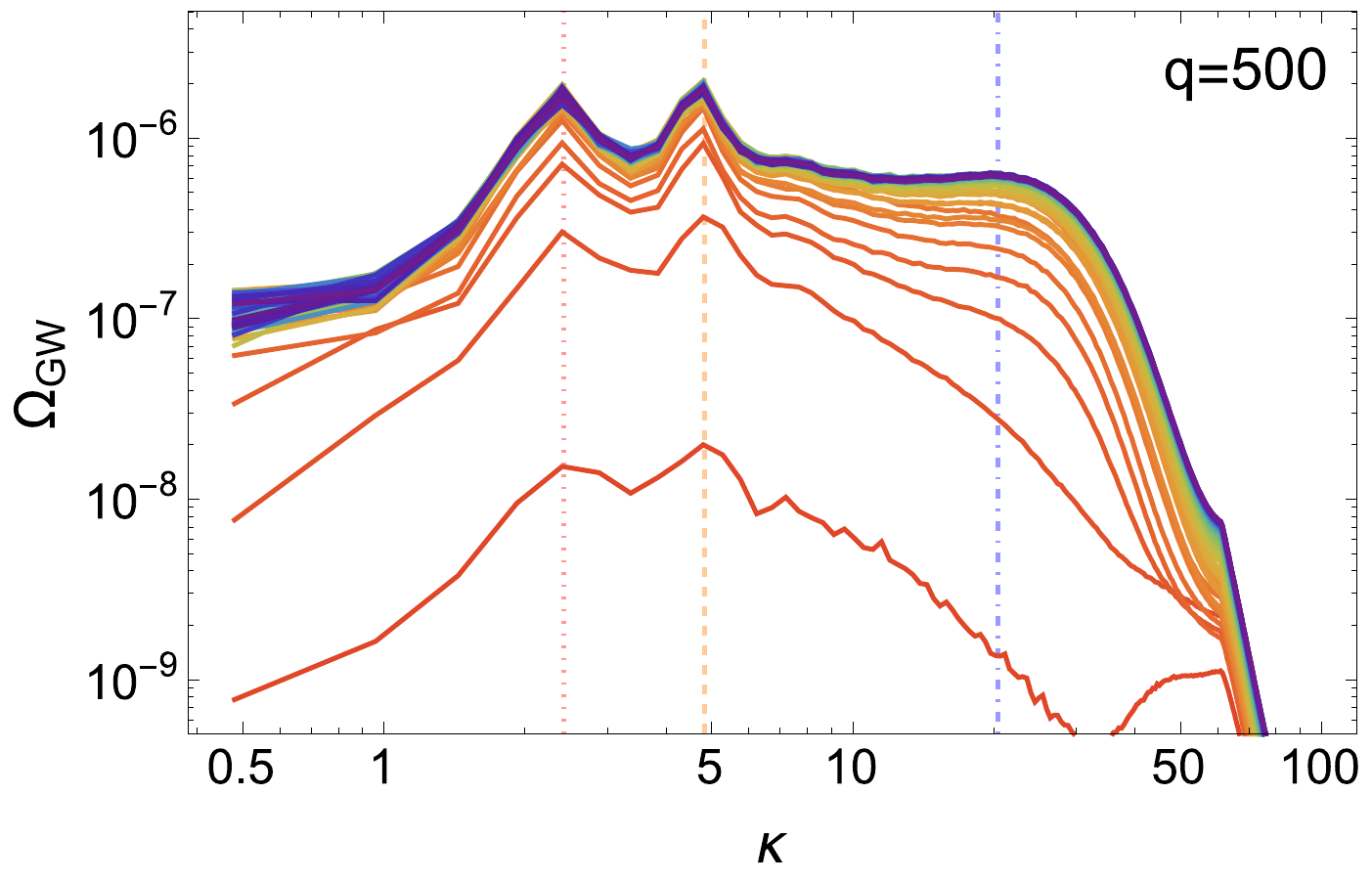} 
        \end{center}
      \caption{Numerical spectra of GWs $\Omega_{\rm GW} (\kappa, z)$, Eq.~(\ref{eq:lphi4-GWspectra}), as a function of the momentum $\kappa \equiv k / (\sqrt{\lambda} \phi_{\rm i})$, for the resonance parameters $q=5$, $q=30$, $q=200$ and $q=500$. In all panels, the spectra go from red at early times, to purple at late times, measured at regular intervals $\Delta z = 10$, from $z = 10$ up to $z = 690$. The vertical lines indicate the position of the peaks in the final saturated spectra, with the red dotted, yellow dashed, and blue dot-dashed lines indicating the position of $\kappa_1$, $\kappa_2$, and $\kappa_{\rm hb}$ respectively (see bulk text). } 
\label{fig:lphi4-GWspec}
 \end{figure}
 
We show in Fig.~\ref{fig:lphi4-GWspec} the time-evolution of the GW spectra for the resonance parameters $q=5,30,200,500$, obtained from our lattice simulations. We observe that the GW spectra grow several orders of magnitude in a short time $\Delta z \sim \mathcal{O}(10)$, saturating eventually at a given time scale $z_{\rm f}$, which signals the end of GW production. We have observed that typically, $z_{\rm br} < z_{\rm f} < z_{\rm dec}$, with the last order-of-magnitude growth of the GW spectrum amplitude taking place when the non-linear effects are becoming noticeable at $z \gtrsim z_{\rm br}$. For the given parameters of the figure, the final amplitude of the GWs after saturation is $\Omega_{\rm GW}^{\rm (f)} \sim \mathcal{O} (10^{-5}) - \mathcal{O} (10^{-6})$ approximately. 

\begin{figure}
      \begin{center}
       \includegraphics[width=15cm]{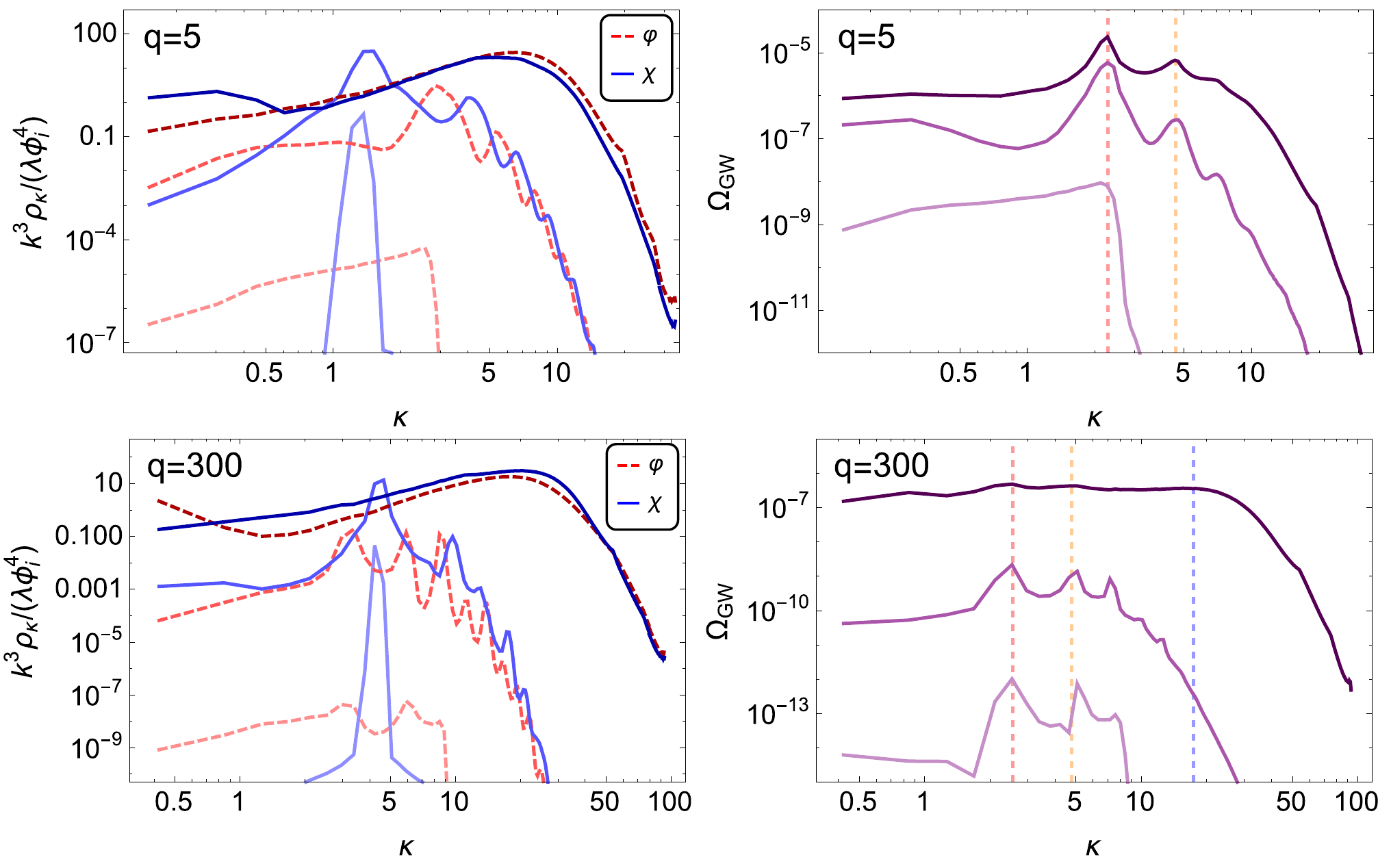} 
           \end{center} 
      \caption{In the top-left panel we show, for $q=5$ the energy density spectra of the inflaton and daughter fields [Eq.~(\ref{eq:energy-density})] at times $z=70, 105, 340$. The same quantities are shown in the bottom-left panel for the same times, but for the resonance parameter $q=300$. The right panels show the corresponding GW spectra at the same times. We also show here with dashed vertical lines the position of the peaks in the final saturated spectra: $\kappa_1$ in red, $\kappa_2$ in orange, and $\kappa_{\rm hb}$ in blue (explanations for these quantities are given in the bulk text).} 
            \label{fig:lphi4-peaksorigin}
 \end{figure}
 
Let us note that certain peaks emerge in the GW spectra during its evolution towards saturation, and some of these remain as features in the final saturated spectra (whereas others disappear). As seen in the EOM of the GWs [Eq.~(\ref{eq:lphi4-GWeom})], GWs are sourced by the matter fields, or more specifically, by their gradients. Therefore, one should be capable, in principle, to explain the origin of the peaks in the GW spectra, in terms of the dynamics of the matter fields in momentum space. To understand better the emergence of the peaks in the GW spectra, let us look at Fig.~\ref{fig:lphi4-peaksorigin}. There, we show the matter and GW spectra for two different resonance parameters, $q=5$ and $q=300$, at different times. 

Let us focus first in the case $q=5$ shown in the top two panels of Fig.~\ref{fig:lphi4-peaksorigin}. The lowest spectra shown in the left-top panel is measured at $z = 70$ ($< z_{\rm br} \approx 80$), when the backreaction effects from the daughter field have not yet affected significantly the inflaton homogeneous condensate. As expected, the daughter field is excited inside its main resonance band, while the inflaton fluctuations are still sub-dominant. At this time, as shown by the lowest spectra in the top-right panel, the GW amplitude is of the order $\Omega_{\rm GW} \sim \mathcal{O} (10^{-9})$. The middle spectra (in both left and right panels) are measured at the time $z = 105$ ($> z_{\rm br} \approx 80$), some time after the onset of backreaction, when the dynamics of the system is already fully non-linear. At that moment, we can observe two important features in the field spectra: first, the spectral amplitude of the inflaton and daughter fields have become comparable for all modes, and second, a detailed structure of peaks have appeared in both spectra. Such structures get imprinted in the GW spectra, which also show different peaks correlated in position with the location of the peaks of the matter fields. The GW amplitude has now become much greater, of the other $\Omega_{\rm GW} \sim \mathcal{O} (10^{-6})$ at the largest amplitude. Finally, the highest spectra are measured at time $z = 340$ ($> z_{\rm dec} \approx 160$). At this time, the initial peaks in the matter spectra have disappeared. Due to the mode-to-mode coupling generated by the non-linearities of the system, both spectra have transferred power to higher modes, developing a peak at shorter scales with a characteristic \textit{hunchback} shape. Correspondingly, the GW spectral power also moves towards the UV, and its amplitude gains a final order of magnitude growth, with the highest amplitude reaching up to $\Omega_{\rm GW} \sim \mathcal{O} (10^{-5})$. Let us remark that even though the structure of peaks is partially maintained in the final spectrum of GWs (specifically, the peaks indicated with red and orange vertical dashed lines), it is also partially washed out, as the peaks at the shorter scales are smoothed-out. 
 
\begin{figure}
      \begin{center} \includegraphics[width=9cm]{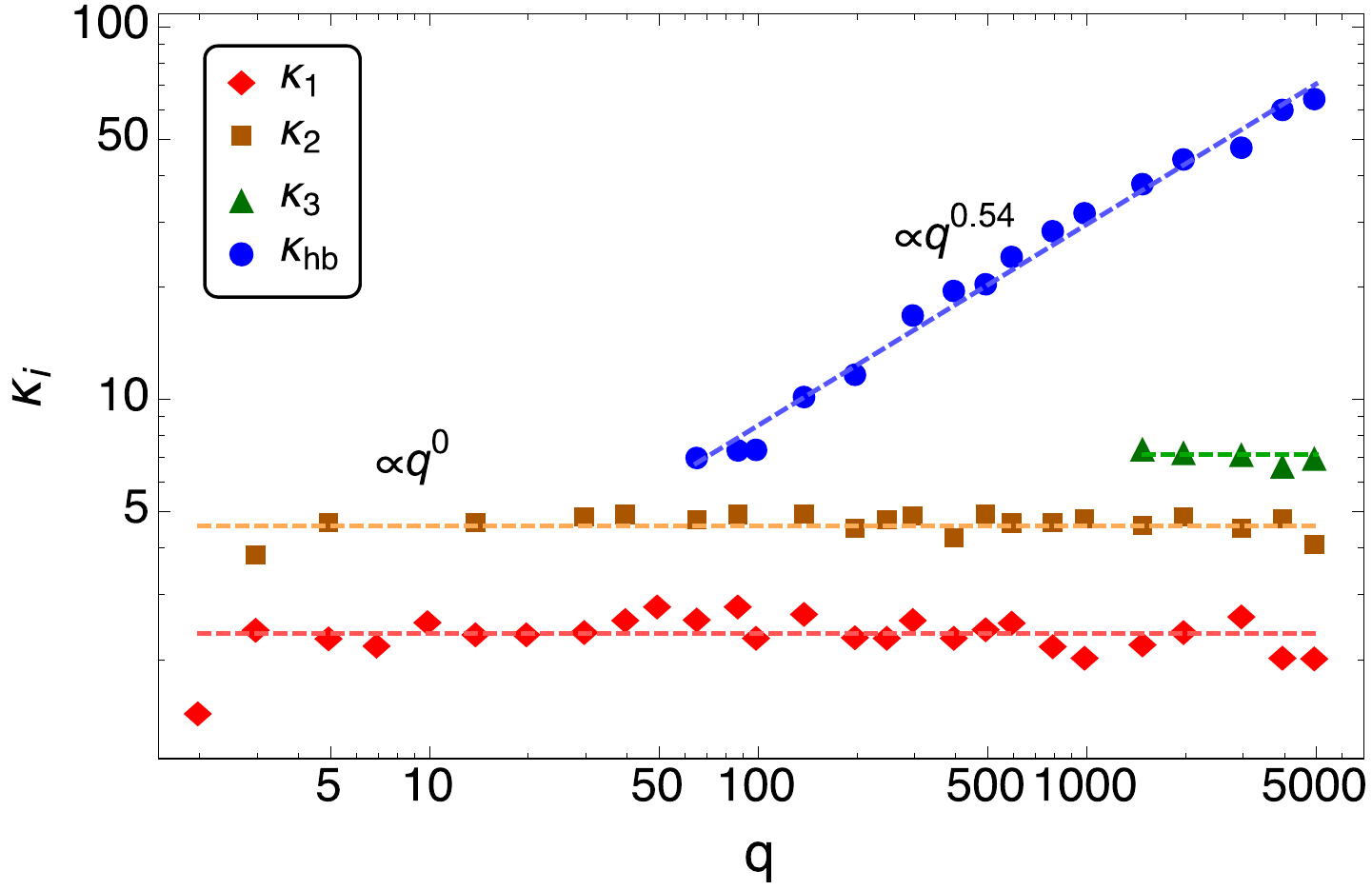} \end{center} 
      \caption{Left: We show the position $\kappa_i \equiv k_i /(\sqrt{\lambda}  \phi_{\rm i} ) $ of the different peaks in the saturated GW spectra, for the peaks $\kappa_1$ (red diamonds), $\kappa_2$ (orange squares), $\kappa_3$ (green triangles), and $\kappa_{\rm hb}$ (blue circles). The dashed lines indicate the fits Eqs.~(\ref{eq:lphi4-fitspectra1})-(\ref{eq:lphi4-fitspectra4}) to these quantities. } 
            \label{fig:lphi4-FitGW}
 \end{figure} 
 
A similar analysis can be done for the case $q=300$, shown in the lower panels of Fig.~\ref{fig:lphi4-peaksorigin}. The highest amplitude of the final saturated GW spectra [shown again at the time $z = 340$ ($> z_{\rm dec} \approx 160$)] is of the order $\Omega_{\rm GW} \sim \mathcal{O} (10^{-6})$. The peaks indicated with red and orange dashed lines, generated during the initial dynamics of the system, are still maintained in the final spectra. However, the main difference with respect to the previous case $q = 5$, is that now the 'displacement' of the matter spectra towards the UV, creates an additional peak in the GW spectra at short scales, with the same \textit{hunchback} shape as for the matter fields. This peak is indicated with a blue dashed line in the GW spectra shown in the lower panels of Fig.~\ref{fig:lphi4-GWspec} and right-bottom panel of Fig.~\ref{fig:lphi4-peaksorigin}, whilst it is absent for lower values of $q$, like those in the top panels of Fig.~\ref{fig:lphi4-GWspec} and right-top panel of Fig.~\ref{fig:lphi4-peaksorigin}. The hunchback peak is clearly generated during the late dynamics of the system, due to the mode-to-mode coupling between short and long modes, when the system is fully non-linear. The hunchback peak only appears visibly in simulations for $q \gtrsim 60$.

The location of the hunchback peak grows monotonically with $q$, and hence only when $q$ is sufficiently large, it becomes a well separated peak from the other more IR peaks. Phenomenologically, we have found that such threshold is precisely $q \gtrsim 60$. Remarkably, the IR peaks in the GW spectra are always placed at the same position, independently of $q$ (see location of red and yellow dashed vertical lines in all panels of Fig.~\ref{fig:lphi4-GWspec}). We think that the appearance of IR peaks at fixed scales, plus an extra peak in the UV at a $q$-dependent scale, is due to a combination of effects: on the one hand, the daughter field spectrum is peaked at $\kappa \sim q^{1/4}$, as we already discussed and quantified in Eq.~(\ref{eq:lphi4fits-MatterSpectra}). As we also discussed and quantified in Eq.~(\ref{eq:lphi4fits-MatterSpectra}), the large and rapidly growing amplitude of the daughter fluctuations 'drags', via the interaction term, the inflaton fluctuations at the same scale $\kappa \sim q^{1/4}$. Secondly, at the same time the inflaton, due to its own self-interactions, posseses a self-resonance for $q = 3$~\cite{Greene:1997fu}. Thus, the inflaton fluctuations start growing at some fixed IR scales due to its self-resonance, and the inflaton spectrum develops a structure of peaks, located always at the same scales, independently of $q$. The rate of growth of the inflaton fluctuations due to it self-resonance is however much slower than the rate of the broad resonant modes of the daughter field (for $q$ sufficiently large), as the resonance for $q = 3$ is much weaker in general than any other broad resonance for $q \gg 1$. However, since the two fields are coupled, the interaction term between them, leads eventually to the development of the same pattern of peaks in both daughter and mother field spectra. This happens mostly when the system becomes fully non-linear around $z \gtrsim z_{\rm br}$, so it is hard to develop an analytical description of it. However we note that, phenomenologically, we always observe this effect, independently of the value of $q$. For instance, this is clearly seen in the spectra at time $z \approx 105$, plotted in the top-left and bottom-left panels of Fig.~\ref{fig:lphi4-peaksorigin}. As a consequence, the GW spectrum ends also exhibiting some peaks in the IR at fixed positions. If the resonance parameter is sufficiently large ($q > 60$), then also a hunchback peak appears in the GW spectrum, at shorter scales. The hunchback peak becomes prominent mostly when the system becomes non-linear at $z \gtrsim z_{\rm br}$. We then expect that, due to the non-linear interactions among modes, the $q$-dependence of the location of the hunchback peak may differ from the linear prediction: presumably, given that it grows out of the initial peak developed at $\kappa \sim q^{1/4}$, it will still depend on $q$. However, given that it evolves significantly during to the non-linear stages of the system, some new $q$-dependence will most likely arise (not necessarily a small correction with respect to the linear expectation). Only by fitting the outcome of our simulations we can figure out the final $q$-dependence of the hunchback peak. 

\begin{figure}
 \begin{center}
       \includegraphics[width=13cm]{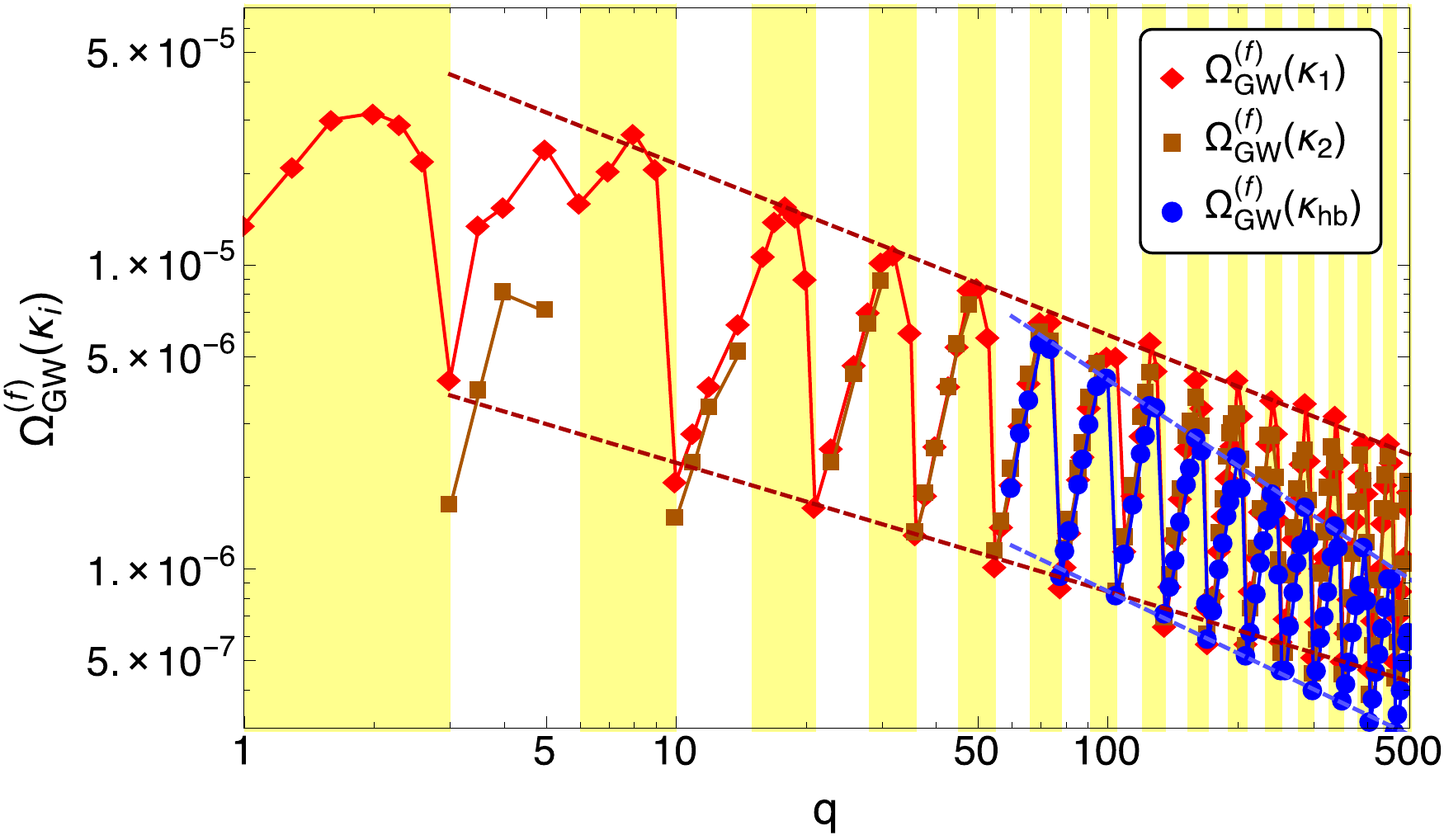} \end{center} 
      \caption{We show the amplitude of the GW spectra $\Omega_{\rm GW}^{\rm (f)}$ at peaks $\kappa_1$ (red diamonds), $\kappa_2$ (orange squares), and $\kappa_{\rm hb}$ (blue circles), as a function of $q$, in the interval $1 < q < 500$. The yellow vertical bands indicate the values of $q$ in which the main resonance band of the corresponding \emph{Lam\'e} equations is of the type $0 < \kappa < \kappa_+$. The diagonal dashed lines indicate the upper and lower bounds, whose fit we provide in Eqs.~(\ref{eq:lphi4-fitAmp1})-(\ref{eq:lphi4-fitAmp2}).} 
            \label{fig:lphi4-FitGWB}
 \end{figure}

In light of the discussion above, we proceed to parametrize the peaks in the final GW spectra $\Omega_{\rm GW}^{\rm (f)}$, as a function of the resonance parameter $q$. Our main results are presented in Figs.~\ref{fig:lphi4-FitGW} and \ref{fig:lphi4-FitGWB}. Let us start with Fig.~\ref{fig:lphi4-FitGW}, where we show the position of the peaks in the GW spectra, as obtained from the lattice simulations. Note that here we are indicating only the peaks that remain present at the final spectra after saturation. As seen, other peaks can appear in the intermediate, time-evolving spectra, but are washed out afterwards. We first observe two peaks, the location of which we denote as $\kappa_1$ and $\kappa_2$, whose position is clearly independent on the choice of $q$. These peaks appear for the whole range of resonance parameters simulated ($q \in  [1,5000]$), although in some cases the scales or the two peaks are so near that only one of them can be distinguished. These peaks are formed during the initial linear regime of the system, as described in our discussion above. An additional third peak is also observed in the (few) simulations done for $q \gtrsim 1000$, whose position is also independent on the particular choice of $q$. We denote the location of this peak as $\kappa_3$. We have fitted the position of these IR peaks as
\bea  \kappa_1 &\approx & 2.4 \pm 0.3 \label{eq:lphi4-fitspectra1} \ , \\
 \kappa_2 &\approx & 4.6 \pm 0.3  \label{eq:lphi4-fitspectra2}\ , \\
\kappa_3 &\approx & 7.1 \pm 0.3  \label{eq:lphi4-fitspectra3} \ , \hspace{0.3cm} (q \gtrsim 1000)  \ ,  \eea
with the error indicating some random scattering with $q$. 

On the other hand, for $q \gtrsim 60$ we observe an additional peak in the spectra, with its location growing monotonically with $q$. This is the peak with a \textit{hunchback} shape that we reported before, for instance for the case $q = 300$ shown in the right bottom panel of Fig.~\ref{fig:lphi4-peaksorigin}. This peak emerges visibly in the GW spectrum during the non-linear dynamics of the system. We denote its position as $\kappa_{\rm hb}$, for which we find the following power-law fit in the range $q \gtrsim 60$,
\be \kappa_{\rm hb} \approx 8.5 \left( \frac{q}{100} \right)^{0.54} \ , \hspace{0.3cm} (q \gtrsim 60) \, . \label{eq:lphi4-fitspectra4} \ee
As said, for $q \lesssim 60$, we cannot differentiate this peak from the others. The location of the hunchback peak depends on $q$, but as expected, it does not scale accordingly to the linear theory as $\sim q^{1/4}$. It rather scales as $\kappa_{\rm hb} \sim q^{1/2}$, demonstrating -- as argued above -- that the non-linear dynamics changes this peak location in a non-trivial way. The monotonic dependence on $q$ implies that the GW spectra exhibit a clear separation between IR and UV scale features, which grows with the strength of the interaction coupling. This is in fact one of the main reasons why it is unfeasible to simulate systems with arbitrarily large resonance parameter above $q \gtrsim 10^4$. Besides, the reason for $\kappa_3$ to only appear when the resonance parameter is sufficiently large, becomes now clear: only for a sufficiently large resonance parameter ($q > 10^3$ in this case), does the hunchback peak emerge at sufficiently separated (short) scales, hence preventing its IR tail to exceed the amplitude of the peak at $\kappa_3$. In other words, had we been able to simulate arbitrarily large resonance parameters, we would have observed a series of additional peaks, $\kappa_4$, $\kappa_5$, ..., at fixed positions independent of $q$, as long a the hunchback peak was separated enough in the UV (which would be always the case for sufficiently large $q$).

Let us now analyze how the amplitude of these peaks depend on the resonance parameter. In Fig.~\ref{fig:lphi4-FitGWB} we show the GW amplitudes $\Omega_{\rm GW}^{\rm (f)} (\kappa_1)$,  $\Omega_{\rm GW}^{\rm (f)} (\kappa_2)$ and  $\Omega_{\rm GW}^{\rm (f)} (\kappa_{\rm hb})$ as a function of $q$ in the interval $1 < q < 500$, obtained directly from lattice simulations. We have also indicated, with yellow bands, the values of $q$ for which the main resonance band is of the type $0 < \kappa < \kappa_+$, where the resonance is stronger (i.e. $q \in [1,3], [6,10], \cdots $). 

First, we observe that $\Omega_{\rm GW}^{(\rm f)} (\kappa_{\rm hb})$ follows a clear oscillatory pattern, with a dependence on $q$ correlated with the structure of resonance bands of the Lam\'e equation. This was clearly expected, as the strength of the resonance of the daughter field, determines the strength source of the GWs (i.e.~the gradients of the daughter field in this case), and consequently the strength of the GW final amplitude. More interesting is that both $\Omega_{\rm GW}^{(\rm f)} (\kappa_1)$ and $\Omega_{\rm GW}^{(\rm f)} (\kappa_2)$ also follow the same oscillatory pattern, correlated again with the structure of resonance bands of the Lam\'e equation. This is consistent with the picture we developed before: inside a resonance band, the larger the value of $q$, the wider the resonance band and the larger the Floquet index. Hence the broader and stronger the excitation of the daughter field is (initially peaked at $\kappa \sim q^{1/4}$). At the same time, a structure of peaks at fixed IR scales emerge, first in the inflaton spectrum, and later on, due to the interactions, also in the daughter field spectrum. Quite remarkably, the IR structure of peaks developed in the GW spectrum is then such that: on the one hand, the location of the peaks is fixed (as determined initially by the inflaton resonance bands for $q = 3$), while on the other hand, the amplitude of the peaks is modulated by the strength of the resonance of the daughter field, as dictated by the \textit{Lam\'e} equation (for the given resonance parameter $q$). In other words, for the IR peaks, the GW production becomes stronger (larger amplitude) the stronger the resonance of the daughter field is. Let us note that, even though $\Omega_{\rm GW}^{(\rm f)} (\kappa_2)$ follows the same oscillatory pattern as $\Omega_{\rm GW}^{(\rm f)} (\kappa_1)$, for values $q \lesssim 50$ it can be difficult to differentiate the two peaks, and hence the smaller number of data points associated to $\kappa_2$ in both Figs.~\ref{fig:lphi4-FitGW}, \ref{fig:lphi4-FitGWB}.

In all cases, the peak amplitudes $\Omega_{\rm GW}$ decay with $q$. In particular, the upper and lower envelopes of the oscillatory pattern of the $\Omega_{\rm GW}$ peak amplitudes, can be fitted as a decaying power-law with $q$. The decaying behavior of the amplitude is expected from the analytical prediction in Eq.~(\ref{eq:PeakAmplitudeV2}). The exponent of the power-law decay differs however from the analytical result $\Omega_{\rm GW} \propto q^{-1/2}$. We have fitted the upper and lower envelopes of the amplitude oscillations, from the numerical data measured at the relative maxima and minima. The fits are
 \bea
  8.4 \cdot 10^{-7} \left(\frac{q}{100} \right)^{-0.42} &\lesssim & \Omega_{\rm GW}^{\rm (f)} (\kappa_{1},\kappa_{2}) \lesssim 5.9 \cdot 10^{-6} \left(\frac{q}{100} \right)^{-0.56}  \ , \hspace{0.3cm} (q > 1) \ , \label{eq:lphi4-fitAmp1}\\  
 8.4 \cdot 10^{-7} \left(\frac{q}{100} \right)^{-0.68} & \lesssim & \Omega_{\rm GW}^{\rm (f)} (\kappa_{\rm hb}) \lesssim 4.2 \cdot 10^{-6} \left(\frac{q}{100} \right)^{-0.94}  \ , \hspace{0.3cm} (q > 60) \label{eq:lphi4-fitAmp2} \ .\eea
Note that we find $\Omega_{\rm GW}^{\rm (f) } (\kappa_2) \approx  \Omega_{\rm GW}^{\rm (f) } (\kappa_1 )$ (when $\kappa_2$ can be distinguished from $\kappa_1$), while the amplitude of the peak $\kappa_3$ is observed to be $\Omega_{\rm GW}^{\rm (f) } (\kappa_3 ) \approx 10^{-7}$, i.e.~always sub-dominant with respect the peaks at $\kappa_1$ and $\kappa_2$. As commented above,  if $q$ was sufficiently large, we would expect a richer structure of IR peaks (say at some scales $\kappa_4$, $\kappa_5$, ...). We also expect a monotonically decreasing pattern as $\Omega_{\rm GW}^{\rm (f) } (\kappa_{n} ) < \Omega_{\rm GW}^{\rm (f) } (\kappa_{n+1})$, as such is the pattern for the peaks in the matter fields spectra, see for example Fig~\ref{fig:lphi4-peaksorigin}.

As the analytical prediction ${d\log\Omega_{\rm GW}\over d\log q} = -{1\over2}$ is based on the linear regime analysis, it is not surprising that the real dependence of the GW amplitudes at the saturation time, bounded by Eqs.~(\ref{eq:lphi4-fitAmp1})-(\ref{eq:lphi4-fitAmp2}), differs from it. Yet, it is nice to observe that the GW amplitudes follow, at least, a decaying power-law with $q$. The deviation of the measured exponents $-0.42 \lesssim {d\log\Omega_{\rm GW}\over d\log q} \lesssim -0.94$ with respect to the linear prediction $-0.5$ is attributed to the non-linear dynamics, and could have not been predicted {\it a priori} without numerical simulations.

We can now redshift the amplitude and position of the GW peaks. Using Eq.~(\ref{eq:ftoday})
we obtain the following frequencies
today
\be f_p = \kappa_p \times 6 \cdot 10^6 \ {\rm Hz}\label{eq:lphi4fpkp} \ . \ee
Substituting Eqs.~(\ref{eq:lphi4-fitspectra1})-(\ref{eq:lphi4-fitAmp2}) into Eq.~(\ref{eq:lphi4fpkp}), we obtain that the exact frequencies of the peaks today are
\bea
f_1 &\approx &
1.5 \cdot 10^7 \ {\rm Hz} \ , \\
f_2 & \approx &
2.8 \cdot 10^7 \  {\rm Hz} \ , \\
f_3 & \approx &
4.5 \cdot 10^7 \  {\rm Hz}   \ , \hspace{0.4cm} ({\rm only\,\,for\,\,} q \gtrsim 10^3) \ , \\
f_{\rm hb} &\approx  &  \left( \frac{q}{100} \right)^{0.54}
\times 5.3 \cdot 10^7 \ {\rm Hz}  \ , \hspace{0.4cm} ({\rm only\,\,for\,\,} q \gtrsim 60) \ .
\eea
Using Eq.~(\ref{eq:GWtoday}) we also find that the redshifted amplitude(s) today of this background is $h^2 \Omega_{\rm GW}(f_p) \simeq 4\cdot 10^{-6} \Omega_{\rm GW}^{\rm (f)}(\kappa_p)$. This translates into the following (interval of) amplitudes for the measured peaks,
\bea \label{eq:BoundedIRtodayPhi4}
  3.4 \cdot 10^{-12} \left( \frac{q}{100} \right)^{-0.42} &\lesssim & h^2 \Omega_{\rm GW} (f_{1,2}) \lesssim 2.4 \cdot 10^{-11}  \left( \frac{q}{100} \right)^{-0.56}  \ , \\
  \label{eq:BoundedHBtodayPhi4}
  3.4 \cdot 10^{-12}  \left( \frac{q}{100} \right)^{-0.68} & \lesssim & h^2 \Omega_{\rm GW} (f_{\rm hb}) \lesssim 1.6 \cdot 10^{-11}  \left( \frac{q}{100} \right)^{-0.94}  \ . \hspace{0.3cm} 
\eea
These amplitudes are in perfect agreement with the background amplitudes computed in the past for this scenario in the case $g^2/\lambda = 120$, where it was obtained~\cite{Easther:2006vd,GarciaBellido:2007af,Easther:2007vj,Price:2008hq} $h^2\Omega_{\rm GW} \sim 10^{-11}$ at the peak amplitude\footnote{Let us note that the amplitudes obtained in the cited papers may differ by a factor $\sim 1-2$ with respect the $rhs$ inequality amplitude of $\Omega_{\rm GW} (f_{\rm hb})$ in Eq.~(\ref{eq:BoundedHBtodayPhi4}). This is mostly due to different assumptions in the ratio of relativistic species entering in the redshifting amplitude formula Eq.~(\ref{eq:GWtoday}), as $\left(g_{o}/g_{\rm RD}\right)^{1/3} \simeq 0.22-0.10$ for $g_{o}/g_{\rm RD} = 0.01-0.001$. We have made the most conservative choice $g_{o}/g_{\rm RD} = 0.001$, and hence our bounded amplitudes are a factor $\simeq 0.5$ smaller than if we had chosen the Standard Model like value $g_{o}/g_{\rm RD} \sim 0.01$.}.

Even though our analytical prediction in Eq.~(\ref{eq:PeakAmplitudeToday}) was based on the linear analysis, we can still calibrate it based on the numerical outcome. In particular, we can use the highest GW signal, ocurring at the local maxima of the oscillatory pattern in Fig.~\ref{fig:lphi4-FitGWB}, to extract the parameters $C^2$ and $\delta$ characterizing the theoretical prediction. In particular, as $\epsilon_{\rm i} = 1$, $\omega_*^2 \equiv \lambda \phi_i^2$,
and $\rho_{\rm i} \approx {\lambda\over4}\phi_i^4$, 
from equating
\begin{equation}
\Omega_{\rm GW}\big|_{\rm th} \simeq 10^{-9} \times \epsilon_iC^{2}\frac{\omega_*^{6}}{\rho_i m_{p}^{2}}\,q^{-\frac{1}{2}+\delta} ~~~ = ~~~ \Omega_{\rm GW}\big|_{\rm num} \simeq 1.6 \cdot 10^{-11}  \left( \frac{q}{100} \right)^{-0.94} \ ,
\end{equation}
we deduce
\begin{eqnarray}
\delta \gtrsim -0.44\,,~~~~~~{\rm and}~~~~~~ C \simeq {0.61\over \lambda}\left(\frac{m_{p}}{\phi_i}\right) \ .
\end{eqnarray}

\subsection{Lattice simulation of preheating with quadratic potential}
\label{sec:m2phi2-results}

We switch now to study the production of GWs during preheating with quadratic potential,
\be V(\phi) = \frac{1}{2} m^2 \phi^2 \ , \ee
in the case when the inflaton is coupled to another scalar daughter field with coupling $g^2 \phi^2 X^2$.
The natural frequency of oscillation in this model is obviously $\omega_* \equiv m$. We can then define new dimensionless field, spacetime, and momentum variables as
\be \varphi = \frac{1}{\phi_{\rm i}}a^{3/2} \phi \ , \hspace{0.3cm} \chi = \frac{1}{\phi_{\rm i}} a^{3/2} X \ , \hspace{0.3cm} z \equiv m (t - t_{\rm i}) \ , \hspace{0.3cm} \vec{y} \equiv m \vec{x} \ , \hspace{0.3cm}  \kappa = \frac{k}{m} \label{eq:m2phi2-newvariables} \ ,\ee
where $\phi_{\rm i } \equiv \phi (t_i)$, and the time $t_{\rm i}$ is defined below. In these variables, the equations of motion of the mother and daughter fields (\ref{eq:generic-eom}) become
\be \varphi '' + \left( - \frac{3}{4} \frac{a'^2}{a^2} - \frac{3}{2} \frac{a''}{a} \right) \varphi - \frac{1}{a^2} \nabla_y^2 \varphi + \left( 1 +  \frac{4}{a^{3}} q \chi^2 \right)  \varphi  = 0 \ , \label{eq:m2phi2-eom1}\ee
\be \chi'' + \left( - \frac{3}{4} \frac{a'^2}{a^2} - \frac{3}{2} \frac{a''}{a} \right) \varphi - \frac{1}{a^2} \nabla_y^2 \chi + \frac{4}{a^3} q \varphi^2 \chi = 0 \ , \label{eq:m2phi2-eom2}\ee
where $' \equiv d / dz$, $\nabla_y^2 f \equiv \sum_i (\partial f / \partial y_i)^2$, and the resonance parameter is defined as 
\be q \equiv \frac{g^2 \phi_{\rm i}^2}{4 m^2} \label{eq:m2phi2-resq}\ . \ee
Note that in this Section we use $z$ to represent (dimensionless) cosmic time, instead of conformal time as we did in Eq.~(\ref{eq:naturalVariables}). In cosmic time the inflaton oscillation period is exactly constant. Remember also that here, the resonance parameter defined in Eq.~(\ref{eq:m2phi2-resq}) includes an extra $1/4$ factor with respect that in Eq.~(\ref{eq:ResParam}), to match the usual description of the linear stage of parametric resonance described by the {\it Mathieu} equation.

As discussed in Section \ref{sec:Analytics}, towards the end of inflation the Hubble parameter becomes of the order of the inflaton mass $m$, and preheating starts. The inflaton starts oscillating around the minimum of its potential with frequency $\omega_* \equiv m$. This induces a time-dependent effective mass to all particles coupled to it, generating a strong particle production through parametric resonance. We define the time $t_{\rm i}$ when the condition $H (t_{\rm i}) = m$ holds exactly. Solving numerically and simultaneously the homogeneous inflaton field dynamics and the Friedmann equations, we find
\be \phi_{\rm i} \equiv \phi (t_i) \approx 2.32 m_p \ , \hspace{0.4cm} \dot{\phi} (t_{\rm i}) \approx -0.78 m m_p \ .\ee
We take this time $t_i$ as the moment when we start our lattice simulations.

Before describing and parametrizing the production of GWs, let us briefly review some results about the dynamics of preheating in this model. As discussed in Section \ref{sec:Analytics}, the inflaton can be taken as a homogeneous field during its first oscillations, when the backreaction effects from the daughter field can be neglected. In this regime, the inflaton solution in the absence of expansion (i.e. if $a=1$) is simply $\varphi \propto \cos (z)$,  and the equation of motion of the $\chi$ field modes takes the form of a \textit{Mathieu} equation, see e.g.~\cite{Kofman:1997yn} or {\it Paper I}. This is a very well studied equation with well understood solutions. In particular, the solutions exhibit a structure of resonance bands, similar to the ones for the \textit{Lam\'e} equation described in Section \ref{sec:lphi4-results}. For certain regions in the $(\kappa,q)$ parameter space, the solution of the $\chi$ field modes is exponential, $\chi_k \sim e^{\mu_k t}$, with a real Floquet index $\mathfrak{Re} [\mu_k] > 0$. Two regimes can be distinguished in this model, according to the structure and width of the resonance bands in the \textit{Mathieu} equation. If $q<1$, only some very narrow sets of momenta experience exponential creation, and due to this, its dynamics cannot be well captured with lattice simulations. On the other hand, if $q > 1$ we are in broad resonance, and in this case a wide band of momenta within $\kappa = 0$ and a maximum scale $\kappa_{\rm M} \propto q^{1/4}$, is excited. As the resonance bands in the $q \gg 1$ case are so wide (hence the name 'broad parametric resonance'), the system can be simulated in the lattice.

\begin{figure}
      \begin{center} \includegraphics[width=12cm]{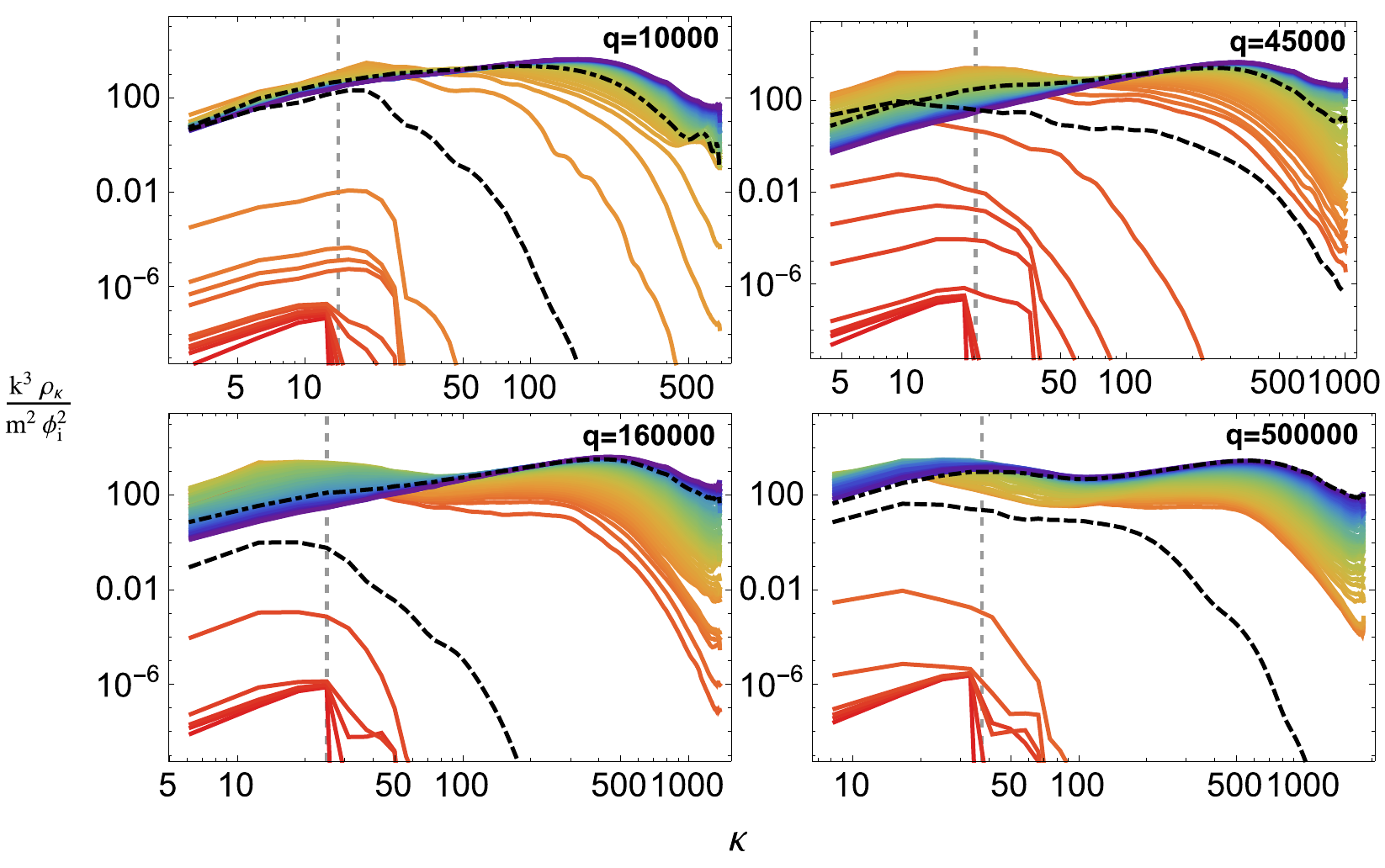} \end{center} 
      \caption{We show the time-evolution of the daughter-field energy density spectra $ \rho_{\kappa, \chi}$ [Eq.~(\ref{eq:m2phi2-energy-density})] as a function of the momentum, for the resonance parameters $q = 10000, 45000, 160000, 500000$. The spectra are measured at equally spaced times $z = 10, 20, \dots$, with red lines corresponding to early times, and purple lines to late times (after a stationary regime has been established). The gray, vertical dashed lines indicate the position of the maximum momentum excited according to the linear analysis, which scales as $\propto q^{1/4}$. We also show with black dashed and dot-dashed lines the spectra at times $z \approx z_{\rm br}$ and $z \approx z_{\rm dec}$.} 
            \label{fig:m2phi2-spectra}
 \end{figure}

However, when the expansion of the Universe is included, the dynamics becomes more complicated. The effective resonance parameter in Eq.~(\ref{eq:m2phi2-eom2}) becomes $q_{\rm eff} = q a^{-3}$, continuously decreasing with time. Due to this, a system that starts in broad resonance with $q_{\rm eff} > 1$, eventually reaches a regime of narrow resonance with $q_{\rm eff} < 1$. In this process, a particular comoving mode $k$ of the daughter field $\chi_k$ scans many resonance bands within one inflaton oscillation, and as a result, the effective resonance is of stochastic nature~\cite{Kofman:1997yn}. In broad resonance only modes below a given cut-off are excited,
\be\label{eq:kM} \kappa \lesssim \kappa_M \sim q^{1/4} \ .\ee
In order to trust or lattice simulations, we must ensure that before the system shifts to narrow resonance, the backreaction effects from the daughter field onto the inflaton condensate have already induced its decay. In \textit{Paper I}, we observed that in practice, this requirement implies that systems with $q \lesssim 6000$ cannot be appropriately simulated. In {\it Paper I} we identified and parametrized two main time scales in the preheating process. The first one, $z_{\rm br}$, indicates the time when the inflaton energy density starts decaying noticeably due to the backreaction effects from the daughter field. From lattice simulations in the range $6000 \lesssim q \lesssim 2.5 \cdot 10^6$, we found that $z_{\rm br} \in [40,130]$, with a pattern of stochastic oscillations within that interval of values. At the time $z \simeq z_{\rm br}$, and independently\footnote{In reality we expect a logarithmic dependence, see {\it Paper I}.} of $q$, only $\sim 1\%$ of the total energy has been transferred to the $\chi$ field (mainly kinetic energy), while the other $\sim 99 \%$ is still in the $\varphi$ field (kinetic and potential energies). The second relevant time scale is $z_{\rm dec}$, indicating the onset of a stationary regime after the inflaton decay ceases. Our simulations showed that it behaves as $z_{\rm dec} \propto q^{0.27}$. We refer the reader to {\it Paper I} for a more extended discussion about the parametrization and technical definition of these scales.

In this work, we have done real-time classical lattice simulations of the preheating process with a quadratic potential, computing the associated GW background created during the process. Our simulations are made in regular boxes with $N^3 = 256^3$ points, with a box size chosen so that the minimum momentum is $\kappa_{\rm min} \sim \mathcal{O} (1) < q^{1/4}$ , guaranteeing in this way that the lattice captures the relevant momenta modes for the dynamics. We have run simulations varying the resonance parameter within the broad interval $q \in [6 \cdot 10^3, 10^6]$. The reasons for the lower bound have already been explained, while the upper bound is mainly due to the inability of covering appropriately the UV dynamics with increasing $q$. Besides, it also becomes more and more expensive in term of computational time to simulate large resonance parameters, as the decay time grows monotonically as $z_{\rm dec} \sim q^{1/4}$. The interested reader can find an extended explanation of these restrictions in Appendix B of {\it Paper I}.

In Fig.~\ref{fig:m2phi2-spectra} we plot the time-evolution of the energy density spectra of the daughter field as a function of the momentum, 
\be k^3 \rho_{k, \chi} = \frac{m^2 \phi_{\rm i}^2}{2} \kappa^3 a \left(  | \chi^{'}_{\kappa} - \frac{a'}{2 a } \chi_{\kappa} |^2  + \omega_{\kappa,\chi}^{2} | \chi_{\kappa} |^2 \right) \ , \hspace{0.5cm} \omega_{\kappa,\chi} = \sqrt{\frac{\kappa^2}{a^2} + \frac{4}{a^3} q \varphi^2 
} \label{eq:m2phi2-energy-density} \ , \ee
for four different resonance parameters. We have highlighted the spectra at times $z_{\rm br}$ and $z_{\rm dec}$ with dashed black lines. As expected, we clearly observe that before $z \lesssim z_{\rm br}$, the excitation of the field modes occurs mainly inside the resonance band $\kappa_{\rm M} \propto q^{1/4}$, while for $z \gtrsim z_{\rm br}$ the system becomes non-linear and power is transferred to higher modes in the UV. At $z \approx z_{\rm dec}$, the spectra have already saturated, as the fields have just reached a stationary regime, and hence do not source GWs anymore. During the process, only a single peak emerges in the matter spectra, and consequently, only a single peak is expected in the GW spectra. 

\begin{figure}
      \begin{center}
       \includegraphics[width=7.3cm]{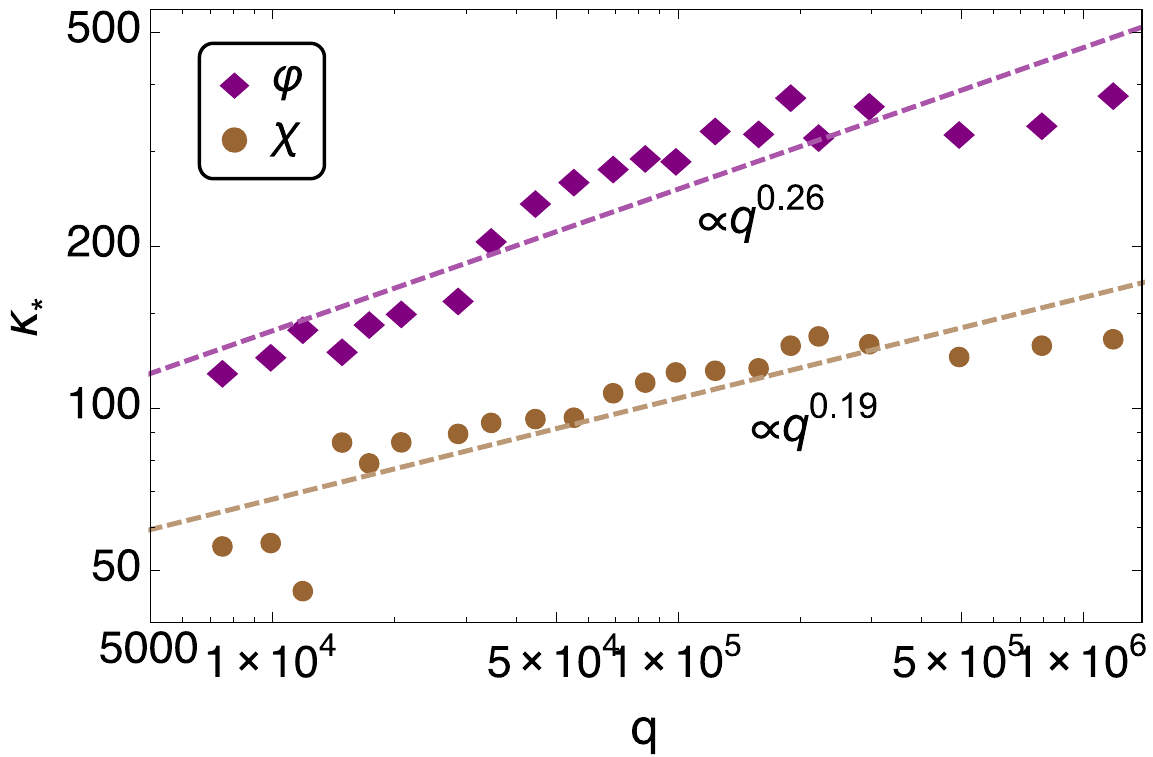} \,\,
        \includegraphics[width=7.4cm]{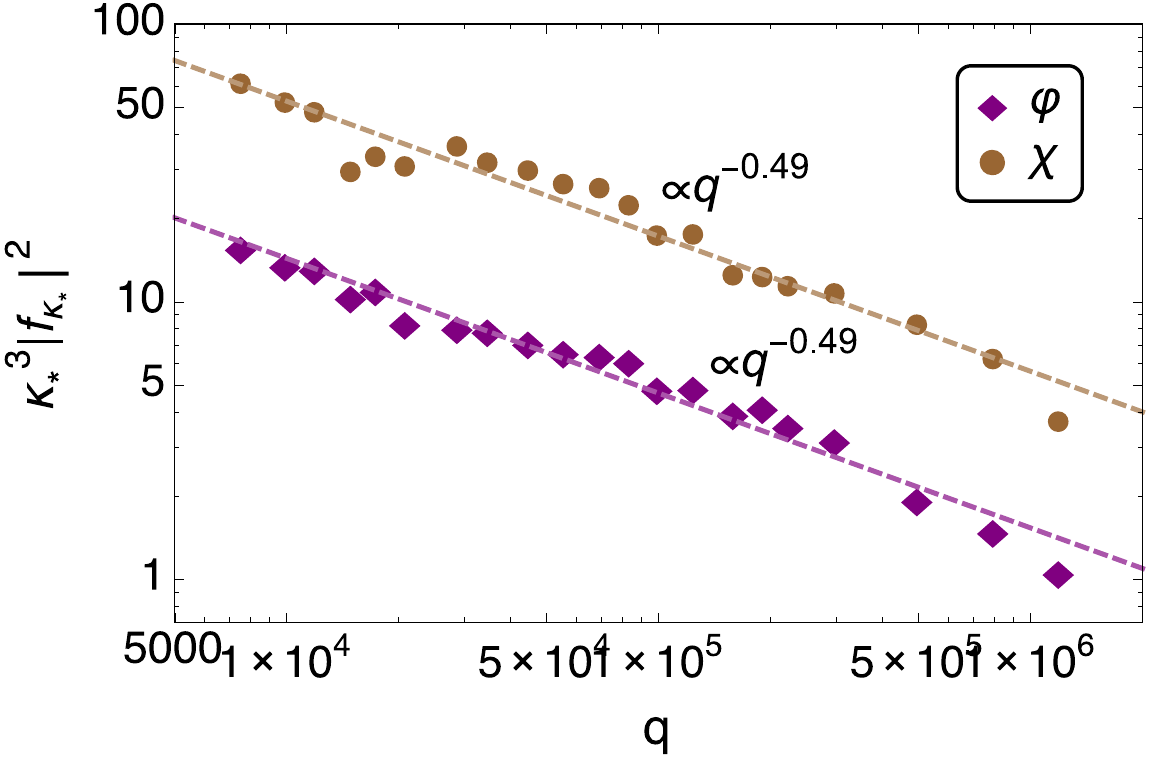} 
        \end{center}
      \caption{We plot, for the inflaton and daughter fields, the position of the peak $\kappa_*$ in the spectra after saturation as a function of $q$ (left panel), and the corresponding amplitude $\kappa_*^3 |f_{\kappa_*}|^2$ ($f= \varphi, \chi$) (right panel). Dashed lines in both panels correspond to the fits in Eq.~(\ref{eq:m2phi2fits-MatterSpectra}).} 
            \label{fig:m2phi2-GWspec}
 \end{figure}

In Fig.~\ref{fig:m2phi2-GWspec}, we show the position $\kappa_*$ where a peak appears in the fields spectra $\kappa^3 |f_{\kappa}|^2$ (left panel), as well as the corresponding peak amplitude $\kappa_*^3 |f_{\kappa_*}|^2$ (right panel), obtained from our lattice simulations for different choices of $q$. We obtain the following fits for these quantities,
 \bea  {\rm Daughter~field}~\chi : \hspace{0.5cm} \kappa_* & \approx & 69 \left(q \over 10^4 \right)^{0.19}   \ , \hspace{0.64cm} \kappa_*^{3}{ |\chi_{\kappa_*} |^2} \approx  53 \left(q \over 10^4 \right)^{-0.49} \ , \nonumber \\ 
 {\rm Mother~field}~\varphi: \hspace{0.5cm}\kappa_* &\approx & 136 \left(q \over 10^4 \right)^{0.26} \ , \hspace{0.5cm} \kappa_*^{3} {| \varphi_{\kappa_*} |^2} \approx  14 \left(q \over 10^4 \right)^{-0.49}
  \ . \label{eq:m2phi2fits-MatterSpectra}
 \eea
Like in the case of a quartic inflationary potential, the expected power-law scaling $\propto q^{-1/2}$ for the daughter spectral peak holds also quite well (within the sampling), with a deviation of the mean exponent with respect the theoretical prediction of only $100\times{|0.49-0.5|\over0.5} \sim 2\%$. The theoretical location of the daughter field's peak at $\kappa_* \sim q^{1/4}$ is however only realized with a correction (of the exponent) of $100\times{|0.19-0.25|\over0.25} \sim 24\%$. The fact that the location of the daughter spectra deviate to some extent from the theoretical expectation is actually expected, as strictly speaking such prediction is only valid when the linear regime applies. The spectra fitted in Eq.~(\ref{eq:m2phi2fits-MatterSpectra}) are however measured at $z \simeq z_{\rm dec}$, after the sytem went non-linear. The fact that the amplitude of the spectrum follows then so well the theoretical scaling as $\propto q^{-1/2}$ is again, certainly remarkable. 

Analogously to the quartic case, the mother field $\varphi$ also follows a similar scaling as the daughter field. The (mean value of the) peak position of the mother field spectra is actually located almost exactly at the theoretical expectation $\sim q^{1/4}$, whereas the exponent of the spectral peak amplitude presents only a deviation of $100\times{(0.49-0.5)\over0.5} \sim 2\%$. This can only be explained, again, due to the coupling between the daughter and the mother fields: slightly before $z \simeq z_{\rm br}$, when the daughter field mode amplitudes have grown significantly (following the linear analysis resonance), the modes of the mother field become excited through the interaction term. This 'dragging' effect excites exactly the same inflaton modes as in the daughter spectra.

\subsubsection{Gravitational wave parametrization}

We now proceed to study the GW production in the quadratic potential model. To do so, let us define a rescaled GW field as $\bar{h}_{ij} \equiv a^{3/2} h_{ij}$. The EOM of the GWs, Eq.~(\ref{eq:GWeom}), can then be written as
\be \bar{h}_{ij}'' - \nabla_y^2 \bar{h}_{ij} - \left( \frac{3}{4} \frac{a'^2}{a^2} + \frac{3}{2} \frac{a''}{a} \right) \bar{h}_{ij} = \frac{2 \phi_{\rm i}^2}{m_p^2 a^{7/2} } ( \partial_i \varphi \partial_j \varphi + \partial_i \chi \partial_j \chi)^{\rm TT} \ .\ee
The total energy $\rho_t$ of the system contributed by the matter fields (the contribution from the GWs is negligible) is
\be \rho_t = \frac{m^2 \phi_{\rm i}^2}{2 a^3} \times \left[  \sum_{f=\varphi,\chi} \left(  f' - \frac{3}{2} \frac{a'}{a} f \right)^2 + \frac{1}{a^2} \sum_{f=\varphi,\chi}  | \nabla_y f |^2 + \left( 1 + \frac{4 q}{a^3} \chi^2 \right) \varphi^2  \right] \equiv \frac{m^2 \phi_{\rm i}^2}{2 a^3} E_t \ . \ee
The amplitude of the stochastic background of GW Eq.~(\ref{eq:ThetaGW}) can then be written as
\be \Omega_{\rm GW} (k, z) = \frac{1}{\rho_t} \frac{d \rho_{{\rm GW}}}{d \log k} (k,z) =  \frac{m_p^2}{\phi_i^2}\frac{(m\kappa)^3}{4 \pi^2 V E_t(z)} \int \frac
{d \Omega_k}{4 \pi}  \bigl\lvert \bar{h}'_{ij} - \frac{3}{2} \mathcal{H} \bar{h}_{ij} \bigr\rvert^2   \ . \label{eq:m2phi2-GWspectra}\ee
For the discrete version of the GW spectrum, see Appendix~\ref{app:Lattice-formulation}.

 \begin{figure}
      \begin{center}
       \includegraphics[width=7.5cm]{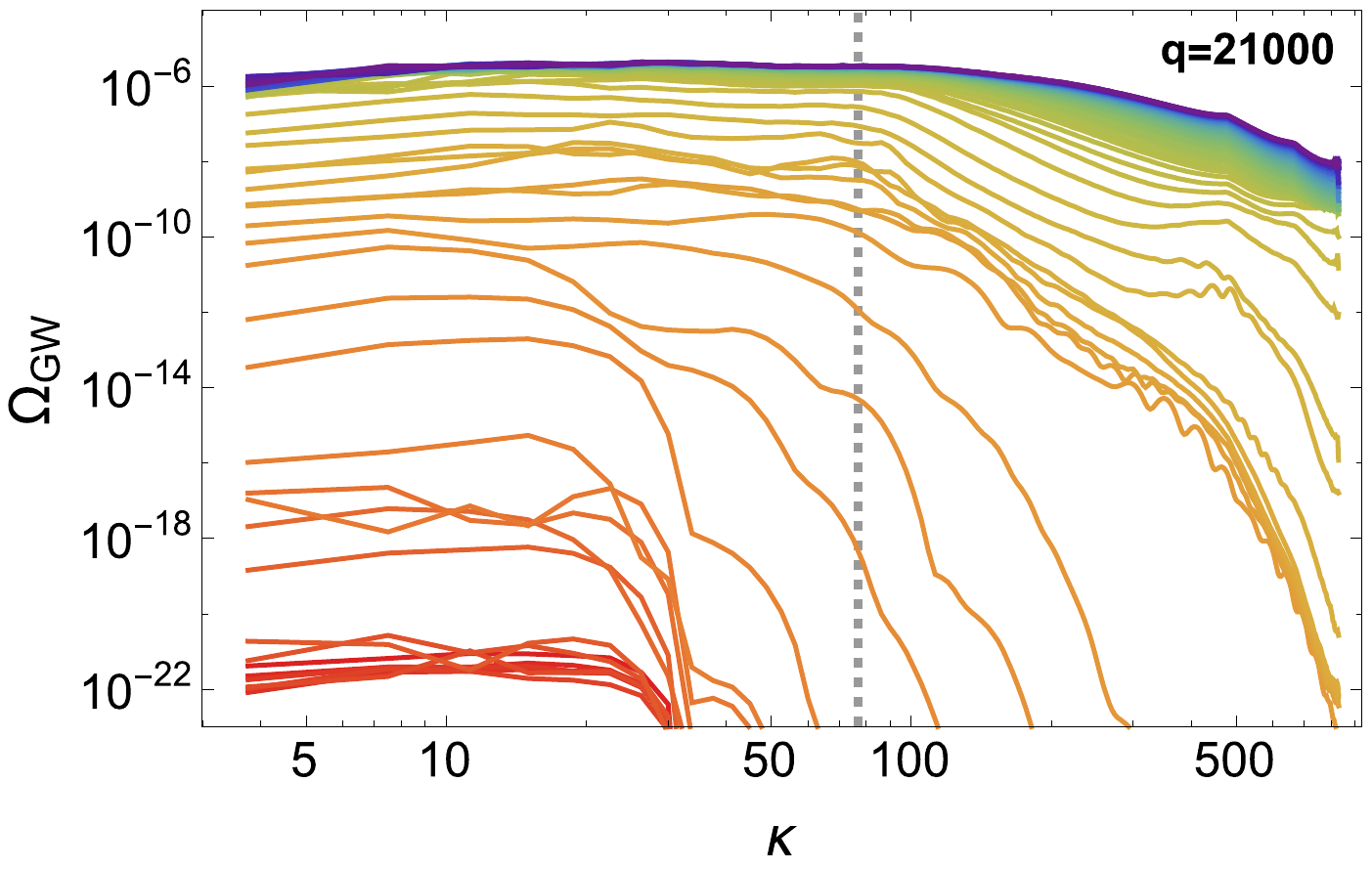} 
              \includegraphics[width=7.5cm]{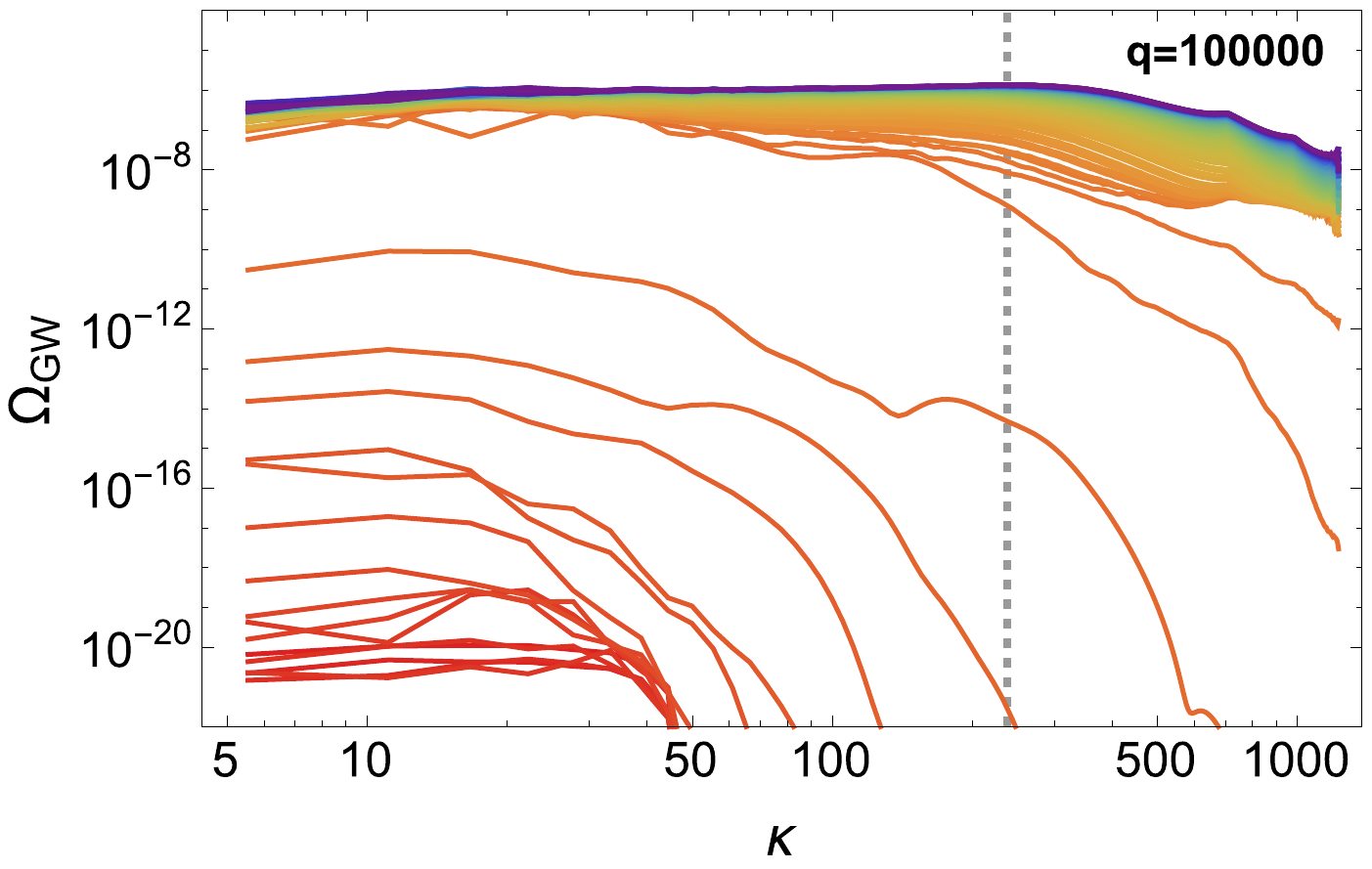} 
                     \includegraphics[width=7.5cm]{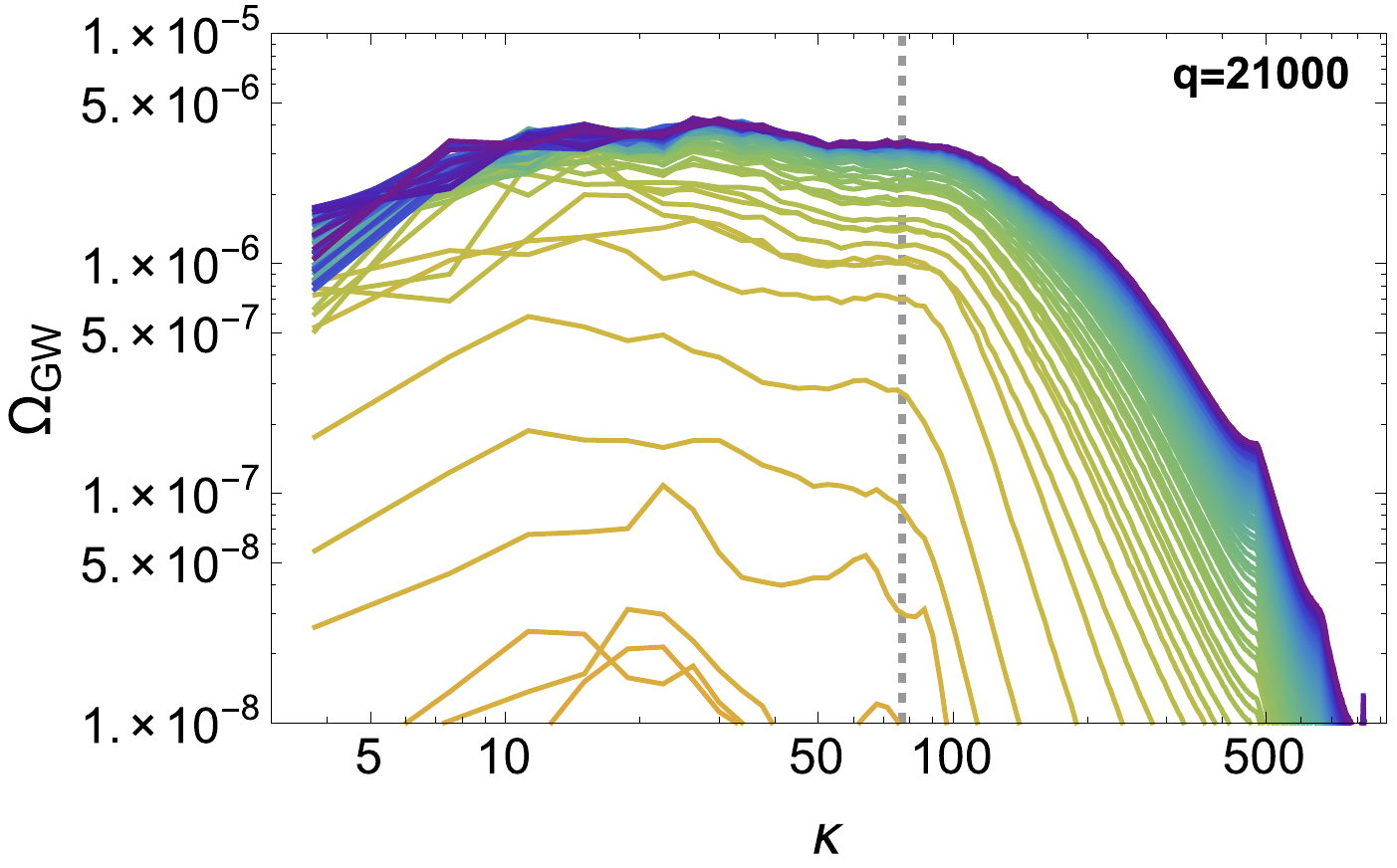} 
              \includegraphics[width=7.5cm]{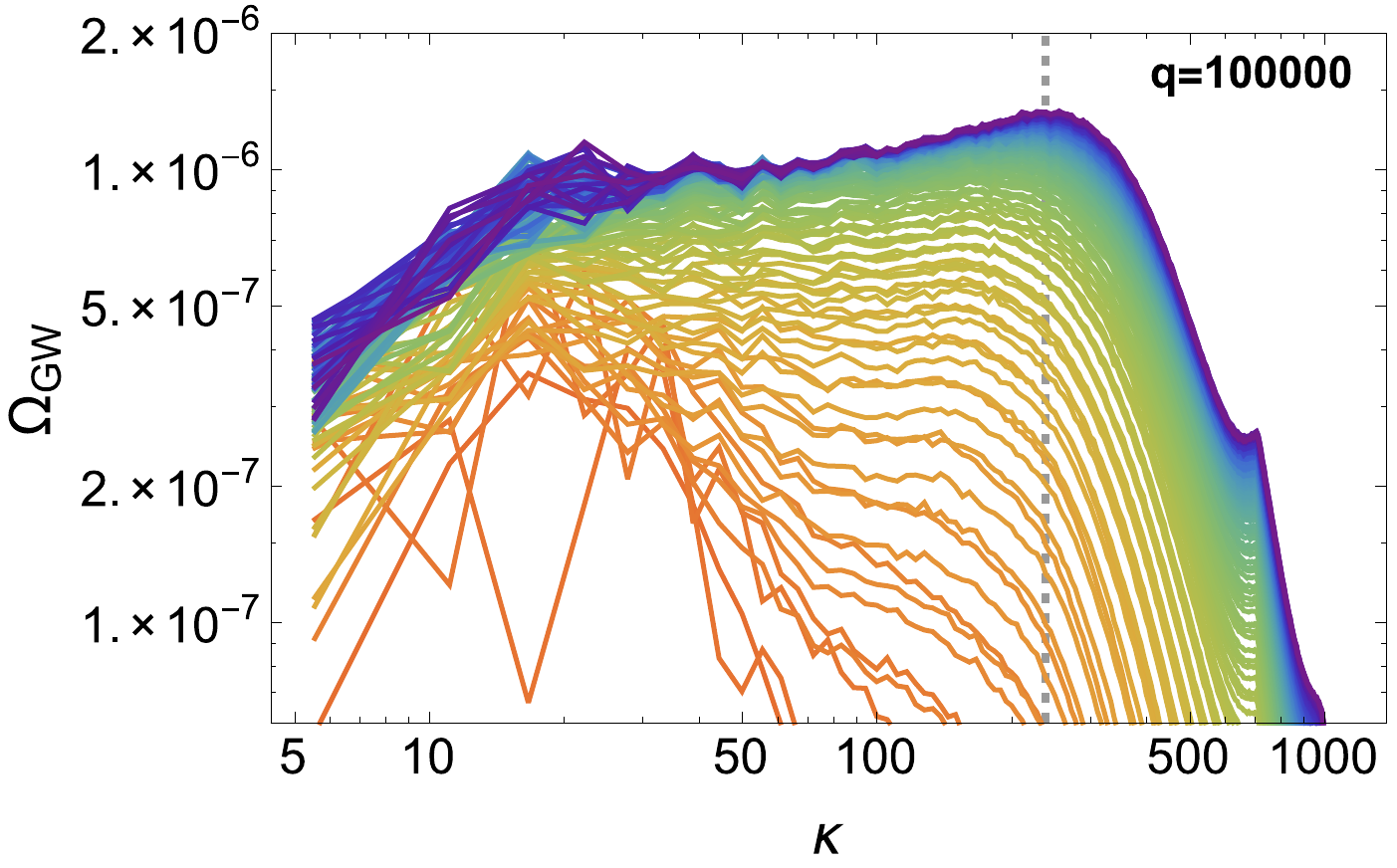} 
        \end{center}
      \caption{The top panels show the time-evolution of the GW spectra $\Omega_{\rm GW} (\kappa, z)$ for the quadratic preheating model, for both $q=21000$ (top-left) and $q=100000$ (top-right). The spectra are depicted at times $z = 0,5,10,\dots$, and go from red (early-times) to blue (late-times). The bottom panels show the same quantities, but zoomed to observe the peak better.} 
            \label{fig:m2phi2-GW}
 \end{figure}

In the top panels of Fig.~\ref{fig:m2phi2-GW} we show the time-evolution of the {\rm GW} spectra produced during preheating, for the cases $q=2.1\cdot 10^{4}$ and $q = 10^5$. We observe that the initial fluctuations imposed in the matter fields generate an initial GW amplitude of the order $\Omega_{\rm GW} \sim 10^{-22}$. During the subsequent preheating stage, the amplitude grows fifteen orders of magnitude, saturating at a final amplitude of the order $\Omega_{\rm GW}^{\rm (f)} \sim 10^{-6}$. During the GW creation there is a significant population of higher modes beyond the initial cut-off scale Eq.~(\ref{eq:kM}). Therefore, a significant displacement of the {\rm GW} spectra towards UV scales occurs, as higher modes of the GW are also populated. During this displacement, a peak forms at a given scale $\kappa_p > \kappa_M$. We will refer to the final amplitude of this peak as $\Omega_{\rm GW}^{\rm (f)} (\kappa_p)$. As the position of this peak cannot be properly observed in the top panels of Fig.~\ref{fig:m2phi2-GW}, we have plotted the same spectra in the bottom panels, zooming in the last stages of GW production. The position $\kappa_{\rm p}$ clearly indicates the transition from short to large momenta, so that for $\kappa > \kappa_p$, the amplitude of the GW spectra starts decreasing significantly.  It constitutes therefore an estimate of the maximum momenta attained by the GW spectra, due to the population of UV modes outside the initial radius $\kappa \lesssim \kappa_M$, when the system becomes non-linear at $z \gtrsim z_{\rm br}$.

\begin{figure}
                      \includegraphics[width=7.3cm]{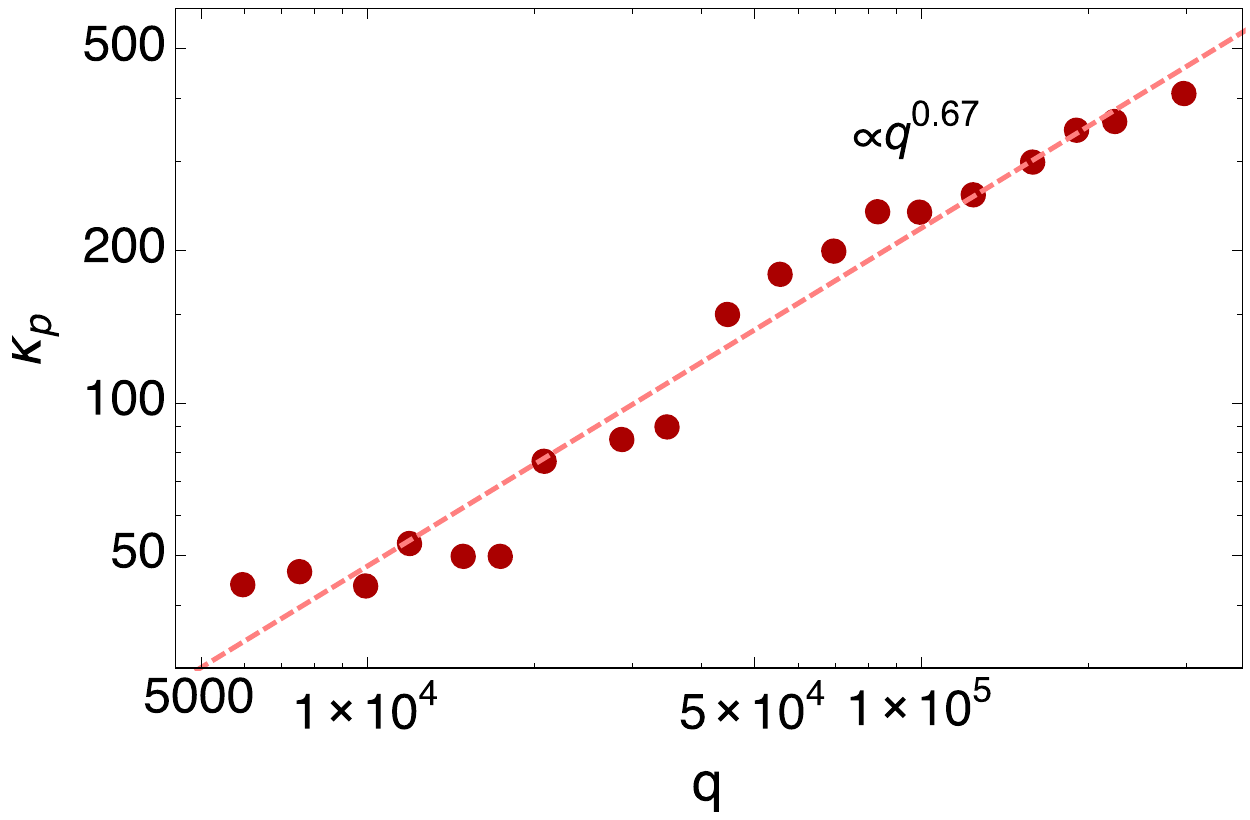} 
         \includegraphics[width=7.8cm]{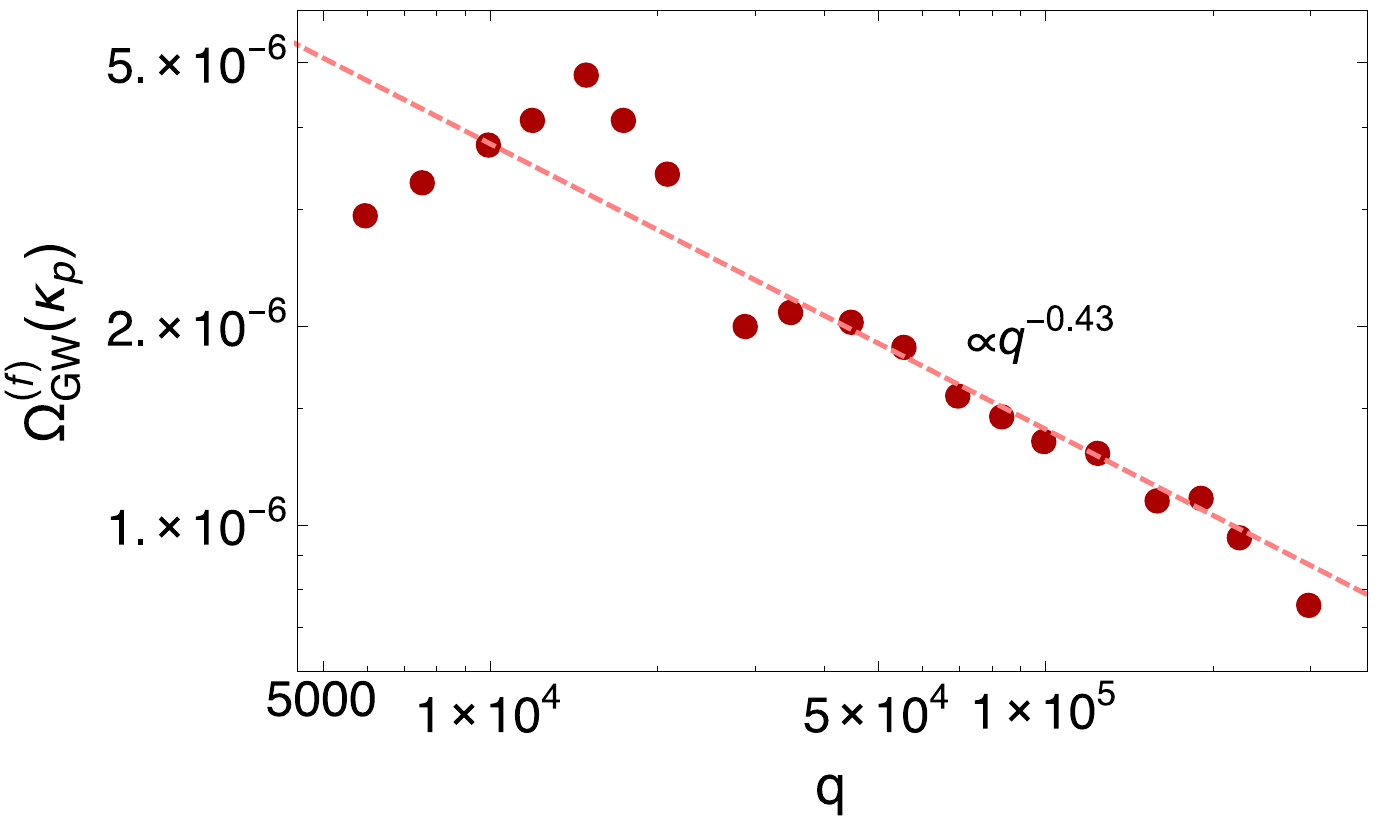} 
               \caption{We show, for the quadratic preheating model, the position of the peak $\kappa_p$ in the saturated GW spectra (bottom-left) as a function of $q$, as well as its corresponding amplitude $\Omega_{\rm GW}^{\rm (f) } (\kappa_p)$. Each point corresponds to a lattice simulation. The fits to both quantities [Eq.~(\ref{eq:fit-GWm2phi2})] are shown with dashed lines.}  \label{fig:m2phi2-GWfit}
 \end{figure}

In Fig.~\ref{fig:m2phi2-GWfit} we show the position $\kappa_p$ and amplitude $\Omega_{\rm GW} (\kappa_p)$ of the peak in the GW spectra, extracted from our lattice simulations for different values of $q$. We observe that as we increase $q$, the position of the peak $\kappa_p$ in the saturated spectra moves to the UV, while the amplitude of the peak decreases. We have found the following fits to the peak amplitude and position,
\be \kappa_{p} \approx 48  \left( \frac{q}{10^4} \right)^{0.67} \ , \hspace{0.5cm}  
\Omega_{\rm GW}^{\rm (f)} (\kappa_p ) \approx 3.8 \times 10^{-6} \left( \frac{q}{ 10^4 } \right) ^{-0.43} \ . \label{eq:fit-GWm2phi2}\ee
Let us remark that the fit for $q \lesssim 25000$ should be taken with a 'grain of salt', as the position of the peak is not so clearly distinguishable (given that the spectral amplitude flattens out). Not surprisingly, we see that the linear prediction for the peak position at $\kappa_p \sim q^{1/4}$ is not well verified, given that the location indicated in Eq.~(\ref{eq:fit-GWm2phi2}) corresponds to the final peak, measured after the system became non-linear and ceased to source GWs. The mentioned shift of power into shorter scales by the matter fields, translates into a new $q$-dependence of the peak position, which cannot be predicted with the linear theory, as it is the result of the non-linearities in the system. The scaling $\kappa_p \sim q^{2/3}$ reported in Eq.~(\ref{eq:fit-GWm2phi2}), can only be obtained with numerical simulations like ours. At the same time, the amplitude of the peak approaches very well the theoretical scaling predicted by the linear theory ${d \log \Omega_{\rm GW}\over d \log q} = -{1\over2}$, with a deviation of the measured exponent of only $100\times(|0.43-0.5|/0.5) = 14\%$. We believe the reason for this is that the scaling of the GW amplitude with $q$ is set during the linear stage, when the GWs grow exponentially fast due to the resonance of the daughter field. During the non-linear regime, the peak position is modified non-trivially from $\kappa_p \sim q^{1/4}$ to $\kappa_p \sim q^{2/3}$, but the amplitude receives only a boost that is independent of the resonance parameter $q$. This behavior is certainly remarkable, and certainly could not be anticipated by the linear theory.

Let us finally compute the GW spectra redshifted today. In this scenario, the post-inflationary expansion rate corresponds to a matter-dominated Universe~\cite{Turner:1983he}, as the inflaton energy density after averaging over its oscillations, behaves as $\rho_{\phi} \propto 1/a^3$. The equation of state is then $\omega \simeq 0$, so the redshifting factor from Section~\ref{sec:Analytics}, see Eq.~(\ref{eq:epsilonParameter}), becomes 
\begin{eqnarray}
\epsilon_i \equiv \left({a_{\rm i}\over a_{\rm RD}}\right) = \left({a_{\rm i}\over a_{\rm f}}\right)\epsilon_{\rm f}\,, ~~~~~~~~~{\rm with}~~~ \epsilon_{\rm f} \equiv \left({a_{\rm f}\over a_{\rm RD}}\right)\,.
\end{eqnarray}
From our simulations we measure directly the final time of GW production $t_{\rm f}$, and hence determine the pre-factor $(a_{\rm i}/a_{\rm f})$, which is typically of the order of $ \sim \mathcal{O} (10^{-2})$. Therefore, it is better to express the amplitude and frequency today, Eqs.~(\ref{eq:ftoday})- (\ref{eq:GWtoday}), in terms of $\epsilon_{\rm f}$,
\begin{eqnarray}\label{eq:GWphi2today}
f \simeq \epsilon_{\rm f}^{1/4}\left(\frac{k}{\rho_{\rm i}^{1/4}}\right) \times 2.5 \cdot 10^{9}~\mathrm{Hz}\,,~~~~~~~~ h^{2}\Omega_{\GW} \simeq  4\cdot 10^{-6} \epsilon_{\rm f} \times \Omega_{_{\rm GW}}^{({\rm f})} \ .
\end{eqnarray}
Plugging our fits in Eq.~(\ref{eq:fit-GWm2phi2}) into Eq.~(\ref{eq:GWphi2today}), we obtain
\bea \label{eq:m2phi2-finalfitFREQ}
f_p &=&  \epsilon_{\rm f}^{1/4} \left( \frac{q}{10^4} \right)^{0.67} \times 2.0 \cdot 10^{8} \ {\rm Hz}\ , \\
h^2 \Omega_{\rm GW} (f_p) &=&  \epsilon_{\rm f}  \left( \frac{q}{10^4} \right)^{-0.43} \times 1.5 \cdot 10^{-11}   \ , \hspace{0.3cm} (q \gtrsim 6 \cdot 10^3 )   \ . \label{eq:m2phi2-finalfit}
\eea
The longer the Universe takes to reach a RD stage, the smaller the factor $\epsilon_{\rm f}$ is. This means that the longer the post-inflationary matter-dominated expansion phase lasts, the more the GW peak moves to the IR, but the more suppressed its amplitude becomes.

Based on the numerical outcome, we can calibrate our analytical prediction of Section~\ref{sec:Analytics}. Knowing that $\omega_*^2 \equiv m^2$, 
and $\rho_i \simeq {1\over2}m^2\phi_i^2$, we can extract the parameters $C^2$ and $\delta$ characterizing the theoretical GW amplitude in Eq.~(\ref{eq:PeakAmplitudeToday}). 
In particular, equating
\begin{equation}
\Omega_{\rm GW}\big|_{\rm th} \simeq 2\cdot 10^{-11} \times \epsilon_f C^{2}\frac{m^{4}}{\phi_i^2 m_{p}^{2}}\,q^{-\frac{1}{2}+\delta} ~~~ = ~~~ \Omega_{\rm GW}\big|_{\rm num} \simeq \epsilon_f \left( \frac{q}{10^4} \right)^{-0.43} \times 1.5 \cdot 10^{-11}\,,
\end{equation}
we deduce
\begin{eqnarray}
\delta \simeq 0.06\,,~~~~~~{\rm and}~~~~~~ C \simeq 6.3\left(\frac{m_{p}\phi_i}{m^2}\right) \ .
\end{eqnarray}

 \section{Gravitational waves from parametric resonance in spectator field scenarios}\label{sec:SubFields}

As mentioned in the introduction, inflationary preheating is not the only case where parametric resonance can be developed in the early Universe. If a light spectator field is present during inflation, it will form a homogeneous condensate with a large amplitude, due to its quantum fluctuations. Following inflation, the condensate amplitude will oscillate around the minimum of its potential. The paradigmatic example of this is the {\it curvaton} scenario~\cite{Enqvist:2001zp,Lyth:2001nq,Moroi:2001ct,Mazumdar:2010sa}, where the curvaton field may decay via parametric resonance after inflation, transferring abruptly all its energy to the particle species coupled to it~\cite{Enqvist:2008be, Enqvist:2012tc, Enqvist:2013qba, Enqvist:2013gwf}. 

Another example of a spectator field, naturally decaying through parametric resonance after inflation, is the Higgs field of the Standard Model (SM). If the Higgs is weakly coupled to the inflationary sector, the Higgs is always excited with a large amplitude, either during inflation~\cite{Starobinsky:1994bd,Enqvist:2013kaa,DeSimone:2012qr}, or towards the end of it~\cite{Herranen:2015ima,Figueroa:2016dsc}. The Higgs is then 'forced' to decay into the rest of the SM species after inflation, via parametric resonance~\cite{Enqvist:2013kaa,Enqvist:2014tta,Figueroa:2014aya,Kusenko:2014lra,Figueroa:2015rqa,Enqvist:2015sua,Figueroa:2016dsc}. 

Before we specifically estimate the GW amplitude from parametric resonance due to a spectator field, let us recall that the EOM of the GWs Eq.~(\ref{eq:GWeom}), can be written symbolically as
\begin{eqnarray}\label{eq:GWsymbolicEOM}
\Box h_{**} = {2\over m_p^2}\Pi_{**}^{\rm TT}\,,~\hspace*{1cm} \Pi_{**}^{\rm TT} = \left\lbrace{\partial\phi\over\partial x^*}{\partial\phi\over\partial x^*}\right\rbrace^{\rm TT}\,,
\end{eqnarray}
where $\phi$ is some field involved in the process of parametric resonance [one can imagine a sum over fields in the $rhs$ of Eq.~(\ref{eq:GWsymbolicEOM})]. For the shake of the argument, we will first consider $\phi$ as the mother field. The latter will only start oscillating after inflation (with initial amplitude $\Phi_i$), when its (effective) mass becomes of the order of the Hubble rate $\sim H$. We can then 'parametrically' re-scale the source of GWs as 
\begin{eqnarray}\label{eq:GWsymbolicEOMII}
\Pi_{**}^{\rm TT} \sim H^2\Phi_i^2\times\left\lbrace{\partial\varphi\over\partial y^*}{\partial\varphi\over\partial y^*}\right\rbrace^{\rm TT} \sim H^2\Phi_i^2\,,
\end{eqnarray}
with $\vec y \equiv H\vec x$, and $\varphi \equiv \phi/\Phi_i$, and where we have (crudely) estimated that $\lbrace{\partial\varphi\over\partial y^*}{\partial\varphi\over\partial y^*}\rbrace^{\rm TT} \sim (\Delta \varphi/\Delta y)^2 \sim \mathcal{O}(1)$, as within a spatial scale $\Delta y \sim \mathcal{O}(1)$, the field amplitude typically oscillates (in real space), and hence $\Delta\varphi \sim \varphi \sim \mathcal{O}(1)$. As crude as our estimation of $\lbrace{\partial\varphi\over\partial y^*}{\partial\varphi\over\partial y^*}\rbrace^{\rm TT}$ might be, this does not change the fact that the amplitude of the source of the GWs is parametrically controlled by $\propto H^2\Phi_i^2$. Thus, in order to estimate the GW production from a spectator field, we need to determine first the typical amplitude $\Phi_i$ of such a field.

The amplitude of a spectator field excited during (pure $de~Sitter$) inflation is~\cite{Linde:2005ht}
\begin{eqnarray}
\left\langle \phi^2 \right\rangle = {3H^2\over 8\pi^2}\left({H\over m}\right)^2\left(1-\exp\left\lbrace -{2\over3}{m^2\over H^2}N\right\rbrace\right)~~\longrightarrow~~\left\lbrace \begin{array}{l}
{H^2\over 4\pi^2}N\,,~N \ll {H^2\over m^2}\vspace*{0.2cm}\\
{3H^4\over 8\pi^2m^2}\,,~N \gg {H^2\over m^2}
\end{array}\right.\,,
\end{eqnarray}
where we have implicitly assumed the initial field amplitude (say at the onset of inflation) to be zero, and the mass to be bounded as $0 \leq m \ll H$. The typical amplitude of a strictly massless spectator field is then of the order of $\phi_{\rm rms} \simeq \mathcal{O}(1)H(N/100)^{1/2}$. In other words, $\phi_{\rm rms} \sim H$, modulo some mild dependence on the number of e-folds. For a massive but light field with $m < H$, after a (typically large) number of efolds $N \gg (H/m)^2$, the spectator field reaches its saturation amplitude $\phi_{\rm rms} \rightarrow \mathcal{O}(0.1) (H/m) H$. 

Unless $N \ggg 1$ and $m/H \lll 1$, it is fair to say that the typical amplitude of a spectator field is, roughly speaking, $\phi \sim H$. Using this fact, and Eq.~(\ref{eq:GWsymbolicEOMII}), we conclude that the source of GWs, in the case of a spectator field (with initially vanishing amplitude), is bounded as $\Pi_{**}^{\rm TT} \lesssim H^4$ (modulo some mild dependence on the number of efolds). Let us note that this is an upper bound because in reality, the initial source of GWs in any process of parametric resonance, are the daughter field(s) rather than the mother field. The mother field typically contributes to the GW production when it finally develops sizeable time-dependent gradients. This happens when the daughter field backreacts over the mother field, manifesting the truly non-linear nature of the field dynamics due to the coupling between the field species. The daughter fields, however, never become significantly more energetic than the mother field (see {\it Paper I} for a discussion about this). Therefore, even though the parametrization of the GW source due to the daughter field(s) may differ from Eq.~(\ref{eq:GWsymbolicEOMII}), Eq.~(\ref{eq:GWsymbolicEOMII}) should still represent a good estimation of an upper bound for the GW source in a process of parametric resonance. 

As the energy density spectrum of GWs is proportional to $(\Pi_{**}^{\rm TT})^2$, see Eqs.~(\ref{eq:GW_spectra(Pi)}), (\ref{eq:UTC}), we can estimate now the GW production of fields in parametric resonance when the mother field is a spectator field. More specifically, we can parametrically compare it to the GW production when parametric resonance is due to the oscillations of an inflaton. In the latter case, the estimation Eq.~(\ref{eq:GWsymbolicEOMII}) also applies, though in this occasion the typical amplitude of the inflaton at the end of (large-field) inflation is $\Phi_* \sim m_p$. The GW source of parametric resonance during inflationary preheating, using Eq.~(\ref{eq:GWsymbolicEOMII}), is then bounded as $\Pi_{**}^{\rm TT} \lesssim m_p^2H^2$. The ratio of the GW energy density produced by parametric resonance due to the oscillations of a spectator field, $\Omega_{\rm GW}^{\rm (s)}$, to the GW energy density created (for the same daughter-mother coupling) by the oscillations of an inflaton, $\Omega_{\rm GW}^{\rm (i)}$, can be parametrically estimated as
\begin{eqnarray}\label{eq:StoIcomparison}
{\Omega_{\rm GW}^{\rm (s)} \over \Omega_{\rm GW}^{\rm (i)}} \sim {(\Pi_{**}^{\rm (s)})^2\over(\Pi_{**}^{\rm (i)})^2} \sim \left({H\over m_p}\right)^4 \ll 1
\end{eqnarray}
The GW production due to parametric resonance from a spectator field (with initially vanishing amplitude), can then only be much smaller than that of the analogous production from the parametric resonance of an inflaton field (with a large amplitude like in single-field slow-roll inflation). This result is actually expected, as the typical energy of a spectator field is always very sub-dominant compared to the inflaton energy. If the fraction of energy converted in GWs in the process of parametric resonance is fixed by the daughter-mother coupling, it is therefore natural to expect that the absolute GW production from the parametric resonance of a spectator field, is very sub-dominant as compared to the analogous GW production from an inflaton field, as the latter exceeds the energy budget of the spectator field.

This explains, for instance, the small amplitude of the GW background due to the decay of the SM Higgs after inflation~\cite{Figueroa:2014aya,Figueroa:2016ojl}, when the latter is considered to be an inflationary spectator field. In the case of Ref.~\cite{Figueroa:2014aya} the GW production was rather due to the parametric excitation of the SM fermions coupled to the Higgs, whereas in the case of Ref.~\cite{Figueroa:2016ojl}, the GWs were due to the parametric resonance excitation of the electroweak gauge fields. The energy arguments provided above, remain however valid: the GW amplitude from the parametric excitation of daughter field(s) coupled to an oscillating spectator field, is still very suppressed independently of the daughter field(s) spin. As as example we quote below the results from the SM Higgs spectator scenario~\cite{Figueroa:2016ojl}, based on the resonance of the electroweak $Z$ and $W^\pm$ gauge bosons. The final redshifting depends on the inflationary Hubble rate $H$, and on the initial Higgs amplitude parametrized as $\beta \equiv \sqrt{\lambda}|\Phi_i|/H$, with $\lambda$ the Higgs self-coupling. It also depends on the unknown post-inflationary expansion rate, characterized by an equation of state $w$. As the Higgs potential is quartic, the structure of the GW peaks is analogous to the preheating case with $V(\phi) \propto \phi^4$ discussed in Section~\ref{sec:lphi4-results}, i.e.~there are some IR peaks at fixed scales independent of $q$, and a 'hunchback' peak monotonically growing with $q$, which typically represents the highest peak amplitude of the GW spectrum. The resonance parameter of the system is 
\begin{eqnarray}
q \equiv {(g_Z^2 + 2g_W^2)\over 4\lambda}\,,
\end{eqnarray} 
where $g_Z^2, g_W^2$ are the gauge couplings of the SU(2) gauge bosons of the Standard Model. The frequency and amplitude of highest peak today is found to be~\cite{Figueroa:2016ojl}
\begin{gather}
f_p \simeq \epsilon_{i}^{1/4} \left(\frac{H}{H_{\rm max}}\right)^{1\over2}\,\left(\frac{\beta}{0.01}\right)^{p(w)} q^{r}~\times~10^{7}~{\rm Hz} \ , \label{eq:SpectFinalFreqPeak}\\
h^{2}\Omega_{\GW}^{\rm (o)}(f_p) \sim 10^{-33}\times \epsilon_{i}\,\left(\frac{q}{100}\right)^{1.5}\left(\frac{H}{H_{\rm max}}\right)^{4}\left(\frac{\beta}{0.01} \right)^{4+v(w)} \ , \label{eq:SpectFinalAmplitudePeak}
\end{gather}
where we have normalized the Hubble rate to its current upper bound $H \lesssim H_{\rm max} \simeq 8.5\cdot 10^{13}$ GeV~\cite{Ade:2015xua}, and defined
\be \lbrace\, p(w),v(w),r \,\rbrace \equiv \left\lbrace\,\frac{1 + 3 \omega}{3 (1 + \omega)},{2(3w-1)\over 3(1+w)},r\,\right\rbrace = \left\{ \begin{array}{cccccl}
        \lbrace\, {2/3} &, & {2/3} &, & 0.44 \,\rbrace & ~\mbox{for}~w = 1\\
        \lbrace\, {1/2} &, & 0 &, & 0.59 \,\rbrace & ~\mbox{for}~w = 1/3\\
        \lbrace\, 1/3 &, & -1/3 &, & 0.82 \,\rbrace & ~\mbox{for}~w = 0 \end{array} \right. \,.\nonumber
    \label{eq:k1fitb}  \ee
    
We quote below the cases when the expansion rate is radiation-dominated (RD, $w = 1/3$) and kinetion-dominated (KD, $w = 1$). When the inflationary Hubble rate saturates its upper bounds, $H = H_{\rm max} \simeq 8.5\cdot 10^{13}$ GeV~\cite{Ade:2015xua}, and we take a large initial Higgs amplitude $\beta = 0.1$, Ref.~\cite{Figueroa:2016ojl} obtains
\bea \label{eq:spectRD}
{\rm RD}: \,\,\,\,\, h^2 \Omega_{\rm GW} (f_p) &\lesssim & 10^{-29} \ , \,\,\,\,\,\,\,\,\, f_p \lesssim 3 \cdot 10^8 \,\,{\rm Hz} \ , \\
\label{eq:spectKDv2}
{\rm KD}: \,\,\,\,\, h^2 \Omega_{\rm GW} (f_p) &\lesssim & 10^{-16} \ , \,\,\,\,\,\,\,\,\, f_p \lesssim 3 \cdot 10^{11} \,\,{\rm Hz} \ .
\eea
For a matter-dominated ($MD$, $w \simeq 0$) universe, the GW amplitude is even more sub-dominant than in RD. For the KD case, $\epsilon_i$ becomes a boosting factor as $\epsilon_i \equiv (a_i/a_{\rm RD})^{1-3w} = (a_{\rm RD}/a_i)^2 \gg 1$, and hence the relatively large amplitude quoted in Eq.~(\ref{eq:spectKDv2}). However this also shifts the background towards higher frequencies, explaining the large frequency in Eq.~(\ref{eq:spectKDv2}). For smaller Hubble rates, these amplitudes are suppressed as $\propto (H/H_{\rm max})^4$, making the final amplitude even smaller. For more details, see~\cite{Figueroa:2016ojl}. For a spectator field with a quadratic potential $V(\phi) \propto \phi^2$ we also expect very tiny GW amplitudes, so we find pointless to make an analogous parameter fit study for a quadratic spectator field. 

As a last remark, let us perhaps note that in the case of a curvaton, the amplitude of the GWs due to its decay via parametric resonance, may be larger than the estimation given by Eq.~(\ref{eq:StoIcomparison}). The reason is that a curvaton field can have an amplitude $\phi_*$ during inflation, larger than its fluctuations $\delta \phi \sim H$. Typically one requires the curvaton~\cite{Enqvist:2001zp,Lyth:2001nq,Moroi:2001ct,Mazumdar:2010sa} to have an amplitude and a mass bounded as $\phi_*/m_p < 1$ and $m/H < \mathcal{O}(0.1)$. Thus, the curvaton energy could be, in principle, smaller than that of the inflaton, but not extremely suppressed. It is therefore conceivable that the process of parametric resonance due to the oscillations of a curvaton field after inflation, may produce a larger background of GWs, than the parametric resonance from an arbitrary spectator field with initially vanishing amplitude at the onset of inflation. However, this is only expected for somehow extreme values in the parameter space of the curvaton scenario, and in any case such background will still be sub-dominant compared to the GW background from a parametric resonance in inflationary preheating.

\section{Collection of fitted formulas} \label{sec:Summary}

We collect in this section the fitted formulas obtained from our numerical simulations. They constitute one of the the main results of our present work. This way, the interested reader can find rapidly any necessary equation and quote it, without the necessity of going through all the details in the paper. In this work, we have considered preheating scenarios with quartic potential $V (\phi) = \frac{\lambda}{4} \phi^4$ and quadratic potential $V (\phi) = \frac{m^2}{2} \phi^2$. We take the parameters $\lambda$ and $m$ as fixed by the amplitude of the CMB anisotropies, see discussion below Eq.~(\ref{eq:inflation-potentials}). In all scenarios, we have considered the mother field, i.e.~the inflaton, coupled to the daughter field with coupling $g^2\phi^2X^2$. We list the frequencies and amplitudes of the maxima (peaks) of the redshifted spectrum today.
 
\begin{itemize}
\item \textbf{Preheating with inflationary potential $V (\phi) = \frac{1}{4} \lambda \phi^4$}:

Results are given as a function of the resonance parameter, defined as $q\equiv \frac{g^2}{\lambda}$, with $\lambda = 9 \cdot 10^{-14}$ and $q > 1$. Several peaks appear located at frequencies
\bea
f_1 &\approx & 1.5 \cdot 10^7 \ {\rm Hz} \ ,  \\
f_2 & \approx & 2.8 \cdot 10^7 \  {\rm Hz} \ , \\
 f_{\rm hb} &\approx  &   \left( \frac{q}{100} \right)^{0.54} \times 5.3 \cdot 10^7 \ {\rm Hz}  \ , \hspace{0.4cm} ({\rm only\,\,for\,\,} q \gtrsim 1000) \ ,
\eea
and with an amplitude
\bea   \label{eq:BoundedIRtodayPhi4v2}
  3.4 \cdot 10^{-12} \left( \frac{q}{100} \right)^{-0.42} &\lesssim & h^2 \Omega_{\rm GW} (f_{1,2}) \lesssim 2.4 \cdot 10^{-11}  \left( \frac{q}{100} \right)^{-0.56}  \ , \\
  \label{eq:BoundedHBtodayPhi4v2}
  3.4 \cdot 10^{-12}  \left( \frac{q}{100} \right)^{-0.68} & \lesssim & h^2 \Omega_{\rm GW} (f_{\rm hb}) \lesssim 1.6 \cdot 10^{-11}  \left( \frac{q}{100} \right)^{-0.94}  \ . \hspace{0.3cm} 
\eea 
An additional IR peak appears in simulations with $q \gtrsim 1500$, with frequency $f_3 \approx 5 \cdot 10^7 {\rm Hz}$, and a sub-dominant amplitude $h^2 \Omega_{\rm GW} (f_{3}) \simeq 4\cdot 10^{-13}$.

\item \textbf{Preheating with inflationary potential $V (\phi) = \frac{1}{2} m^2 \phi^2$}:

The results for this model are given in terms of the initial resonance parameter $q \equiv \frac{g^2 \phi_{\rm i}^2}{4 m^2}$. Only one peak appears in the GW spectrum, at a frequency and amplitude
\bea 
f_p &=&   \epsilon_{\rm f}^{1/4} \left( \frac{q}{10^4} \right)^{0.67} \times 2.0 \cdot 10^{8} \ {\rm Hz}\,, \\
h^2 \Omega_{\rm GW} (f_p) &=&  \epsilon_{\rm f}  \left( \frac{q}{10^4} \right)^{-0.43} \times 1.5 \cdot 10^{-11}   \ , \hspace{0.3cm} (q \gtrsim 6 \cdot 10^3 )   \ , \label{eq:m2phi2-finalfitv2}
\eea
where the ratio $\epsilon_{\rm f} \equiv (a_{\rm f} / a_{_{\rm RD}})^{1-3w}$ quantifies the unknown period (with equation of state $w$) between the end of GW production and the onset of a RD universe.

\item \textbf{Spectator fields}: 

As discussed in Section~\ref{sec:SubFields}, the GW background generated during the oscillations of a spectator field, is very subdominant with respect the background expected from preheating. We list below the results for the Standard Model Higgs taken from~\cite{Figueroa:2016ojl}. The final redshifting depends on the post-inflationary expansion rate, which can be characterized by an unknown equation of state $w$. We quote here the cases when the expansion rate is radiation-dominated (RD, $w = 1/3$) and kinetion-dominated (KD, $w = 1$). When the inflationary Hubble rate saturates its current upper bound $H \lesssim H_{\rm max} \simeq 8.5\cdot 10^{13}$ GeV~\cite{Ade:2015xua}, and the Higgs amplitude is large as $\Phi_i \simeq 0.1 H/\sqrt{\lambda}$ (where $\lambda$ is the Higgs self-coupling), Ref.~\cite{Figueroa:2016ojl} obtains for the highest peak amplitude
\bea \label{eq:spectRD}
{\rm RD}: \,\,\,\,\, h^2 \Omega_{\rm GW} (f_p) &\lesssim & 10^{-29} \ , \,\,\,\,\,\,\,\,\, f_p \lesssim 3 \cdot 10^8 \,\,{\rm Hz} \ , \\
\label{eq:spectKD}
{\rm KD}: \,\,\,\,\, h^2 \Omega_{\rm GW} (f_p) &\lesssim & 10^{-16} \ , \,\,\,\,\,\,\,\,\, f_p \lesssim 3 \cdot 10^{11} \,\,{\rm Hz} \ .
\eea

For a matter-dominated universe ({\rm MD}, $w \simeq 0$), the GW amplitude is even more sub-dominant than in RD. For the KD case, $\epsilon_i$ becomes a boosting factor as $\epsilon_{\rm i} \equiv (a_i/a_{\rm RD})^{1-3w} = (a_{\rm RD}/a_{\rm i})^2 \gg 1$, and hence the relatively large amplitude quoted in Eq.~(\ref{eq:spectKD}). However this also shifts the background towards higher frequencies, and hence the high frequency quoted. For smaller Hubble rates, these amplitudes are suppressed as $\propto (H/H_{\rm max})^4$, making the final amplitude even more tiny. For details, see~\cite{Figueroa:2016ojl}.
\end{itemize}

\section{Discussion}
\label{sec:discussion}

Preheating in the early Universe is expected to generate a large amount of gravitational waves (GWs), see e.g.~\cite{Khlebnikov:1997di,Easther:2006gt,Easther:2006vd,GarciaBellido:2007af,Dufaux:2007pt,Dufaux:2008dn,Figueroa:2011ye,Bethke:2013aba,Bethke:2013vca}. The non-equilibrium dynamics of the fields after inflation develop energy gradients, which source very efficiently tensor perturbations. When the fields relax into a stationary state, the GW production ceases, and GWs decouple and travel freely ever since, redshifting until now. One of the most paradigmatic situations is when the inflaton field exhibits a monomial potential as $V(\phi) \propto \phi^{n}$ after the end of inflation. Following the end of inflation, when the inflaton (mother field) oscillates around the minimum of its potential, it provides a non-adiabatic time-dependent mass to all species (daughter fields) coupled to it. As a result, the fluctuations of such species grow exponentially in the process known as parametric resonance. This sources a significantly large background of GWs.

In this work we have studied and parametrized the production of GWs during parametric resonance in standard preheating scenarios. The dynamics of the matter fields is characterized in terms of the dimensionless resonance parameter $q$, which depends on the coupling strength, as well as on the initial amplitude and curvature potential of the mother field. In Section~\ref{sec:Analytics} we derived an analytical estimate based on the linear theory, for the position and amplitude of the main peak in the GW spectra. We find that the peak amplitude should scale theoretically as $\Omega_{\rm GW} \propto q^{-1/2}$. We then carried out in Section \ref{sec:lattice} lattice simulations of two main scenarios where parametric resonance takes place: preheating with quadratic $V(\phi) \propto \phi^2$ and quartic $V(\phi) \propto \phi^4$ potentials. We computed and parametrized the spectra of both GWs and matter fields, and confronted the numerical results with our theoretical formulae. In Section~\ref{sec:SubFields} we also discussed briefly the GW production from spectator-fields, showing that in general, the resulting GW background is always very sub-dominant with respect the one produced in preheating. We have collected all relevant formulas in Section~\ref{sec:Summary}.

In Section~\ref{sec:lphi4-results} we focused in the quartic case. We observed that there are two types of peaks imprinted in the GW spectra: infrared peaks located at fixed scales independently of $q$, and a higher frequency peak located at a scale $\kappa \sim q^{1/2}$. In all cases, the amplitude deviates from the theoretically linear expectation ${d\log\Omega_{\rm GW}\over d\log q} = -{1\over2}$, with a characteristic oscillatory pattern between $-0.42 \lesssim {d\log\Omega_{\rm GW}\over d\log q} \lesssim -0.94$, depending on the strength of the resonance (which is determined by $q$). See Eqs.~(\ref{eq:lphi4-fitAmp1})-(\ref{eq:lphi4-fitAmp2}). The amplitude decays in fact as a power-law with $q$, but faster than predicted by the linear theory; a behavior that could have not been anticipated {\it a priori} without numerical simulations capturing the non-linear dynamics of the system. In the range explored numerically of resonance parameters, $q \in [1, 5000]$, we find all peaks at around $f_p \approx \mathcal{O} (10^7) - \mathcal{O} (10^8) \ {\rm Hz}$, and the amplitude today as $h^2 \Omega_{\rm GW} \approx \mathcal{O} (10^{-11} ) - \mathcal{O} (10^{-13})$. See Eqs.~(\ref{eq:BoundedIRtodayPhi4})-(\ref{eq:BoundedHBtodayPhi4}). In Section~\ref{sec:m2phi2-results} we focused in the quadratic case. In this scenario we observe just a single peak in the GW spectrum, with an amplitude scaling with the resonance parameter as $\propto q^{-0.43}$. This behavior is in relative good agreement with the analytical estimate $\propto q^{-1/2}$, with only a $\sim 14\%$ deviation of the power-law index. This is a remarkable result, given the fact that the analytical prediction is based on the linear regime, whereas the numerical outcome is obtained after the field dynamics became non-linear. See Eqs.~(\ref{eq:m2phi2-finalfitFREQ}),(\ref{eq:m2phi2-finalfit}). The final position and amplitude of the spectrum today are however more uncertain than in the quartic case, as there is a dependence on the unknown duration of the period following the end of GW production, during which the universe maintains an expansion rate different than RD. Assuming that such period does not last for long after GW generation ceases, the redshifted amplitude can reach amplitudes today up to $h^2 \Omega_{\rm GW} \approx \mathcal{O}(10^{-11})-\mathcal{O}(10^{-13})$ (for the simulated range $6000 \lesssim q \lesssim 2.5 \cdot 10^6$). For larger $q$'s, as the amplitude decays as $\sim q^{-1/2}$, the signal becomes weaker and weaker.

One of the most remarkable results we have obtained, is precisely the fact that the peak amplitudes of the GW background decay with increasingly larger resonance parameters $q$. Naively, one would expect the opposite, as the larger the $q$, the broader the resonance. However, as explained in Section~\ref{sec:Analytics}, although more (daughter field) modes are excited for larger values of $q$, there is also less power transferred per mode: the daughter field spectrum may be wider, but it is also lower in amplitude. The two effects combine in such a way, that both the spectra of the fields, and of the GWs, decrease in amplitude with increasingly bigger values of $q$. This is to be contrasted with the case when the daughter fields experiencing a parametric excitation (due to the oscillations of some coherent field) are either gauge fields~\cite{Bezrukov:2008ut,GarciaBellido:2008ab,Bezrukov:2014ipa,Dufaux:2010cf,Deskins:2013lfx,Adshead:2015pva,Figueroa:2015rqa,Enqvist:2015sua,Lozanov:2016pac,Figueroa:2016dsc} or fermionic species~\cite{Greene:1998nh,Greene:2000ew,Peloso:2000hy,Berges:2010zv,Enqvist:2012im,Figueroa:2013vif,Figueroa:2014aya}. For both gauge fields and fermions, it is found that the corresponding GW background scales as $\Omega_{\rm GW} \sim q^{3/2 + \delta}$~\cite{Figueroa:2016ojl,Figueroa:2013vif,Figueroa:2014aya}, with $\delta \ll 1$ some small correction. In the case of gauge bosons this can be easily explained: even though they experience the same dynamics\footnote{This is demonstrated explicitly in Ref.~\cite{Figueroa:2015rqa} for Abelian gauge fields. Non-Abelian gauge fields may however exhibit a different behavior given the intrinsic non-linearities of the non-Abelian gauge structure of the interactions.} as scalar fields when coupled to an oscillatory (homogeneous) field, their anisotropic stress (i.e.~the source of GWs) has a different structure than in the scalar field case. One can show in fact, that analytical calculations based on a linear analysis, similar to those displayed in Section~\ref{sec:Analytics}, lead theoretically to expect $\Omega_{\rm GW} \sim q^{3/2}$. In the case of fermionic daughter fields, the theoretical analysis also predicts that $\Omega_{\rm GW} \sim q^{3/2}$~\cite{Figueroa:2013vif,Figueroa:2014aya}. The reason in this case is different than for the gauge fields: as fermion fluctuations are Pauli blocked, for larger values of $q$ there is a larger range of fermion modes excited, but this does not imply a lowering of the power per mode, as the spectral amplitude is typically saturated by the exclusion principle. Besides, the anisotropic stress from fermions has also a different structure than for scalar fields. All together, a scaling as $\Omega_{\rm GW} \sim q^{3/2}$ emerges, see~\cite{Figueroa:2013vif} for a detailed derivation.

As a final remark, let us note that there are scenarios of preheating where our analysis cannot be applied. The case of trilinear or non-renormalizable interactions between the mother and the daughter field(s)~\cite{Dufaux:2006ee,Croon:2015naa,Antusch:2015vna,Enqvist:2016mqj}, are not captured well by our fitted formulae. The case of oscillations of a multi-component field is neither captured by our analysis\footnote{In the case of super-symmetric flat directions, it may well happen that the flat directions are never really excited in first place~\cite{Enqvist:2011pt}, and therefore it makes no sense to speak about oscillations after inflation.}, see e.g.~\cite{Tkachev:1998dc,Olive:2006uw,Gumrukcuoglu:2008fk,DeCross:2015uza,Ballesteros:2016euj,Ballesteros:2016xej}. Besides, there are also scenarios where the mechanism responsible for the particle production is not parametric resonance, e.g.~tachyonic preheating~\cite{Felder:2000hj,Felder:2001kt,Copeland:2002ku,GarciaBellido:2002aj,GarciaBellido:2007dg,GarciaBellido:2007af,Dufaux:2008dn,DiazGil:2007dy,DiazGil:2008tf,Dufaux:2010cf,Tranberg:2017lrx}, in which case our analysis does obviously not apply.

\acknowledgments
We thank Juan Garc\'ia-Bellido for collaboration on related projects. This work is supported by the Research Project of the Spanish MINECO  FPA2015-68048-C3-3-P and the Centro de Excelencia Severo Ochoa Program SEV-2016-0597. F.T. is supported by the FPI-Severo Ochoa Ph.D. fellowship SVP-2013-067697. We acknowledge the use of the IFT Hydra cluster for the development of this work.

\appendix

\section{Lattice formulation} \label{app:Lattice-formulation}
In this work, we have solved the discrete field equations of motion in lattice cubes of different sizes. Let us denote the length of the cube by $L$, so that the volume is $V = L^3$, and the number of points per length dimension by $N$. The separation between lattice points is then ${\rm dx} \equiv L /N$. The minimum and maximum momenta covered by such lattice is 
\be p_{\rm min } \equiv \frac{2 \pi}{L} \ , \hspace{0.5cm} p_{\rm max} = \frac{\sqrt{3} N}{2} p_{\rm min} \ . \label{eq:pminpmax}\ee

 \begin{figure}
 	\begin{center}
                      \includegraphics[width=7.3cm]{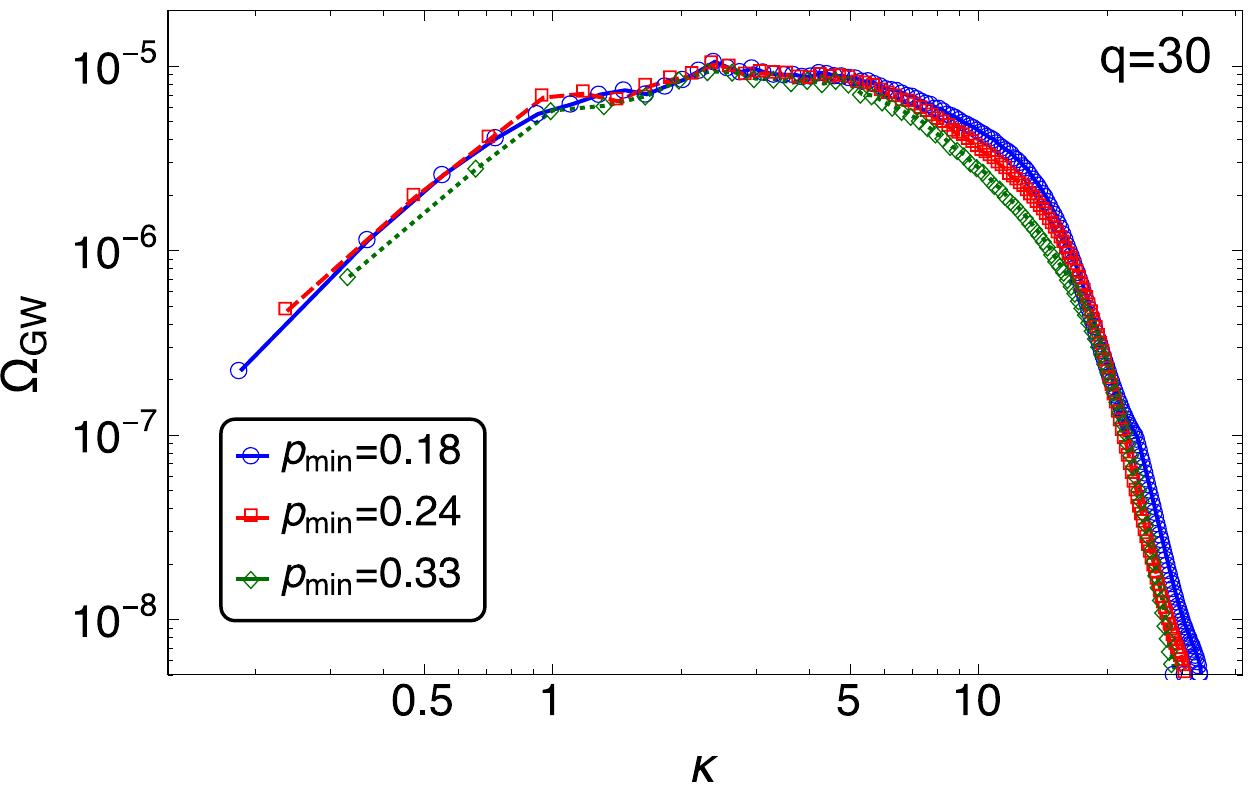} 
         \includegraphics[width=7.3cm]{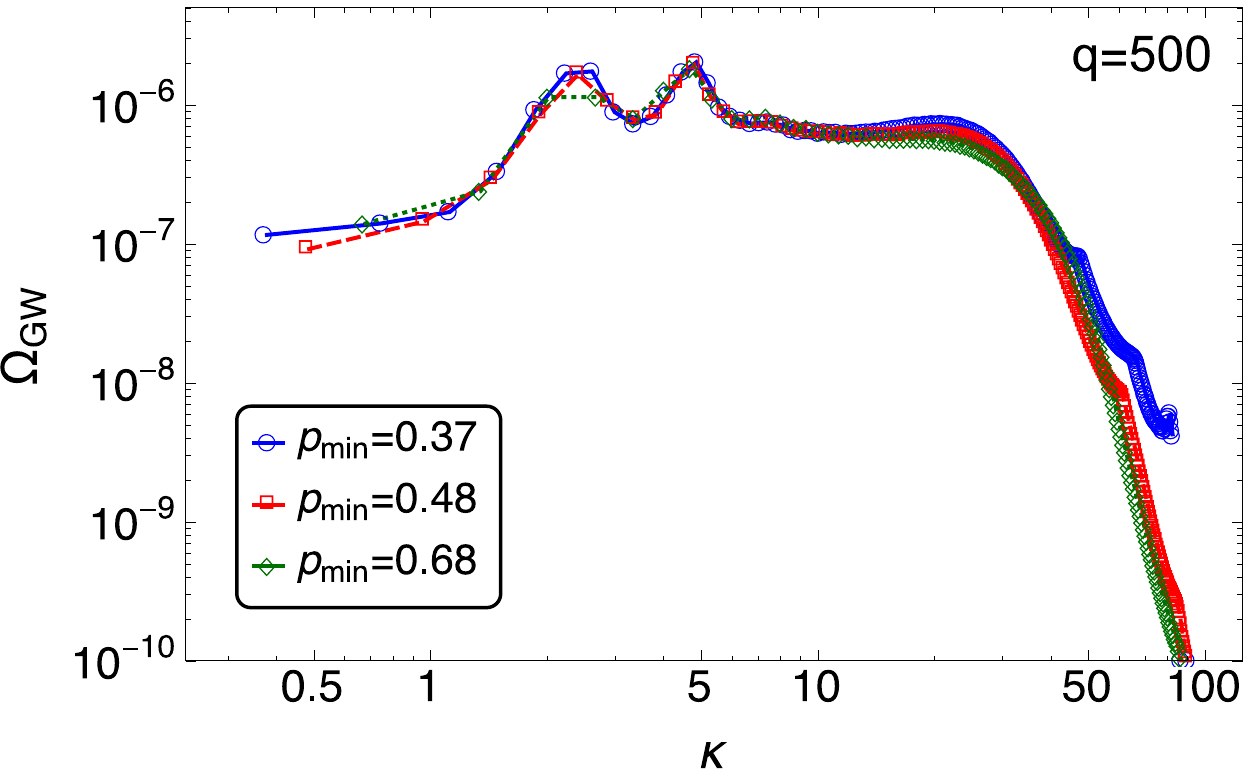} 
                               \includegraphics[width=7.6cm]{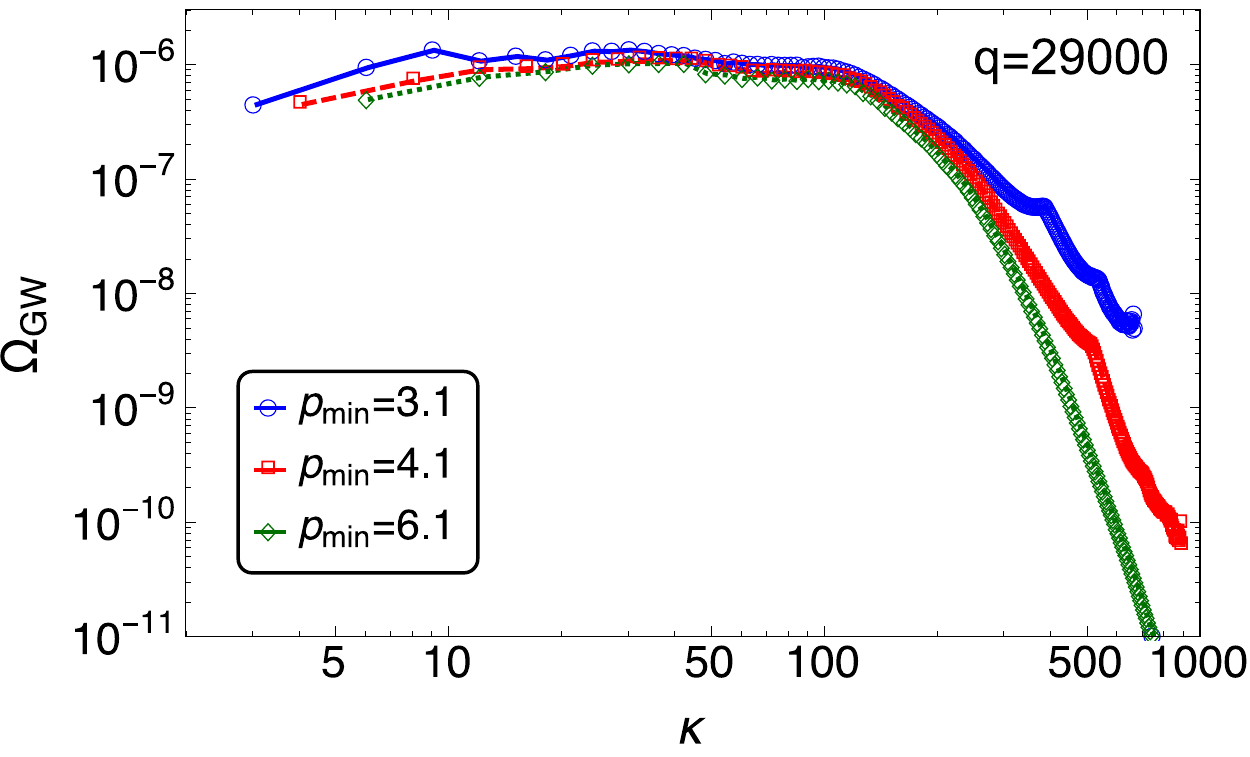} 
         \includegraphics[width=7.3cm]{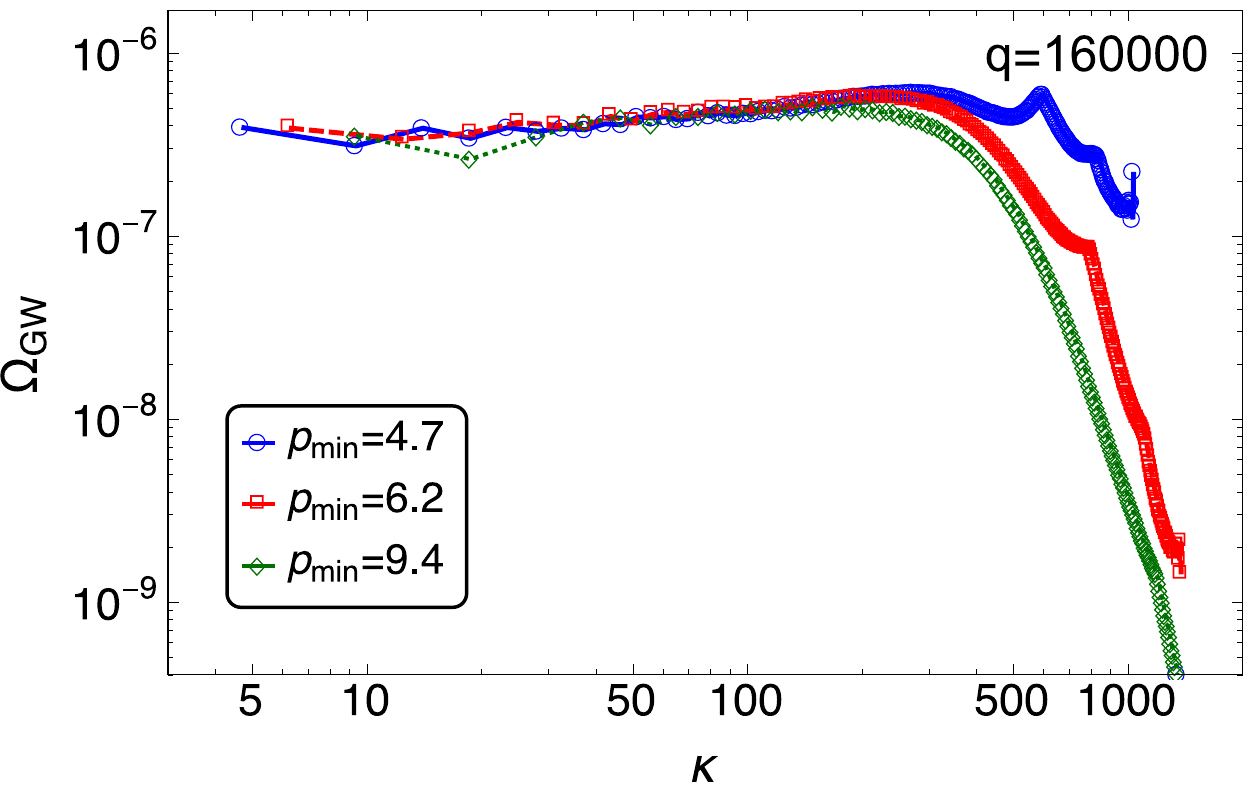} 
         \end{center}
               \caption{We compare the GW spectra obtained at a certain time for different lattice sizes. We take in all cases $N=256$, and show the spectra for different $p_{\rm min}$ in Eq.~(\ref{eq:pminpmax}). Top panels show, for the $\lambda \phi^4$ case, the GW spectra at time $z \approx 1500$, for the resonance parameters $q=30, 500$ [defined in Eq.~(\ref{eq:lphi4-qresdef})]. Bottom panels show, for the $m^2 \phi^2$ case, the GW spectra at time $z \approx 210, 270$, for the resonance parameters $q = 29000,160000$ [defined in Eq.~(\ref{eq:m2phi2-resq})].} \label{fig:lat-dependence}
 \end{figure}

In appendix B of {\it Paper I} we presented a detailed discussion about the technical aspects of the numerical simulation of the matter (mother and daughter) fields dynamics. We discussed the need to control well IR and UV scales in appendix B.1, the discretization technique for the lattice equations of motion in appendix B.2, and the initial conditions set-up for the field modes in appendix B.3. Therefore, we do not repeat that discussion here, and refer to the reader to appendix B of {\it Paper I}. We would simply like to emphasize here that the relevant features of the field dynamics must not be sensitive to the choice of $p_{\rm min}$ and $p_{\rm max}$. This is of course obtained if the relevant range of momenta for the dynamics of the system is well contained within $p_{\rm min} < k < p_{\rm max}$. In the parametric resonance cases analyzed in this work, this means simply two conditions. First, we must have $p_{\rm min} \lesssim \mathcal{O} (0.1) \kappa_{\rm +}$, with $\kappa_{\rm +}$ the maximum momenta of the daughter field's initial resonance band. Secondly, $p_{\rm max}$ must be large enough to cover well the propagation of power into the spectra at short scales (large momenta), triggered by the non-linearities of the system after the onset of backreaction. In \textit{Paper I}, we compared daughter field spectra for different choices of $(N, p_{\rm min})$, both quadratic and quartic preheating, varying $N$ from 128 to 256, and changing $p_{\rm min}$ by a factor of $\sim 2$. There, we observed that the main features of these spectra, namely the position and amplitude of the peaks, were quite insensitive to the choice of $(N, p_{\rm min})$ within the mentioned ranged, see Fig.~12 of {\rm Paper I}. 

Once the dynamical range is well captured for the matter fields, it follows naturally that the dynamical range for the GWs is also well captured, as the characteristic scales of the tensor modes are essentially the same as those of the matter fields. In order to demonstrate this explicitly, we show different GW spectra in Fig.~\ref{fig:lat-dependence}, for both quartic preheating (top panels) and quadratic preheating (bottom panels), varying in each panel the range of momenta covered by the lattice. We have fixed in all simulations $N=256$, and chosen three different values of $p_{\rm min}$ in each panel [and correspondingly three different values of $p_{\rm max}$, according to Eq.~(\ref{eq:pminpmax})]. We can observe in all cases, that all spectra agree with great accuracy in the position and amplitude of the peaks. This agreement is fulfilled even when the UV coverage is clearly insufficient, such as in the $p_{\rm min}=4.7$ case, at bottom-right panel. Therefore, this confirms the robustness of the numerical results presented in the main text.

We discuss now the prescription we used to obtain the spectrum of GWs in the lattice, following Ref.~\cite{Figueroa:2011ye}. The spectrum of the energy density of a (statistically) homogeneous and isotropic GW background in the continuum, in the limit of a very large volume $V$ encompassing all relevant wavelengths, is
\begin{eqnarray}\label{eq:GWrhoContSpectrum}
\frac{d\rho_{GW}}{d\log k} = \frac{m_p^2k^3}{8\pi^2 V} \int {d\Omega_k\over 4\pi}\,\dot h_{ij}^{\rm TT}(k,\hat\bk,t)\dot h_{ij}^{\rm TT^*}(k,\hat\bk,t)\,,
\end{eqnarray}
where $d\Omega_k$ represents a solid angle element in $\bk$-space, and $h_{ij}^{\rm TT}$ are transverse-traceless (TT) metric perturbations representing GWs.

In a lattice we simply need the volume $V = L^3$ ($L = Ndx$) to encompass sufficiently well the characteristic wavelengths of the simulated GW background. This is precisely equivalent to our previous discussion above, about the need to capture well the relevant modes within the range $[p_{\rm min},p_{\rm max}]$. To derive an analogous discrete expression to Eq.~(\ref{eq:GWrhoContSpectrum}) but valid in a lattice of volume $V = L^3$, we need first to specify our discrete Fourier transform (DFT) convention. We use
\begin{eqnarray}
f({\bf n}) = \frac{1}{N^3}\sum_{\tilde n}e^{-\frac{2\pi i}{N}\tilde{\bf n} {\bf n}}\,\tilde f(\tilde{\bf n})\,,\hspace*{1cm} \,\tilde f(\tilde{\bf n}) = \sum_{n}e^{+\frac{2\pi i}{N}\tilde{\bf n} {\bf n}} f({\bf n})\,,
\end{eqnarray}
where the index ${\bf n} = (n_1,n_2,n_3)$, with $n_i = 0,1,...,N-1$, labels our lattice sites in configuration space, whereas the index $\tilde{\bf n} = (\tilde n_1, \tilde n_2, \tilde n_3)$ labels the reciprocal lattice, with $\tilde n_i = -\frac{N}{2}+1, -\frac{N}{2}+2,...$ $-1,0,1,..., \frac{N}{2}$. Following Eqs.~(4.1)-(4.5) from Ref.~\cite{Figueroa:2011ye}, one arrives at
\begin{equation}\label{eq:GWrhoDiscreteSpectrum}
\left(\frac{d\rho_{GW}}{d\log k}\right)(\tilde{\bf n}) ~\equiv~ \frac{m_p^2|k(\tilde{\bf n})|^3}{8\pi^2\,L^3}\,\,
\left\langle\left[dx^3\dot h_{ij}^{\rm TT}(|\tilde{\bf n}|,t)\right]\left[dx^3\dot h_{ij}^{\rm TT}(|\tilde{\bf n}|,t)\right]^*\right\rangle\,,
\end{equation} 
where $\left\langle\dot h_{ij}^{\rm TT}(|\tilde{\bf n}|,t)\dot h_{ij}^{{\rm TT}^*}(|\tilde{\bf n}|,t)\right\rangle$ is an average over configurations with lattice momenta $\tilde{\bf n}' \in [\,|\tilde{\bf n}|,|\tilde{\bf n}|+\delta\tilde n\,]$. In the continuum limit, one identifies $\,DFT\lbrace f({\bf n})dx^3 \rbrace \rightarrow CFT\lbrace f({\bf x}) \rbrace$, where $DFT$ and $CFT$ stand for discrete and continuous Fourier transforms, respectively. The expression~(\ref{eq:GWrhoDiscreteSpectrum}) matches therefore, in the continuum limit, the expression~(\ref{eq:GWrhoContSpectrum}). Besides, expression~(\ref{eq:GWrhoDiscreteSpectrum}) highlights that the natural momenta in terms of which to express the lattice GW spectrum, is the discretized version of the continuum one ${\bf k} = \tilde{\bf n} k_{\rm IR}$, and not any of the lattice-momenta that one can defined based on the choice of a lattice derivative. As $\left(\frac{d\rho_{GW}}{d\log k}\right)(\tilde{\bf n})$ has dimension of $(energy)^4$, one can finally write the GW spectrum for each specific case of parametric resonance, in terms of its natural (dimensionless) variables $\kappa(\tilde{\bf n}) \equiv k(\tilde{\bf n})/\omega_*$, $d\tilde x = \omega_*dx$, $\tilde t = \omega_* t$,
\begin{equation}\label{eq:GWrhoDiscreteSpectrumII}
\left(\frac{d\rho_{GW}}{d\log k}\right)(\tilde{\bf n},\tilde t) ~\equiv~ \omega_*^2m_p^2\frac{|d\tilde x\kappa(\tilde{\bf n})|^3}{8\pi^2\,N^3}\,\,
\left\langle h_{ij}^{{\rm TT}'}(|\tilde{\bf n}|,\tilde t) {h_{ij}^{{\rm TT}'}}^*(|\tilde{\bf n}|,\tilde t)\right\rangle\,,
\end{equation} 
where $\omega_*$ is the specific natural frequency of oscillations of a given mother field's model, and $'$ denotes time derivatives with respect $\tilde t$. 

In the continuum, the transverse-traceless (TT) metric perturbations follow the equation $\Box h_{ij}^{\rm TT} = (2/m_p^2)\Pi_{ij}^{\rm TT}$. However, as originally observed in~\cite{GarciaBellido:2007af}, the TT perturbations can be obtained in Fourier space from a simple projection like $h_{ij}^{\rm TT}({\bf k},t) = \Lambda_{ij,lm}(\hat{\bf k})u_{ij}({\bf k},t)$, where $\Lambda_{ij,lm}(\hat{\bf k})$ is the standard TT-projector given in Eq.~(\ref{eq:projector}), and $u_{ij}({\bf k},t)$ is an auxiliary tensor perturbation, solution of the equation $\Box u_{ij} = (2/m_p^2)\Pi_{ij}$. Besides, given the property $\Lambda_{ij,lm}(\hat{\bf k})\Lambda_{lm,pq}(\hat{\bf k}) = \Lambda_{ij,pq}(\hat{\bf k})$, we can always express the main argument of the GW spectrum as $\dot h_{ij}^{\rm TT}({\bf k},t)\dot h_{ij}^{\rm TT^*}({\bf k},t)$ = $\dot u_{ij}({\bf k},t)\Lambda_{ij,lm}(\hat{\bf k})\dot u_{lm}({\bf k},t)$. 

In a discrete grid, however, one needs to be careful with the construction of a lattice projector $\Lambda_{ij,lm}^{(L)}$ that provides correctly a (lattice version) of the 'transversality' and 'tracelessness' conditions, whenever contracted with some tensor~\cite{Huang:2011gf,Figueroa:2011ye}. It turns out that constructing a correct lattice projector is not as trivial as one may think. Several lattice projectors can actually be built, one for each spatial discrete derivative one may imagine. In Ref.~\cite{Figueroa:2011ye} it is shown explicitly that different TT projectors can give rise to some discrepancies in the very UV part of the GW numerical spectrum. For the particular case of GWs from preheating driven by parametric resonance, Ref.~\cite{Figueroa:2011ye} showed that the total energy in the GW backgrounds computed there (integrating the spectrum over its Fourier modes), amounted only to $\sim \%$ differences. Therefore, as it is in principle irrelevant which lattice projector to use, we decided to obtain all our GW spectra with the projector
\begin{eqnarray}
\Lambda_{ij,lm}^{(L)}(\tilde{\bf n}) \equiv P^{(L)}_{il}(\tilde{\bf n})P^{(L)}_{jm}(\tilde{\bf n})-\frac{1}{2}P^{(L)}_{ij}(\tilde{\bf n})P^{(L)}_{lm}(\tilde{\bf n})\,,\\
P^{(L)}_{ij}(\tilde{\bf n}) = \delta_{ij} - \frac{k^{(L)}_{i}k^{(L)}_{j}}{|k^{(L)}|^2}\,,~~~~~ k^{(L)}_{i}= 
2\frac{\sin(\pi \tilde{n}_i/N)}{dx}\,,
\end{eqnarray}
based on a symmetric nearest-neighbors spatial derivative, see Eq.~(3.2) in Ref.~\cite{Figueroa:2011ye}. We then built the argument of the discrete energy density spectrum of GWs as ${{\dot h}_{ij}^{\rm TT}}(|\tilde{\bf n}|,t) {\dot h}_{ij}^{{\rm TT}^*}(|\tilde{\bf n}|,\tilde t)$ = ${\dot u}_{ij}(|\tilde{\bf n}|,t)\Lambda_{ij,lm}^{(L)}(\hat{\bf n}) {\dot u}_{lm}^{*}(|\tilde{\bf n}|,t)$, with $u_{ij}(|\tilde{\bf n}|,t)$ the Fourier transform of the solution to the discrete version of the equation $\Box u_{ij} = 2m_p^{-2}(\partial_i\phi\partial_j\phi + \partial_iX\partial_jX)$. 
\bibliography{FitGWParamRes}
\bibliographystyle{h-physrev4}

\end{document}